\documentclass{article}
\usepackage{titlesec}
\usepackage{titling}
\usepackage{graphicx} 
\usepackage{amsmath}
\usepackage[margin=1.0in]{geometry} 

\titleformat{\section}{\normalfont\large\bfseries}{\thesection}{1em}{}
\titleformat{\subsection}{\normalfont\normalsize\bfseries}{\thesubsection}{1em}{}
\titleformat{\subsubsection}{\normalfont\normalsize\itshape}{\thesubsubsection}{1em}{}

\titlespacing*{\section}{0pt}{1.5ex}{0.5ex} 
\titlespacing*{\subsection}{0pt}{1.5ex}{0pt}

\usepackage{chngcntr} 

\newcommand{\startSupplementary}{
    \setcounter{section}{0} 
    \setcounter{table}{0}   
    \setcounter{figure}{0}  
    \renewcommand{\thesection}{S\arabic{section}} 
    \renewcommand{\thetable}{S\arabic{table}} 
    \renewcommand{\thefigure}{S\arabic{figure}} 
}
\usepackage{tocloft}
\addtocontents{toc}{\cftpagenumbersoff{chapter}}
\addtocontents{toc}{\cftpagenumbersoff{section}}
 
\usepackage{float}

\usepackage{lscape}

\usepackage{palatino}

\usepackage{verbatim}

\usepackage{authblk}

\usepackage{hyperref}

\usepackage{soul}

\usepackage[numbers,sort&compress]{natbib}

\usepackage[labelfont=bf]{caption}
\usepackage{threeparttable}
\usepackage{threeparttablex}
\usepackage{booktabs}
\usepackage{array}

\usepackage{longtable}
\usepackage{bm}
\usepackage{xcolor}
\definecolor{DeclineRed}{HTML}{B84545}
\definecolor{IncreaseGreen}{HTML}{3D7A4A}
\definecolor{MixedBlue}{HTML}{4A6FA5}
\definecolor{StableGray}{HTML}{666666}
\definecolor{metgray}{HTML}{666666}
\newcommand{\nociteinv}[1]{}
\usepackage[bottom]{footmisc}

\title{Robust Evidence for Declining Disruptiveness: Assessing the Role of Zero-Backward-Citation Works\thanks{We thank the National Science Foundation (grant nos. 1829168, 1932596, 2318172, and 2449660 to R.J.F. and no. 1829302 to E.L),  Wellcome Leap Foundation (grants to R.J.F. and M.P.), and Alfred P. Sloan Foundation (grant no. G-2024-25123 to R.J.F.) for financial support of work related to this project. The funders had no role in study design, data collection and analysis, or preparation of the manuscript. For helpful comments, we thank Thomas Gebhart, Julia Lane, Jason Owen-Smith, Lutz Bornmann, Christian Leibel, Aks Zaheer, Joe Nahm, Kara Kedrick, Matt VanEseltine, Raviv Murciano-Goroff, Huquan Kang, Linhui Wu, Shuping Wu, Xiangting Wu, Yeon Jin Kim, Dennie Kim, and Zhen Ge. Address correspondence to R.J.F. (\href{mailto:rfunk@umn.edu}{rfunk@umn.edu}).}}
\author[1]{Michael Park}
\author[2]{Erin Leahey}
\author[3]{Russell J. Funk}
\affil[1]{{\small{Organisational Behaviour, INSEAD}}}
\affil[2]{{\small{School of Sociology, University of Arizona}}}
\affil[3]{{\small{Carlson School of Management, University of Minnesota}}}
\date{}

\begin{document}

\maketitle

\begin{abstract}
We respond to Holst et al.'s critique that the decline in scientific disruptiveness documented in Park et al.\ (\textit{Nature}, 2023) is an artifact of including works with zero backward citations. Using their advocated dataset, metric, and exclusion criteria, we find declines equivalent to major benchmark transformations in science. Their own regression model---designed to address their concerns about zero-citation works---yields large and significant declines for both papers and patents (p$<$0.001), a result found in their supplementary tables yet left unaddressed, despite directly contradicting their central claim. Their critique is further undermined by severe quality issues in their data, which contain three times more zero-citation works than ours. We trace this excess to their inclusion of at least 2.8 million editorials, obituaries, and comments, 1.5 million books and proceedings, and 254,000 product and artistic reviews---in all, 20\% of their sample is non-research content that almost by definition lacks backward citations. Simple keyword searches confirm the problem's severity, identifying among others 456 For Dummies guides, 50 Dr.\ Seuss and Curious George books, and the Captain Underpants series---all zero-citation entries in their sample. Applying granular document type classification to their data reveals that such non-research content fell from 40\% to 8\% of their sample between 1945 and 2010---a shift sufficient to generate the decline in zero-citation prevalence they attribute to metadata errors in our study. Standard practice excludes such content to guard against the metadata quality concerns at the center of their critique---concerns their dataset exemplifies rather than addresses. Declining disruptiveness has been documented in nearly 100 studies across multiple databases, metrics, and non-citation-based measures. The weight of evidence does not support an artifact-based explanation.
\end{abstract}

\pagebreak

We thank Holst et al.\ (HATWG) for their engagement with our (PLF) work. Their critique highlights important questions about the influence of citation practices and metadata quality on scientometric analyses. In our study, we shared these concerns and conducted many robustness checks to assess the impact of such factors. These included replications on multiple datasets, use of alternative disruptiveness measures, adjustments for shifting citation practices, and verification using non-citation metrics (e.g., text analysis). 

Since publication, evidence of declining disruptiveness has been documented in nearly 100 studies spanning multiple databases (e.g., Web of Science, OpenAlex, SciSciNet, Dimensions, PubMed, USPTO), disruption metrics, and domains (Table~\ref{table:LitReview})\cite{wu2026inventory}. Many of these studies directly evaluate robustness to the types of methodological concerns HATWG raise---including sensitivity to zero-citation works, citation inflation, and metadata quality---employing alternative measures and robustness checks designed to address such issues (e.g., \cite{lin2026disruption, li2025science, yu2025wp, wu2019large, li2025awp,li2024pnas,deng2025wp,zeng2023wp,shu2023stylization}). Declining disruption has also been documented using measures that do not rely on citation data at all, including text-based analysis, product similarity networks, and economic indicators of firm displacement \citep{bessen2020wp,bessen2022disruption,boot2025jar,bowen2019wp,peters2022wp,huang2024wp,he2024bwp,jeong2025s}---providing evidence against the possibility that the trend is an artifact of citation-based measurement. Notably, several studies document \emph{increasing} disruptiveness in young or rapidly evolving fields such as AI and fetal surgery \cite{wu2023joi, vaughn2025jps,tang2024ic,jiang2025jsp}, patterns consistent with theoretical expectations and inconsistent with a purely mechanical decline. While reasonable debate continues about measurement and interpretation, the weight of evidence on the central empirical trend does not support an artifact-based explanation.

HATWG argue that zero-backward-citation (0-bcite) works are a primary driver of the observed decline. We find no support for this hypothesis. First, using their advocated dataset, metric, and exclusion of 0-bcite works, we find statistically and practically significant declines that equal major benchmark transformations in science. These robust results hold using HATWG's own regression model, designed specifically to mitigate perceived biases of 0-bcite works.

Second, we identify fundamental flaws in HATWG's methodology and dataset that undermine their critique. Support for their hypothesis is based on visual inspection, with no assessment of statistical or practical significance. This is concerning, as their own regression model (shown in their Supplementary Information [SI], Table S1), which they designed to address perceived biases from 0-bcite papers, demonstrates a highly significant decline in disruptiveness for both papers and patents (p$<$0.01)---a finding that directly contradicts their central claim, yet that they do not acknowledge or interpret.

Moreover, their critique is undermined by severe quality issues in their SciSciNet data. Their data contain nearly three times more 0-bcite papers than the datasets used in our study. We sought to understand why. Because SciSciNet does not provide granular document type indicators, we linked HATWG's analytical sample to external bibliometric databases. Conservative estimates reveal that their sample includes at least 2.8 million editorial and commentary items (letters to the editor, obituaries, corrections, errata, news), 1.5 million books and proceedings, and 254,000 product and artistic reviews (Table~\ref{table:HATWGNonResearch}). In total, 20\% of their matched sample consists of non-research content---material that almost by definition contains no backward citations. Such non-research content dropped from 40\% to 8\% of their sample between 1945 and 2010---a shift sufficient to generate the decline in 0-bcite prevalence they attribute to metadata error in our study.

Even cursory inspection of titles confirms the severity of the problem. Simple keyword searches identified 456 \textit{For Dummies} guides (including \textit{Massage For Dummies}, \textit{Sushi For Dummies}, and \textit{Existentialism For Dummies}), 88 \textit{Complete Idiot's Guides}, 26 Dr.\ Seuss children's books including \textit{The Cat in the Hat Comes Back}, 24 \textit{Curious George} picture books, and works from the \textit{Captain Underpants}, \textit{Goosebumps}, and \textit{Berenstain Bears} children's series—all zero-backward-citation entries in HATWG's analytical sample (Table~\ref{table:HATWGNonScientific}). Standard scientometric practice excludes such content—children's books and \textit{For Dummies} guides do not cite prior literature. These practices exist precisely to ensure robustness against the metadata quality concerns at the center of HATWG's critique---concerns their own dataset exemplifies rather than addresses.

Below, we primarily consider the implications of 0-bcite works empirically. However, contrary to HATWG's characterization of 0-bcite works as `hidden outliers,' our inclusion of these works was deliberate and theoretically informed by established literature. Many foundational contributions contain no backward citations precisely because they opened new domains, derived results from first principles, or reported observations with no precedent to cite.\footnote{In patents, well-known examples include foundational biotechnology inventions such as the Ptashne patents on protein synthesis (\#4,332,892, \#4,418,149), methods for human growth hormone production (\#4,363,877), the Axel patent on cotransformation (\#4,634,665), the landmark \emph{Diamond v.\ Chakrabarty} patent (\#4,259,444), the Milstein-K\"{o}hler patent on monoclonal antibodies (\#4,172,124), and Caruthers' DNA synthesis method (\#4,458,066). In scientific papers, examples include the EPR paper on quantum entanglement \citep{einstein1935can}, Gabor's invention of holography \citep{gabor1948new}, Kanner's clinical definition of autism \citep{kanner1943autistic}, Fisher's foundational work in population genetics \citep{fisher1937wave}, Harlow's work on attachment in primates \citep{harlow1958nature}, and Skinner's experiments on conditioning \citep{skinner1948superstition}.} Wholesale exclusion of such works from analyses of disruptive innovation would be deeply problematic. We discuss these considerations in Sec.~\ref{sec:0BciteTheory}.

\addtocontents{toc}{\protect\setcounter{tocdepth}{0}}
\section{Practical Significance: Comparing Disruptiveness Trends to Major Benchmarks}
\addtocontents{toc}{\protect\setcounter{tocdepth}{2}}

To evaluate whether the practical significance of declining disruptiveness persists after excluding 0-bcite papers, we analyze trends using HATWG's advocated data source and metric---the precomputed disruption scores in SciSciNet \citep{lin2023sciscinet}, where 0-bcite papers are excluded. We then compare the magnitude of trends in disruptiveness to major benchmark transformations in science identified in Wang and Barabasi's \emph{Science of Science} textbook \citep{wang2021science}. Benchmarks include the age of cited references  \citep{evans2008electronic, lariviere2008long, chu2021slowed, verstak2014shoulders, sinatra2015century}, team members with career age $>$20 years \citep{cui2022aging, jones2011age, jones2010age, blau2017us, alberts2015addressing}, female team members \citep{nsf2021, huang2020historical}, countries per team \citep{wagner2017growth, lin2023remote, adams2013fourth, ribeiro2018growth, leydesdorff2008international, luukkonen1992understanding}, and team size \citep{wuchty2007increasing, jones2021rise, milojevic2014principles}. Metrics are percentile-normalized to allow comparison across measures with varying scales.

Even after excluding all 0-bcite papers, the decline in disruptiveness is evident across all fields (Fig.~\ref{fig:BenchmarkMeasures}) and is comparable in magnitude to major transformations in how science is conducted. Between 1945 and 2010, average disruptiveness dropped by 15.53 percentile points. The decline is comparable in magnitude to the shift toward citing older papers (13.21 points) and the rise in female participation (13.56 points). The decrease in disruptiveness far exceeds the growth in international collaboration (7.08 points) and is second only to the surge in team size (27.96 points). All trends are statistically significant (p$<$0.001, Table \ref{table:RegressionBenchmarkMeasures}).

To further evaluate the robustness of our findings to the exclusion of 0-bcite works, in Sec.~\ref{sec:IndependentMeasuresDecline}, we report analyses using four additional measures of disruptiveness, applied to both SciSciNet and WoS data. We find statistically and practically significant declines across all measures and both data sources, indicating these results are sensitive neither to our operationalization of disruptiveness nor the data source selected.

In summary, by adopting HATWG's preferred dataset and metric, and applying their recommended exclusion of 0-bcite papers, we replicate our original findings using their own advocated methodology.

\addtocontents{toc}{\protect\setcounter{tocdepth}{0}}
\section{Statistical Significance: Decline Persists Under HATWG's Regression Framework}\label{sec:StatisticalSignificance}
\addtocontents{toc}{\protect\setcounter{tocdepth}{2}}

In this section, we statistically test HATWG's hypothesis that 0-bcite works explain observed declines in disruptiveness. We use their own regression model specification, which was designed to address perceived biases from 0-bcite papers. The model predicts paper-level disruptiveness (CD$_5$) using publication-year indicators and builds on PLF's regression framework, which included controls for changing citation practices and metadata quality (see PLF Methods). To this specification, HATWG add an indicator for 0-bcite papers. Our analysis is of the original PLF data from Web of Science (papers) and Patents View (patents).

Even after applying HATWG's adjustments (Table \ref{table:RegressionAdjustment}, Fig.~\ref{fig:HolstRegressionAdjustment}), we find highly significant declines in disruptiveness. For papers, from 1945 to 2010, the predicted disruptiveness declines by $\beta$=-0.082 (p$<$0.001). The magnitude of decline is important, equalling the difference between an average paper (CD$_5$=0.040) and a Nobel-winner (CD$_5$=0.131) \citep{li2019dataset}, or moving a median-ranked paper to the 93rd percentile. For patents, the 1980-2010 decline is even more pronounced ($\beta$=-0.155, p$<$0.001), and is comparable to the gap between an average patent (CD$_5$=0.123) and the average of the 37 landmark (1980 onwards) patents identified by Kelly et al. \citep{kelly2021measuring} (CD$_5$=0.270), or a median-ranked patent rising to the 84th percentile. 

Our results mirror what HATWG find but overlook in their own analyses. Their regressions (their Table S1) show significant declines for papers and patents (p$<$0.01), but instead of engaging with these findings, HATWG rely on visual inspection of predicted values to argue their adjustments eliminate the decline. This represents a concerning departure from scientific rigor---their analysis produces statistical evidence that directly contradicts their main hypothesis, yet they neither acknowledge nor interpret these results.

In summary, by applying HATWG's proposed methodological adjustments, we demonstrate that the decline in disruptiveness persists, remaining statistically and practically significant.

\addtocontents{toc}{\protect\setcounter{tocdepth}{0}}
\section{Simulations: Robust Evidence of Declining Disruptiveness}\label{sec:Simulations}
\addtocontents{toc}{\protect\setcounter{tocdepth}{2}}

In this section, we examine HATWG's re-analysis of Monte Carlo simulations presented in PLF. HATWG plot the average disruptiveness of papers over time in the observed and rewired networks. They find that disruptiveness declines in both and conclude declining disruptiveness is an artifact of 0-bcite papers.\footnote{We identify a logical issue in HATWG's inferences about 0-bcite papers from their simulations. HATWG assert that similar disruptiveness trends between observed and rewired networks, combined with the ``one-to-one correspondence between zero reference papers within the original and rewired networks\ldots [constitutes] proof that the reported decline in disruptiveness is merely showing a relative decrease of zero reference papers over time'' (HATWG, Fig.~S10). However, their rewiring process preserves \emph{multiple} network features beyond the number of 0-bcite papers (e.g., degree distributions). Because these features are also consistent between the observed and rewired networks, the correspondence between the two cannot isolate 0-bcite papers as the causal factor. This is akin to claiming that because both A and B are preserved and effect C is observed, A must cause C---while disregarding B's potential role.}

However, this conclusion stems from a misunderstanding of the design of PLF's rewiring analysis and the mathematical properties of the CD index (our disruptiveness measure). HATWG's approach of comparing the average CD index in the observed and random networks is inappropriate because the rewiring algorithm effectively preserves a component of the CD index, $n_K$ (future papers citing the focal paper's references but not the focal paper) (see Fig.~\ref{fig:RandomNetworkComponents}). Therefore, similar trajectories between the average CD in the observed and rewired networks are expected, as we show formally in Sec.~\ref{sec:MathematicalPropertiesLargeRandom}.

In PLF, the rewired networks were generated to `net out' the disruptiveness attributable to structural changes for \emph{individual papers}: a paper making $n$ references and receiving $m$ citations in the observed network will do the same in the rewired networks, so disruptiveness cannot be attributed to the number of citations made and received. To implement this `net-out' approach, PLF calculated a $z$-score for each paper, comparing the observed $\text{CD}$ to the mean $\text{CD}$ for the \emph{same paper} across 10 rewired networks \citep{uzzi2013atypical}. Average $z$-scores declined over time. Because the rewiring preserves $n_K$, this term cannot drive trends in the $z$-scores. If one wishes to compare means between observed and random networks (HATWG's approach), an alternative metric unaffected by the preservation of $n_K$ is needed. The CD$_5^{\smash{noK}}$ index \citep{bornmann2020disruption, bornmann2020disruptive, leibel2024we, leydesdorff2020proposal, deng2023enhancing}, which excludes $n_K$, serves this purpose.\footnote{CD$_5^{\smash{noK}}$ excludes the $n_K$ term (papers citing a focal paper's references but not the focal paper), providing a more local measure of disruption than the original CD index. Both have been independently validated \citep{bornmann2020disruption, leibel2024we, leydesdorff2020proposal, deng2023enhancing, li2024displacing, bornmann2021convergent, bornmann2020disruptive, lin2023remote, wang2021science, wu2019large, li2024productive, funk2017dynamic}.} As Fig.~\ref{fig:RandomNetworkComparisonCDnoK} shows, using CD$_5^{\smash{noK}}$ results in a persistent decline in disruptiveness in the observed networks, while the rewired network trend is flat.

Finally, HATWG suggest the decline in average $z$-scores observed in PLF could stem from a shrinking standard deviation of the CD index in the random networks. We therefore conducted an alternative `net-out' analysis, estimating a variation of HATWG's regression model (with 0-bcite dummy) that also controls for each paper's $\text{CD}_5^{\smash{\text{random}}}$ value in the rewired networks. As Table \ref{table:RandomRegressionAdjustment} and Fig.~\ref{fig:RandomNetworkRegressionPlot} show, we find statistically significant declines for papers and patents (both p$<$0.001), suggesting these changes are substantive rather than structural artifacts, confirming PLF's findings even under exceptionally stringent criteria.

\addtocontents{toc}{\protect\setcounter{tocdepth}{0}}
\section{Data and Methods: Concerns in HATWG's Analysis}
\addtocontents{toc}{\protect\setcounter{tocdepth}{2}}

In this section, we examine HATWG's data, beginning by noting their dataset has an order of magnitude more CD=1 works than PLF. While CD=1 documents comprise 4.3\% of works in WoS and 4.9\% in Patents View, they account for 23.1\% in HATWG's SciSciNet (Fig.~\ref{fig:Distributions})\footnote{HATWG correctly identify a visualization artifact in seaborn 0.11.2 affecting two sub-panels in PLF's Extended Data Figure 1. The complete data, including all CD=1 papers, was used in all PLF's analyses and is shown correctly in other figures throughout PLF.} (see Sec.~\ref{sec:CD1AllGone} on HATWG's concerning handling of CD=1 works). This discrepancy is driven partially by a substantial overrepresentation of 0-bcite works in HATWG's data. Fig.~\ref{fig:HolstDataQuality}\textbf{a},\textbf{b} shows SciSciNet has three times more 0-bcite works than WoS/Patents View. Because 0-bcite works are central to their argument, determining the source of this excess is essential.

Our analysis shows this excess of 0-bcite works is attributable to HATWG's departure from scientometric best practices---practices developed precisely to address the metadata quality concerns they raise. First, HATWG do not subset their data to appropriate document types. Bibliometric databases include document types (e.g., news) that rarely make citations and have lower metadata quality; scientometric research therefore typically focuses on research articles \citep{annalingam2014determinants, tahamtan2016factors, bornmann2013how, nicolaisen2019zero, andersen2023field, bornmann2013problem, mendoza2021differences, hilmer2009determinants, ioannidis2016citation, ozturk2024how, rehn2014bibliometric}. Fig.~\ref{fig:HolstDataQuality}\textbf{e},\textbf{f} shows the importance of appropriate document type selection in the context of HATWG's critique. Due to SciSciNet's limited document type data \citep{hug2017citation}, we match papers to Dimensions.ai via DOIs to obtain detailed indicators. We find research articles are far less likely to have 0-bcites than other document types. While PLF exclude non-research articles (\textbf{e}), HATWG include all document types (\textbf{f}), inflating their 0-bcite proportion. The problem is underscored in \emph{Nature}, where most works are editorials or commentaries (\textbf{f}, left inset) \citep{tunger2021journal}, yet HATWG include them. The lack of granularity in SciSciNet's document type categories also undermines HATWG's attempt to rule out compositional change as an explanation for declining 0-bcite prevalence. Their Fig.\ S10 shows stable document type proportions over time, but this is meaningless when the `Journal' category conflates research articles with editorials, news, letters, and book reviews. Applying proper classification to their data, we find such content declined from 40\% to 8\% of `Journal'---tagged works between 1945 and 2010. An editorial with zero citations is not a data quality problem---it is an editorial. This `hidden' 81\% decline is sufficient to generate the perceived bias from 0-bcite works that HATWG attribute to metadata error.

Second, HATWG do not subset to appropriate fields. Citation practices vary widely by discipline, with fields like the humanities relying heavily on footnotes or endnotes, resulting in lower quality metadata \citep{ball2017bibliometrics, ochsner2021bibliometrics, andersen2023field, tahamtan2016factors, sivertsen2012comprehensive, hug2014framework, bonaccorsi2018towards, ochsner2012indicators, ochsner2014setting, ochsner2016research, ochsner2017future}. Consequently, scientometric studies typically exclude such fields \citep{ahmadpoor2019decoding, gates2019natures, milojevic2015quantifying, sekara2018chaperone, tshitoyan2019unsupervised, uzzi2013atypical, varga2019shorter, wang2017bias, wuchty2007increasing}. HATWG, however, include all fields in their analysis, inflating the proportion of 0-bcite works (Fig.~\ref{fig:HolstDataQuality}\textbf{c},\textbf{d}). In SciSciNet, 0-bcite proportions in the humanities approach 70\%, underscoring the importance of proper subsetting. 

Further evidence of data quality issues in HATWG's data emerges from their own analyses. HATWG checked PDFs for 100 SciSciNet papers recorded as having 0-bcites, finding 93\% made references (their Table S4). However, this exercise inadvertently demonstrates the poor metadata quality of SciSciNet itself. Matching the same papers to WoS (Table \ref{table:HolstPDFWoSLink}), we find of the 48 papers present in both databases, 45 had properly coded references in WoS. The remaining three were excluded in PLF---two were non-English, and one was not a research article. More broadly, Table \ref{table:SciSciNetWoSContingencyTable} shows, backward citations are missing in WoS but present in SciSciNet in just 1.2\% of cases, but present in WoS and missing in SciSciNet in 19.1\%, revealing severe quality issues in HATWG's data. (For a discussion of similar issues in the patent data, see Sec.~\ref{sec:0BcitePatentErrors}.)

\addtocontents{toc}{\protect\setcounter{tocdepth}{0}}
\section{Assessment of HATWG's Web of Science Analysis}\label{subsec:WoSRebuttal}\addtocontents{toc}{\protect\setcounter{tocdepth}{2}}

HATWG's Fig.~S17 presents a supplemental analysis using Web of Science data, which the authors frame as a direct replication of PLF. We address the scientific content and transparency of this analysis.

\emph{\textbf{Scientific assessment.}} Fig.~S17(a) presents predicted values from HATWG's regression model but omits all statistical documentation---no coefficients, standard errors, confidence intervals, or significance tests. This is notable given their own SciSciNet regression (their Table~S1) shows significant declines they do not acknowledge. Applying their specification to WoS data (Table~\ref{table:RegressionAdjustment}), we find a highly significant decline ($\beta$=-0.082, p$<$0.001)---comparable to the gap between an average paper and a Nobel winner (Sec.~\ref{sec:StatisticalSignificance})---contradicting their visual interpretation. HATWG's Fig.~S17(b) compares mean CD values between observed and rewired networks, but as demonstrated in Sec.~\ref{sec:Simulations}, this comparison is inappropriate; using CD$_5^{\smash{noK}}$ instead, observed networks show persistent decline while rewired networks remain flat (Fig.~\ref{fig:RandomNetworkRegressionPlot}).

\emph{\textbf{Independent corroboration.}} The decline in disruptiveness has been replicated in independent analyses of WoS data. Yu et al.~\cite{yu2025knowledge} analyzed 114M papers across WoS, OpenAlex, and SciSciNet, concluding ``we find a systematic decline in disruption over time, consistent with prior findings [citing PLF]''. Yu et al.\ explicitly address confounds from citation inflation and zero-reference works---the precise concerns HATWG raise---and find the decline persists. These findings align with other large-scale replications using WoS~\citep{wu2019large, chu2021slowed, cui2022aging}.

\emph{\textbf{Transparency concerns.}} HATWG describe Fig.~S17 as using ``all the relevant data publicly deposited by [PLF],'' ensuring ``identical preprocessing steps.'' However, the relevant data---paper-level identifiers, regression variables, and rewired network values---were not included in our public repository due to licensing restrictions, and HATWG never requested access. In response to editorial inquiries, HATWG acknowledged incorporating third-party data, which was not disclosed in the commentary. Web of Science coverage is dynamic, with content regularly added and removed; achieving an identical sample size (N=$22,479,429$) from an independent pull years later would be extraordinarily unlikely. HATWG have not clarified which components rely on which data, complicating independent evaluation.

\addtocontents{toc}{\protect\setcounter{tocdepth}{0}}
\section{Discussion}
\addtocontents{toc}{\protect\setcounter{tocdepth}{2}}
In summary, we find no support for HATWG's hypothesis that the observed decline in disruptiveness stems from 0-bcite works. Using their advocated dataset, metric, exclusions, and regression adjustments, we find statistically and practically significant declines that equal major benchmark transformations in science. These results align with HATWG's own unacknowledged findings showing statistically significant declines for papers and patents. While their critique centers on the influence of 0-bcite works, they rely on data containing three times more such works than PLF, an excess attributable to millions of editorials, commentaries, and book reviews, alongside children's literature and \emph{For Dummies} guides. Their analysis thus serves more to highlight the limitations of their own data than to challenge PLF's original conclusions.

\addtocontents{toc}{\protect\setcounter{tocdepth}{0}}
\section{Data Availability Statement}
\addtocontents{toc}{\protect\setcounter{tocdepth}{2}}
This study uses data from PLF (\href{https://zenodo.org/records/7258379}{doi:10.5281/zenodo.7258379}), SciSciNet (\href{https://doi.org/10.1038/s41597-023-02198-9}{doi:10.1038/s41597-023-02198-9}), and Dimensions.ai (\url{https://docs.dimensions.ai/bigquery/}).

\addtocontents{toc}{\protect\setcounter{tocdepth}{0}}
\section{Code Availability Statement}
\addtocontents{toc}{\protect\setcounter{tocdepth}{2}}
Our code builds on PLF (\href{https://zenodo.org/records/7258379}{doi:10.5281/zenodo.7258379}); additional scripts available upon request.

\clearpage
\newgeometry{margin=0.5in}

\thispagestyle{empty}
\begin{figure}[htbp]\centering
\centering
\includegraphics[width=1.0\textwidth]{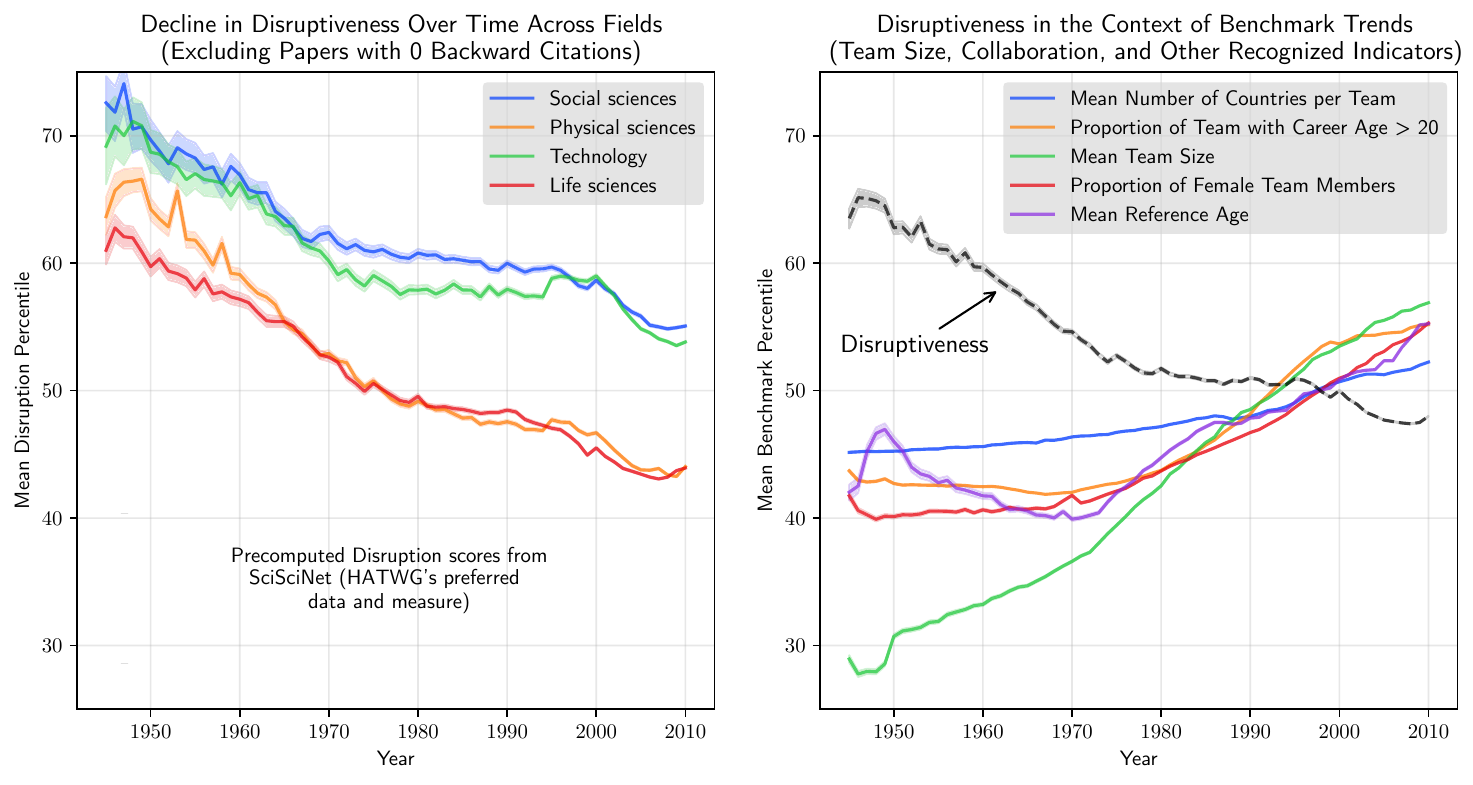}
\caption{\textbf{Declining Disruptiveness Matches Major Benchmark Transformations in Science After Excluding Zero-Backward-Citation Works.} The left panel plots the average percentile values of disruptiveness by field over time, calculated using the precomputed disruption scores in SciSciNet \citep{lin2023sciscinet}, which exclude 0-bcite papers and are consistent with HATWG's advocated methodology. The right panel plots the average percentile values of major benchmark transformations in science, including mean reference age \citep{evans2008electronic, lariviere2008long, chu2021slowed, verstak2014shoulders, sinatra2015century, wang2021science}, proportion of team members with a career age $>$20 years \citep{cui2022aging, wang2021science, jones2011age, jones2010age, blau2017us, alberts2015addressing}, proportion of female team members \citep{nsf2021, huang2020historical}, countries per team \citep{wang2021science, wagner2017growth, lin2023remote, adams2013fourth, ribeiro2018growth, leydesdorff2008international, luukkonen1992understanding}, and mean team size \citep{wuchty2007increasing, jones2021rise, wang2021science, milojevic2014principles}. Disruptiveness (in black) is plotted alongside these benchmarks for comparison. Even after excluding papers that make 0 backward citations, the magnitude of the decline in disruptiveness is comparable to these well-documented trends, underscoring its robust practical significance. Corresponding regression analyses in Table \ref{table:RegressionBenchmarkMeasures} verify that all trends, including disruptiveness overall and within fields, as well as all the benchmarks, are statistically significant at the p$<$0.001 level. Shaded bands correspond to 95\% confidence intervals.}
\label{fig:BenchmarkMeasures}
\end{figure}

\clearpage

\thispagestyle{empty}
\begin{figure}[htbp]\centering
\centering
\includegraphics[width=1\textwidth]{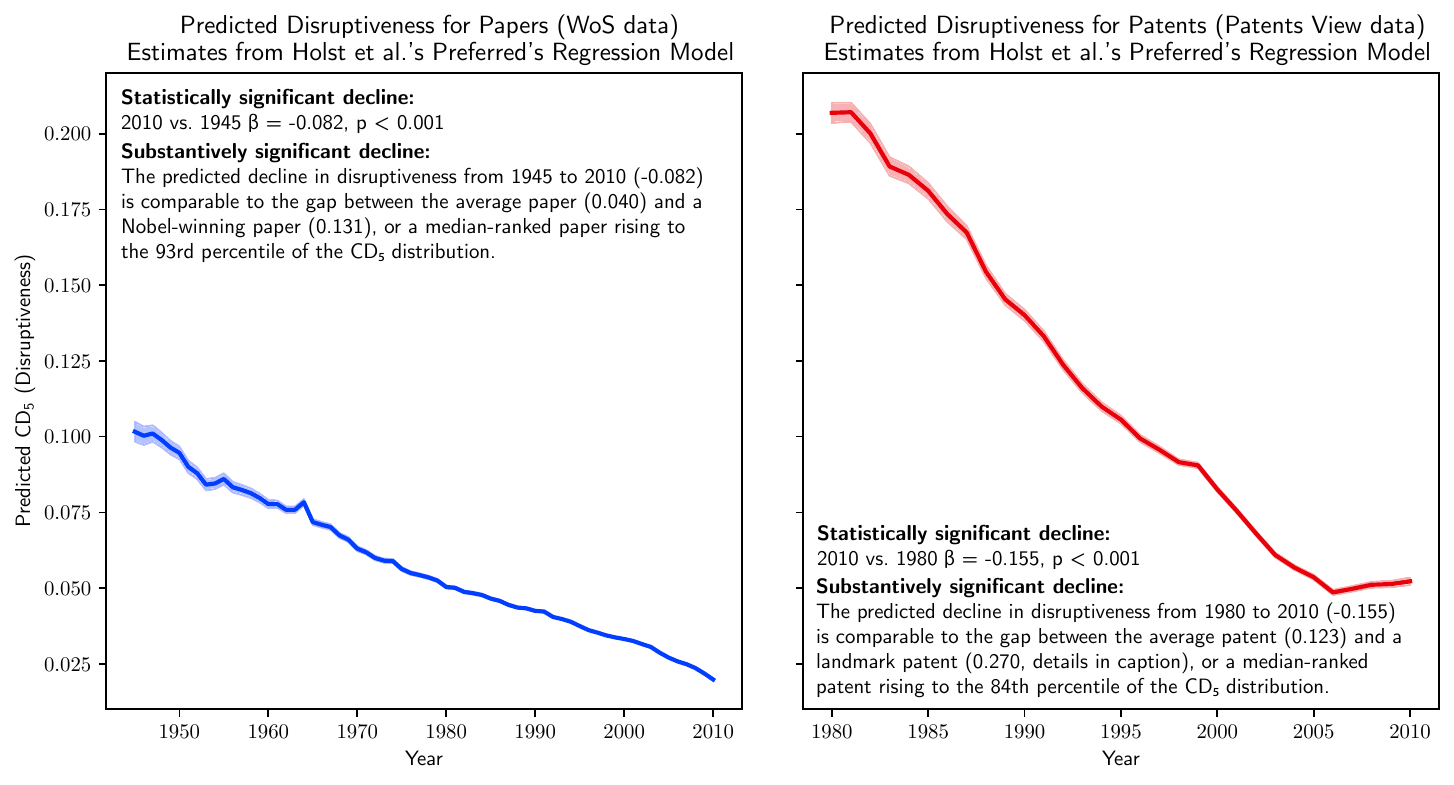}
\caption{\textbf{Persistent Declines in Disruptiveness Using HATWG's Proposed Regression Model.} This figure plots the predicted values of the CD$_5$ index (disruptiveness) for papers in Web of Science (left panel) and patents in Patents View (right panel), obtained from the coefficient estimates in Table~\ref{table:RegressionAdjustment}. Our regressions are based on HATWG's regression model (c.f. HATWG, Table S1), which includes their proposed dummy variable for 0-bcite works, in addition to the full suite of control variables used in PLF (see their Extended Data Figure 8). From 1945 to 2010, the predicted disruptiveness for papers declines by $\beta$=-0.082 (p$<$0.001). The magnitude of decline is important, equalling the difference between an average paper (CD$_5$=0.040) and a Nobel-winner (CD$_5$=0.131) \citep{li2019dataset}, or moving a median-ranked paper rising to the 93rd percentile. For patents, the 1980-2010 decline is even more pronounced ($\beta$=-0.155, p$<$0.001), and is comparable to the gap between an average patent (CD$_5$=0.123) and the average of the 37 landmark (1980 onwards) patents identified by Kelly et al. \citep{kelly2021measuring} (CD$_5$=0.270), or a median-ranked patent rising to the 84th percentile. Thus, even after applying HATWG's adjustments, the decline in disruptiveness over time remains both statistically significant and practically meaningful. Shaded bands correspond to 95\% confidence intervals.}
\label{fig:HolstRegressionAdjustment}
\end{figure}

\clearpage
\thispagestyle{empty}
\begin{figure}[htbp]\centering
\centering
\includegraphics[width=1\textwidth]{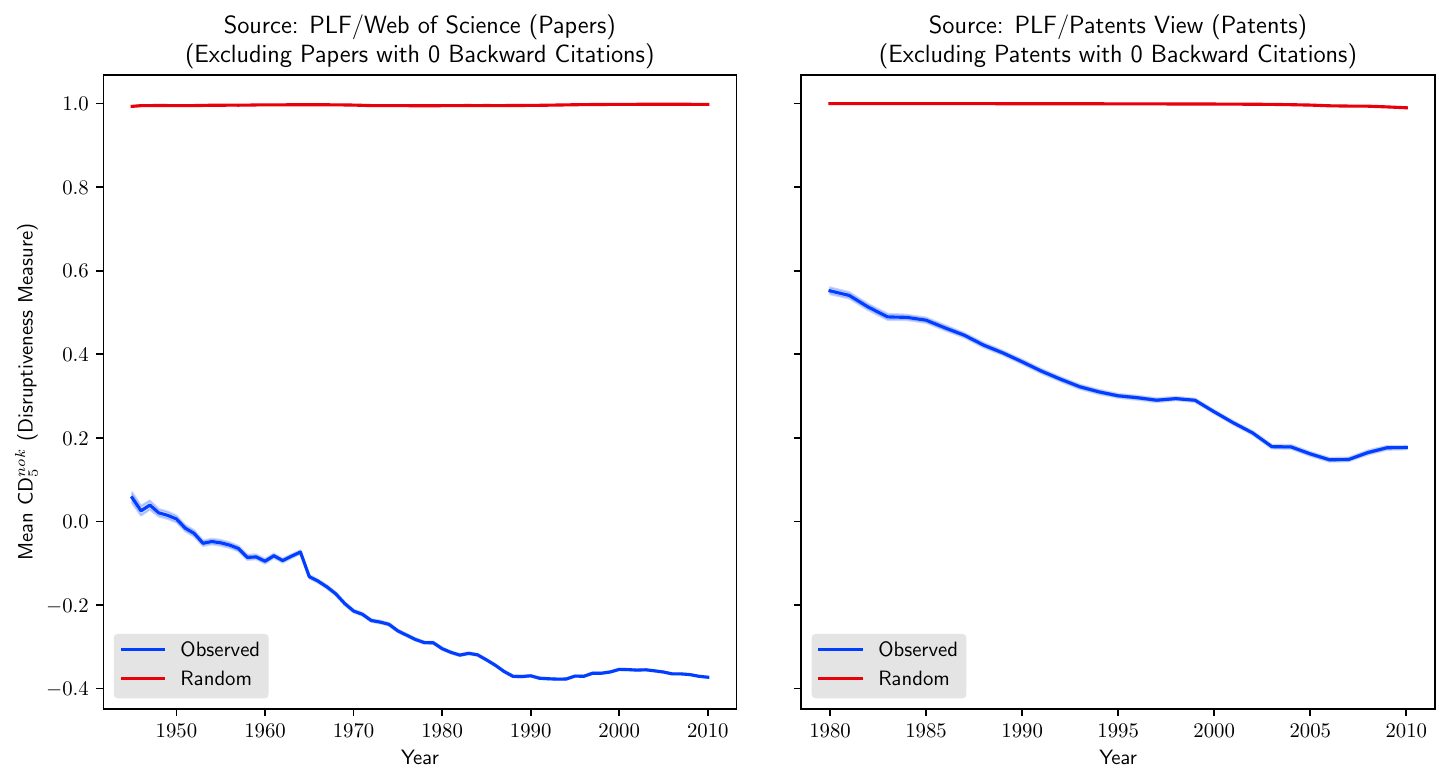}
\caption{\textbf{Persistent Decline in Disruptiveness Relative to Randomly Rewired Citation Networks.} This figure compares the temporal trends in average disruptiveness for papers (Web of Science, left) and patents (Patents View, right), shown alongside average disruptiveness in comparable randomly rewired citation networks. We measure disruptiveness using CD$_5^{noK}$, which has been independently validated as a disruption indicator that excludes the original CD index's $n_K$ term (\citep{bornmann2020disruption, bornmann2020disruptive, leibel2024we, leydesdorff2020proposal, deng2023enhancing}), which is preserved in the rewiring process of HATWG's simulations (see Sec.~\ref{sec:MathematicalProperties} for a formal mathematical demonstration). Because $n_K$ is preserved in the rewired networks, trends in disruptiveness that are attributable to $n_K$ will be present in both the observed and rewired networks, thereby making direct comparisons of the average disruptiveness in the observed and rewired networks using the original CD index (HATWG's approach) inappropriate. The plots reveal that while both papers and patents show persistent declines in CD$_5^{noK}$ in the observed network, the rewired network maintains a flat trend. (Note that the flat trend results from nearly all ``J''-type cites [future works citing the references of the focal work but not the focal work itself] switching to ``I''-type cites [future works citing the focal work itself], leading to an average CD$_5^{noK}$ value of approximately 1; see Section S5.3 for additional mathematical details.) This provides robust evidence that the observed declines in disruptiveness are substantive rather than artifacts of the changing prevalence of 0-bcite papers/patents or other similar factors. For analyses using the original CD index, the appropriate method is to `net out' the level of disruptiveness attributable to structural properties of the citation network by comparing the observed disruptiveness to the disruptiveness in randomly rewired networks at the level of individual papers (or patents). PLF implements this adjustment using $z$-scores (see PLF Extended Data Figure 8). An alternative approach is to estimate a regression model that predicts the observed CD index as a function of time (year dummies) while controlling for the CD index value in the randomly rewired networks for each paper or patent. The results of this approach, shown in Table~\ref{table:RandomRegressionAdjustment} and the corresponding predicted values plot (Fig.~\ref{fig:RandomNetworkRegressionPlot}), provide further support that the observed declines in disruptiveness are not attributable to changes in citation network structure (e.g., the prevalence of 0-bcite papers or patents). Shaded bands correspond to 95\% confidence intervals.}

\label{fig:RandomNetworkComparisonCDnoK}
\end{figure}

\clearpage

\thispagestyle{empty}
\begin{figure}[htbp]\centering
\centering
\includegraphics[width=0.7\textwidth]{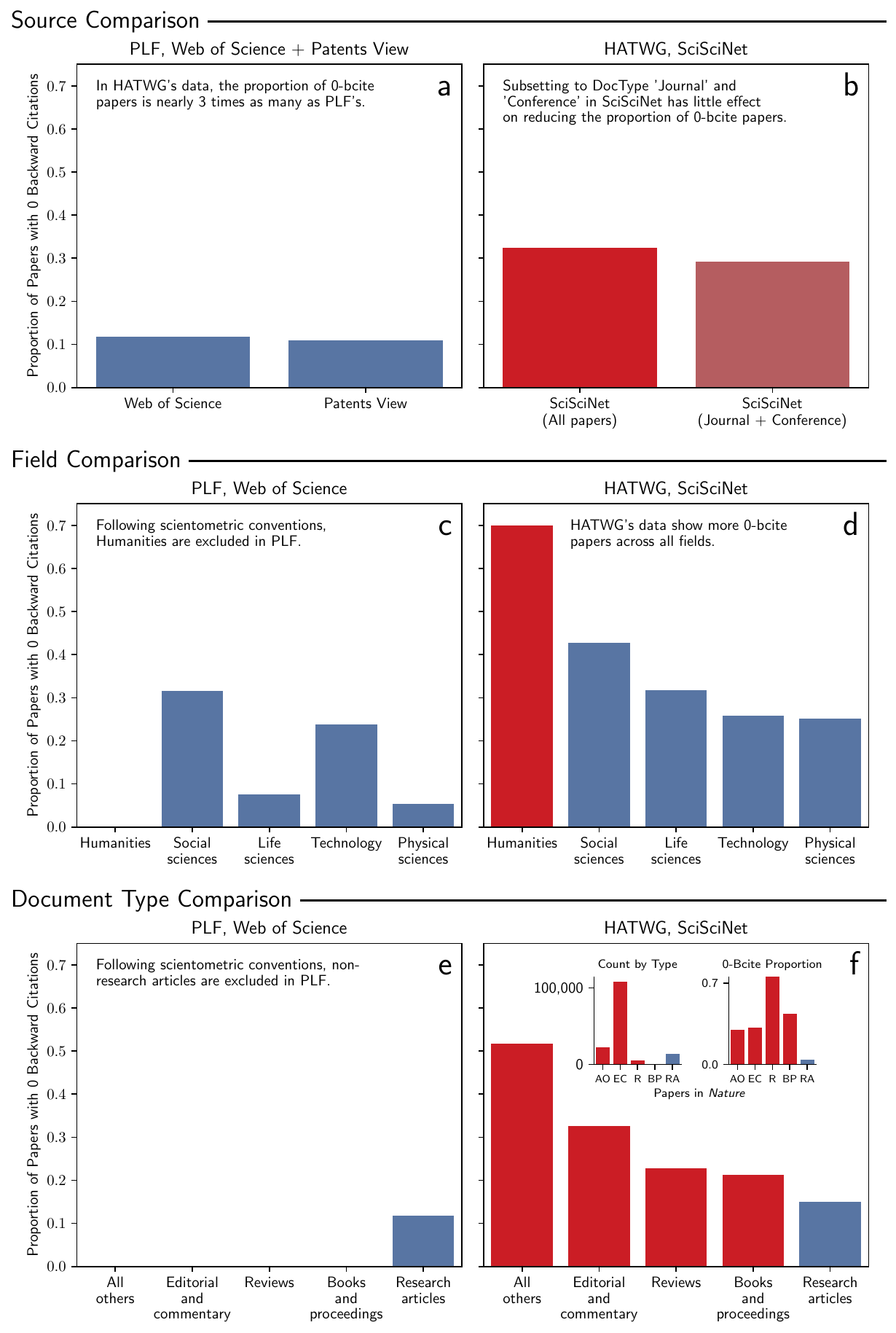}
\caption{\footnotesize \textbf{Departures from Scientometric Best Practices Lead to Severe Overrepresentation of Zero-Reference Works in HATWG's Data.} This figure reveals severe data quality concerns in HATWG's SciSciNet dataset, particularly their handling of 0-bcite documents, which they identify as problematic. Their critique, however, applies more to their own dataset than to PLF's. \textbf{Row 1} (\textbf{a},\textbf{b}): SciSciNet's 0-bcite proportion is approximately three times larger than PLF's for both Web of Science (2.76 times higher) and Patents View (2.98 times higher). HATWG's effort to address this through `Journal' and `Conference' paper subsetting (their Figure S8) is ineffective, with 0-bcite proportions remaining nearly identical, revealing SciSciNet's insufficiently granular document type classification. \textbf{Row 2} (\textbf{c},\textbf{d}): Using a SciSciNet-to-WoS Research Areas crosswalk (Table \ref{table:FieldsMapping}), SciSciNet shows higher 0-bcite proportions across all fields. This disparity reaches its peak in Humanities, approaching 70\%. While PLF excluded this field following scientometric standards, Humanities publications were included in HATWG, which inflated their count of 0-bcite documents. \textbf{Row 3} (\textbf{e},\textbf{f}): The analysis by document type reveals another crucial issue. While PLF adhered to scientometric best practices by including only research articles (yielding fewer 0-bcite documents), HATWG's SciSciNet data incorporated a wide range of document types (e.g., news items, corrections, commentaries, book reviews) that typically lack citations, thereby inflating the proportion of 0-bcite documents in their data. As an example, the left inset of panel \textbf{f} shows that most documents published in \emph{Nature} and coded as `Journal' in SciSciNet are editorial, commentary, or other non-research pieces, yet they were included in HATWG's analysis. The proportion of 0-bcite documents among these unconventional document types vastly exceeds that of research articles (right inset panel), highlighting the critical importance of proper document type selection in the context of HATWG's critique. Granular document type classifications for SciSciNet documents were determined through DOI-based matching with Dimensions.ai (36,530,788 of 45,251,912 papers, or 80.73\% of HATWG's SciSciNet papers were successfully linked), with crosswalk details in Table \ref{table:MetaCategories}.}
\label{fig:HolstDataQuality}
\end{figure}

\clearpage
\newgeometry{margin=1in}

\bibliographystyle{unsrt}
\bibliography{ref}

\clearpage

\clearpage

\begin{center}
    \Large \textbf{Supplementary Information for}
\end{center}

\begin{center}
    \Large {Robust Evidence for Declining Disruptiveness: Assessing the Role of Zero-Backward-Citation Works}
\end{center}

\vspace{10pt}

\begin{center}
    \large Michael Park$^1$, Erin Leahey$^2$, Russell J. Funk$^3$
\end{center}

\vspace{10pt}

\begin{center}
    \small
    \begin{itemize}
        \item[1.] Organisational Behaviour, INSEAD
        \item[2.] School of Sociology, University of Arizona
        \item[3.] Carlson School of Management, University of Minnesota
    \end{itemize}
\end{center}

\vspace{20pt}

\renewcommand{\contentsname}{Table of Contents}
\startSupplementary
\NoHyper
\tableofcontents
\addtocontents{toc}{\protect\thispagestyle{empty}}
\endNoHyper
\pagenumbering{gobble}

\clearpage
\section{Theoretical Considerations for Including Zero-Backward-Citation Works}\label{sec:0BciteTheory}
The exclusion of observations from scientific analysis requires careful methodological and theoretical justification. HATWG's commentary argues for excluding papers and patents with zero backward citations (0-bcite) purely on methodological grounds, without addressing whether such exclusion is theoretically appropriate for studying innovative activity. While we have focused our reply on the methodological issues they raise, their proposal to wholesale exclude 0-bcite works raises broader scientific concerns. Data quality issues alone do not justify excluding observations, particularly when established methodological approaches exist to address such concerns (e.g., proper sample selection) and when the excluded observations may contain theoretically relevant information. 

There are important theoretical reasons why excluding 0-bcite works would be problematic in a study of disruptive innovation. These have been discussed in prior work, and informed our decision to include 0-bcite papers and patents in our study \citep{ahuja2001entrepreneurship, dahlin2005invention, nerkar2007determinants, banerjee2010globally, gerken2012new, chen2021emergence, goldmanirs}. Such works often represent novel directions that forge new paths rather than building directly on existing research. In innovation theory, radical breakthroughs frequently arise independently of existing knowledge bases precisely because they introduce fundamentally new ideas or technologies. By definition, these innovations may cite little or nothing, as they are creating entirely new pathways. Excluding 0-bcite works thus risks overlooking a crucial mechanism of disruption---the emergence of novel knowledge that diverges from established trajectories. These pioneering contributions often become foundational to subsequent advances. Underscoring their importance, previous research has developed innovation metrics that specifically use zero-backward-citations as a proxy for innovative work \citep{ahuja2001entrepreneurship, dahlin2005invention, nerkar2007determinants, banerjee2010globally, gerken2012new, chen2021emergence, goldmanirs}. Their exclusion would bias our understanding of how new contributions shape future knowledge production---the very phenomenon that disruption metrics aim to capture. Indeed, had we excluded these works wholesale, our study would have been rightly criticized for missing potentially transformative innovations.

As an illustration, many early and foundational patents in the emergence of biotechnology---widely considered paradigmatic cases of disruptive innovation---contain no backward citations to prior patents, precisely because of their groundbreaking nature. Well known examples include the Ptashne patents on protein synthesis (patents \#4,332,892, \#4,418,149), methods for human growth hormone production (patent \#4,363,877), the Axel patent on cotransformation (patent \#4,634,665), the landmark Diamond v. Chakrabarty patent (patent \#4,259,444), the Milstein-Kohler patent on monoclonal antibodies (patent \#4,172,124), and Caruthers' DNA synthesis method (patent \#4,458,066).

Prior literature has more extensively examined the relationship between zero backward citations and innovative activity for patents than for papers. The same logic applies, however, and empirically, many landmark scientific contributions contain no backward citations precisely because they opened new domains---defining new phenomena, deriving results from first principles, or reporting observations with no precedent. Examples include the EPR paper on quantum entanglement \citep{einstein1935can}, Gabor's invention of holography \citep{gabor1948new}, Kanner's clinical definition of autism \citep{kanner1943autistic}, Fisher's foundational work in population genetics \citep{fisher1937wave}, Harlow's studies of attachment \citep{harlow1958nature}, and Skinner's experiments on conditioning \citep{skinner1948superstition}. Overmars and Welzl \citep{overmars1985complexity} make this explicit, noting in their references section, ``no references on this topic seem to exist and no useful results could be found''---a candid acknowledgment of what zero backward citations often represent.

While there may be contexts in which excluding works with 0-bcites is conceptually justified, such exclusions require a clear theoretical rationale aligned with the specific aims and scope of the research. Reasonable analysts may disagree on the appropriateness of such exclusions depending on the context. However, the standards established in the innovation literature, along with the conceptual arguments presented above, underscore the importance of providing a robust justification for such decisions. HATWG provide no theoretical justification for their proposal to exclude 0-bcite works. This omission leaves their argument incomplete, particularly given the potential biases and conceptual pitfalls associated with indiscriminate exclusions.

\clearpage
\newgeometry{margin=0.5in}

\section{Persistent Decline in Disruptiveness Matches Major Benchmark Transformations in Science}\label{sec:BenchmarkTrends}

{

\setlength{\tabcolsep}{2pt}

\renewcommand{\arraystretch}{0.9}

\thispagestyle{empty}

\begin{table}[htbp]\centering
\begin{threeparttable}\footnotesize
\caption{Trends in Disruptiveness in the Context of Major Benchmark Transformations}
\label{table:RegressionBenchmarkMeasures}
\begin{tabular}{@{\hskip 0.1in}ll@{\hskip 0.1in}cccccccc@{\hskip 0.1in}}
\toprule
 & & \multicolumn{4}{c}{Year} & \multicolumn{2}{c}{Constant} & &  \\
\cmidrule(lr){3-6} \cmidrule(lr){7-8}
Dependent Variable (Percentile) & Sample & Coefficient & SE & P & CI & Constant & SE & R² & N  \\
\midrule
\multicolumn{10}{l}{\textbf{Benchmarks}} \\
Mean Team Size  & All papers & 0.49 & 0.00 & 0.00 & [0.49, 0.49] & -934.66 & 0.59  & 0.06 & 36987057 \\
Proportion of Team with Career Age $>$ 20  & All papers & 0.31 & 0.00 & 0.00 & [0.31, 0.31] & -568.67 & 0.29  & 0.05 & 59264700 \\
Proportion of Female Team Members  & All papers & 0.32 & 0.00 & 0.00 & [0.32, 0.32] & -593.45 & 0.48  & 0.03 & 47743551 \\
Mean Number of Countries per Team  & All papers & 0.15 & 0.00 & 0.00 & [0.15, 0.15] & -255.16 & 0.38  & 0.01 & 26624698 \\
Mean Reference Age  & All papers & 0.30 & 0.00 & 0.00 & [0.30, 0.30] & -555.26 & 0.87  & 0.02 & 28532742 \\
\midrule
\multicolumn{10}{l}{\textbf{Disruptiveness}} \\
Disruptiveness & All papers & -0.19 & 0.00 & 0.00 & [-0.19, -0.19] & 430.33 & 0.98  & 0.01 & 25022222 \\
Disruptiveness & Life sciences & -0.25 & 0.00 & 0.00 & [-0.25, -0.25] & 548.37 & 1.63  & 0.01 & 9495139 \\
Disruptiveness & Physical sciences & -0.25 & 0.00 & 0.00 & [-0.25, -0.25] & 551.82 & 1.62  & 0.01 & 7733618 \\
Disruptiveness & Social sciences & -0.21 & 0.00 & 0.00 & [-0.21, -0.20] & 470.85 & 2.76  & 0.01 & 3011432 \\
Disruptiveness & Technology & -0.18 & 0.00 & 0.00 & [-0.18, -0.18] & 418.29 & 2.62  & 0.00 & 4782033 \\
\bottomrule
\end{tabular}

\begin{tablenotes}

\item \emph{Notes:} This table reports regression results showing trends in disruptiveness in the context of major benchmark transformations in science over time. Each row corresponds to a regression where the dependent variable is the indicated metric. Disruptiveness is measured using the precomputed disruption scores in SciSciNet \citep{lin2023sciscinet}, which exclude 0-bcite papers, consistent with HATWG's advocated methodology. Benchmark metrics include mean reference age \citep{evans2008electronic, lariviere2008long, chu2021slowed, verstak2014shoulders, sinatra2015century, wang2021science}, proportion of team members with a career age $>$20 years \citep{cui2022aging, wang2021science, jones2011age, jones2010age, blau2017us, alberts2015addressing}, proportion of female team members \citep{nsf2021, huang2020historical}, countries per team \citep{wang2021science, wagner2017growth, lin2023remote, adams2013fourth, ribeiro2018growth, leydesdorff2008international, luukkonen1992understanding}, and mean team size \citep{wuchty2007increasing, jones2021rise, wang2021science, milojevic2014principles}. All metrics are percentile-normalized to allow comparison across measures with varying scales and distributions. The coefficients represent the change in percentile value per year. Statistically significant declines in all trends, including disruptiveness overall and within fields, as well as benchmarks, are observed at the p$<$0.001 level. These results highlight the robustness and practical significance of the observed trends in disruptiveness to the exclusion of 0-bcite works relative to major benchmarks.

\end{tablenotes}

\end{threeparttable}
\end{table}

}%

\clearpage
\newgeometry{margin=1in}

\section{Persistent Decline Across Independently Developed Disruptiveness Measures}\label{sec:IndependentMeasuresDecline}
In this section, we evaluate the robustness of the observed decline in disruptiveness to the exclusion of 0-bcite works using four additional, independently developed measures of disruptiveness, each applied to both SciSciNet and WoS data. The measures include CYG$_5$ (Citation Year Gap), which calculates the average age gap between references cited by citing works relative to the focal paper \citep{brauer2023aggregate, brauer2024searching}; Is D$_5$, a binary indicator for whether the CD$_5$ value (disruptiveness index) is positive \citep{lin2023remote, li2024displacing, yang2024female}; CD$_5^{noK}$, which excludes references-only citations (the ``$n_K$'' term) from the denominator \citep{bornmann2020disruption, bornmann2020disruptive, leibel2024we, leydesdorff2020proposal, deng2023enhancing}; and CD$_5^5$, which introduces a threshold requiring future works to cite multiple references of the focal paper to strengthen bibliographic coupling \citep{li2024displacing, wu2019solo, jiang2024new, bornmann2020disruption, leibel2024we}. All documents with 0-bcites were excluded from this analysis follow HATWG; measures are percentile-normalized to allow comparison across scales and distributions.

\begin{figure}[htbp]\centering
\centering
\includegraphics[width=1\textwidth]{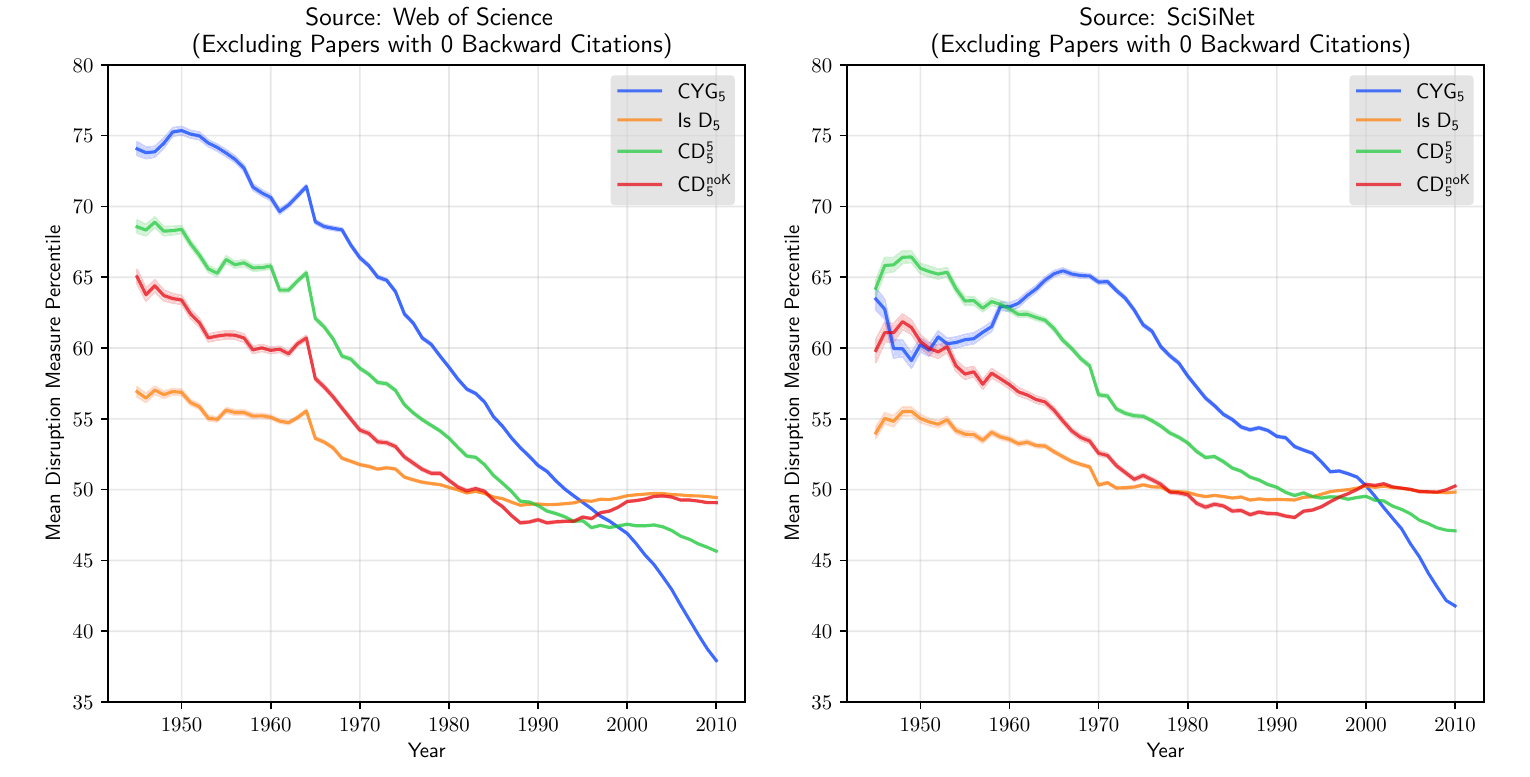}
\caption{\textbf{Persistent Decline Across Independently Developed Disruptiveness Measures.} This figure demonstrates that our finding of a persistent decline in disruptiveness even when 0-bcite papers are excluded is not sensitive to the choice of disruption metric or data set. Specifically, the plots track the average (percentile) values of four independently developed measures of disruptiveness. Both plots exclude all 0-bcite papers. Values of the disruptiveness measures are plotted separately for Web of Science (left panel) and SciSciNet (right panel) across years. The measures include CYG$_5$ (Citation Year Gap), which calculates the average age gap between references cited by citing works relative to the focal paper \citep{brauer2023aggregate, brauer2024searching}; Is D$_5$, a binary variable for whether the CD$_5$ value (disruptiveness index) is positive \citep{lin2023remote, li2024displacing, yang2024female}; CD$_5^{noK}$, which excludes references-only citations (sometimes referred to as the ``$n_K$'' term) from the denominator \citep{bornmann2020disruption, bornmann2020disruptive, leibel2024we, leydesdorff2020proposal, deng2023enhancing}; and CD$_5^5$, which introduces a threshold requiring future works to cite multiple references of the focal paper, emphasizing stronger bibliographic coupling \citep{li2024displacing, wu2019solo, jiang2024new, bornmann2020disruption, leibel2024we}. All measures exclude 0-bcite documents and are percentile-normalized to enable comparison across scales. The figure shows a consistent decline in disruptiveness over time across all measures and datasets---consistent with the findings of PLF and supported by the regression results in Table \ref{table:RegressionAlternativeMeasures}---which demonstrate that the declines are statistically significant at the p$<$0.001 level. Shaded bands correspond to 95\% confidence intervals.}
\label{fig:AlternativeMeasures}
\end{figure}

The results of this analysis are shown in Figure \ref{fig:AlternativeMeasures} and Table \ref{table:RegressionAlternativeMeasures}. Consistent with the findings presented in the main body of our commentary, we observe robust evidence of statistically and practically significant declines in disruptiveness across all four measures and both datasets. This confirms that our results are not sensitive to the operationalization of disruptiveness or the data source used. 

Specifically, in the Web of Science (WoS) dataset, CYG$_5$ values decline by 36 percentile points between 1945 (74.07) and 2010 (37.93), while CD$_5^{noK}$ declines by 16 percentile points over the same period. Similarly, CD$_5^5$ shows a 23 percentile point decline, and Is D$_5$ decreases by 7.5 percentile points. In the SciSciNet dataset, the trends are comparable: CYG$_5$ values drop by 22 percentile points, CD$_5^{noK}$ declines by nearly 10 points, CD$_5^5$ by 17 points, and Is D$_5$ by approximately 4 points. These patterns are consistent across all measures and demonstrate a persistent and practically significant decline in disruptiveness over time, on par with those of the benchmark trends reported in Figure \ref{fig:BenchmarkMeasures} and Table \ref{table:RegressionBenchmarkMeasures}.

These results are also highly statistically significant, as shown in Table \ref{table:RegressionAlternativeMeasures}. All measures decline significantly over time (p$<$0.001), underscoring the robustness of the observed trends.

\clearpage
\newgeometry{margin=0.5in}

\thispagestyle{empty}

\clearpage

{

\thispagestyle{empty}

\begin{table}[htbp]\centering
\begin{threeparttable}\footnotesize

\caption{Trends in Disruptiveness Using Independently-Developed Measures of Disruptiveness\\(Excluding Papers with 0 Backward Citations)}
\label{table:RegressionAlternativeMeasures}
\begin{tabular}{@{\hskip 0.1in}l@{\hskip 0.1in}cccccccc@{\hskip 0.1in}}
\toprule
  & \multicolumn{4}{c}{Year} & \multicolumn{2}{c}{Constant} & & \\
\cmidrule(lr){2-5} \cmidrule(lr){6-7}
 Dependent Variable (Percentile) & Coefficient & SE & P & CI & Constant & SE & R² & N  \\
\midrule
\multicolumn{9}{l}{\textbf{Web of Science}} \\
  CYG$_5$ & -0.652 & 0.000 & 0.000 & [-0.653, -0.652] & 1350.542 & 0.823 & 0.101 & 21113191 \\
  Is D$_5$ & -0.078 & 0.000 & 0.000 & [-0.078, -0.077] & 204.568 & 0.575 & 0.003 & 25956142 \\
  CD$_5^5$ & -0.340 & 0.000 & 0.000 & [-0.341, -0.340] & 728.206 & 0.776 & 0.029 & 25578613 \\
  CD$_5^{noK}$ & -0.158 & 0.000 & 0.000 & [-0.159, -0.157] & 365.228 & 0.894 & 0.006 & 21113191 \\
\multicolumn{9}{l}{\textbf{SciSciNet}} \\
  CYG$_5$ & -0.475 & 0.000 & 0.000 & [-0.476, -0.474] & 998.311 & 0.965 & 0.043 & 22544581 \\
  Is D$_5$ & -0.031 & 0.000 & 0.000 & [-0.032, -0.030] & 111.983 & 0.576 & 0.000 & 30004635 \\
  CD$_5^5$ & -0.255 & 0.000 & 0.000 & [-0.255, -0.254] & 558.157 & 0.807 & 0.014 & 29601318 \\
  CD$_5^{noK}$ & -0.058 & 0.000 & 0.000 & [-0.059, -0.057] & 166.448 & 0.985 & 0.001 & 22544581 \\
\bottomrule
\end{tabular}

\begin{tablenotes}
\item \emph{Notes:} This table demonstrates that our finding of a persistent decline in disruptiveness even when 0-bcite papers are excluded is not sensitive to the choice of disruption metric or data set. Specifically, the table reports regression results showing trends in disruptiveness over time using four independently developed alternative measures of disruptiveness, for both Web of Science and WoS. Each row corresponds to a regression where the dependent variable is one of the measures. CYG$_5$ (Citation Year Gap) measures the average age gap between references cited by citing works relative to the focal paper \citep{brauer2023aggregate, brauer2024searching}. Is D$_5$ is a binary indicator for whether the CD$_5$ value (disruptiveness index) is positive (i.e., whether the paper disrupts prior work) \citep{lin2023remote, li2024displacing, yang2024female}. CD$_5^{noK}$ is a variation of the CD$_5$ index that excludes references-only citations (sometimes referred to as the ``$n_K$'' term) from the denominator \citep{bornmann2020disruption, bornmann2020disruptive, leibel2024we, leydesdorff2020proposal, deng2023enhancing}. CD$_5^5$ introduces a threshold requiring future works to cite multiple references of the focal paper, emphasizing stronger bibliographic coupling \citep{li2024displacing, wu2019solo, jiang2024new, bornmann2020disruption, leibel2024we}. All measures exclude 0-bcite documents and are percentile-normalized to enable comparison across scales. The coefficients represent the change in disruptiveness per year. Statistically significant declines in disruptiveness are observed across all measures and datasets. These results underscore that the observed trend is robust to alternative operationalizations of disruptiveness and persists even when 0-bcite documents are excluded. In our original paper, PLF, we also reported results for two additional measures of disruption, as shown in Extended Data Fig.~7, further supporting the robustness of these findings.
\end{tablenotes}

\end{threeparttable}
\end{table}

}%

\clearpage
\section{Persistent Decline Using HATWG's Proposed Regression Model}\label{sec:HolstRegression}

{


\setlength{\tabcolsep}{10pt}

\renewcommand{\arraystretch}{0.8}

\thispagestyle{empty}

\begin{table}[H]\centering
\scriptsize
\centering
\caption{Trends in Disruptiveness Adjusted Using HATWG's Proposed Regression Model}
\label{table:RegressionAdjustment}
\begin{threeparttable}
\begin{tabular}{lcccccc}
\toprule

& \multicolumn{3}{c}{Papers (Web of Science)} & \multicolumn{3}{c}{Patents (Patents View)} \\\cmidrule(lr){2-4} \cmidrule(lr){5-7}
&	              b   	&	             Robust SE	&	              p-value	&	              b   	&	             Robust SE	&	              p-value\\\midrule
Year=1946                                    	&	        -0.0013   	&	         0.0022	&	         0.5510	&		&		&	\\
Year=1947                                    	&	        -0.0006   	&	         0.0021	&	         0.7775	&		&		&	\\
Year=1948                                    	&	        -0.0027   	&	         0.0020	&	         0.1882	&		&		&	\\
Year=1949                                    	&	        -0.0052** 	&	         0.0020	&	         0.0089	&		&		&	\\
Year=1950                                    	&	        -0.0070***	&	         0.0019	&	         0.0004	&		&		&	\\
Year=1951                                    	&	        -0.0116***	&	         0.0019	&	         0.0000	&		&		&	\\
Year=1952                                    	&	        -0.0137***	&	0.0019	&	0.0000	&		&		&	\\
Year=1953                                    	&	        -0.0175***	&	         0.0019	&	         0.0000	&		&		&	\\
Year=1954                                    	&	        -0.0171***	&	         0.0019	&	         0.0000	&		&		&	\\
Year=1955                                    	&	        -0.0156***	&	0.0019	&	0.0000	&		&		&	\\
Year=1956                                    	&	        -0.0183***	&	0.0019	&	0.0000	&		&		&	\\
Year=1957                                    	&	        -0.0192***	&	0.0018	&	0.0000	&		&		&	\\
Year=1958                                    	&	        -0.0202***	&	0.0018	&	0.0000	&		&		&	\\
Year=1959                                    	&	        -0.0218***	&	0.0018	&	0.0000	&		&		&	\\
Year=1960                                    	&	        -0.0239***	&	0.0018	&	0.0000	&		&		&	\\
Year=1961                                    	&	        -0.0239***	&	0.0017	&	0.0000	&		&		&	\\
Year=1962                                    	&	        -0.0258***	&	0.0017	&	0.0000	&		&		&	\\
Year=1963                                    	&	        -0.0258***	&	0.0017	&	0.0000	&		&		&	\\
Year=1964                                    	&	        -0.0233***	&	0.0017	&	0.0000	&		&		&	\\
Year=1965                                    	&	        -0.0298***	&	0.0017	&	0.0000	&		&		&	\\
Year=1966                                    	&	        -0.0307***	&	0.0017	&	0.0000	&		&		&	\\
Year=1967                                    	&	        -0.0315***	&	0.0017	&	0.0000	&		&		&	\\
Year=1968                                    	&	        -0.0342***	&	0.0017	&	0.0000	&		&		&	\\
Year=1969                                    	&	        -0.0356***	&	0.0017	&	0.0000	&		&		&	\\
Year=1970                                    	&	        -0.0386***	&	0.0017	&	0.0000	&		&		&	\\
Year=1971                                    	&	        -0.0398***	&	0.0017	&	0.0000	&		&		&	\\
Year=1972                                    	&	        -0.0417***	&	0.0017	&	0.0000	&		&		&	\\
Year=1973                                    	&	        -0.0426***	&	0.0017	&	0.0000	&		&		&	\\
Year=1974                                    	&	        -0.0427***	&	0.0017	&	0.0000	&		&		&	\\
Year=1975                                    	&	        -0.0453***	&	0.0017	&	0.0000	&		&		&	\\
Year=1976                                    	&	        -0.0467***	&	0.0017	&	0.0000	&		&		&	\\
Year=1977                                    	&	        -0.0473***	&	0.0017	&	0.0000	&		&		&	\\
Year=1978                                    	&	        -0.0481***	&	0.0017	&	0.0000	&		&		&	\\
Year=1979                                    	&	        -0.0491***	&	0.0017	&	0.0000	&		&		&	\\
Year=1980                                    	&	        -0.0513***	&	0.0017	&	0.0000	&		&		&	\\
Year=1981                                    	&	        -0.0516***	&	0.0017	&	0.0000	&	0.0002	&	0.002	&	         0.9236\\
Year=1982                                    	&	        -0.0529***	&	0.0017	&	0.0000	&	        -0.0068** 	&	0.0021	&	         0.0011\\
Year=1983                                    	&	        -0.0533***	&	0.0017	&	0.0000	&	        -0.0177***	&	0.0021	&	         0.0000\\
Year=1984                                    	&	        -0.0539***	&	0.0017	&	0.0000	&	        -0.0205***	&	0.002	&	         0.0000\\
Year=1985                                    	&	        -0.0551***	&	0.0017	&	0.0000	&	        -0.0257***	&	0.0019	&	         0.0000\\
Year=1986                                    	&	        -0.0558***	&	0.0017	&	0.0000	&	        -0.0335***	&	0.0019	&	         0.0000\\
Year=1987                                    	&	        -0.0572***	&	0.0017	&	0.0000	&	        -0.0395***	&	0.0018	&	         0.0000\\
Year=1988                                    	&	        -0.0580***	&	0.0017	&	0.0000	&	        -0.0525***	&	0.0018	&	         0.0000\\
Year=1989                                    	&	        -0.0583***	&	0.0017	&	0.0000	&	        -0.0616***	&	0.0018	&	         0.0000\\
Year=1990                                    	&	        -0.0592***	&	0.0017	&	0.0000	&	        -0.0668***	&	0.0018	&	         0.0000\\
Year=1991                                    	&	        -0.0594***	&	0.0017	&	0.0000	&	        -0.0738***	&	0.0018	&	         0.0000\\
Year=1992                                    	&	        -0.0612***	&	0.0017	&	0.0000	&	        -0.0832***	&	0.0018	&	         0.0000\\
Year=1993                                    	&	        -0.0618***	&	0.0017	&	0.0000	&	        -0.0909***	&	0.0018	&	         0.0000\\
Year=1994                                    	&	        -0.0627***	&	0.0017	&	0.0000	&	        -0.0970***	&	0.0018	&	         0.0000\\
Year=1995                                    	&	        -0.0641***	&	0.0017	&	0.0000	&	        -0.1013***	&	0.0018	&	         0.0000\\
Year=1996                                    	&	        -0.0655***	&	0.0017	&	0.0000	&	        -0.1075***	&	0.0018	&	         0.0000\\
Year=1997                                    	&	        -0.0663***	&	0.0017	&	0.0000	&	        -0.1112***	&	0.0018	&	         0.0000\\
Year=1998                                    	&	        -0.0673***	&	0.0017	&	0.0000	&	        -0.1153***	&	0.0018	&	         0.0000\\
Year=1999                                    	&	        -0.0679***	&	0.0017	&	0.0000	&	        -0.1164***	&	0.0018	&	         0.0000\\
Year=2000                                    	&	        -0.0684***	&	0.0017	&	0.0000	&	        -0.1243***	&	0.0018	&	         0.0000\\
Year=2001                                    	&	        -0.0691***	&	0.0017	&	0.0000	&	        -0.1313***	&	0.0018	&	         0.0000\\
Year=2002                                    	&	        -0.0701***	&	0.0017	&	0.0000	&	        -0.1388***	&	0.0018	&	         0.0000\\
Year=2003                                    	&	        -0.0711***	&	0.0017	&	0.0000	&	        -0.1460***	&	0.0018	&	         0.0000\\
Year=2004                                    	&	        -0.0730***	&	0.0017	&	0.0000	&	        -0.1501***	&	0.0019	&	         0.0000\\
Year=2005                                    	&	        -0.0746***	&	0.0017	&	0.0000	&	        -0.1533***	&	0.0019	&	         0.0000\\
Year=2006                                    	&	        -0.0758***	&	0.0017	&	0.0000	&	        -0.1583***	&	0.0019	&	         0.0000\\
Year=2007                                    	&	        -0.0767***	&	0.0017	&	0.0000	&	        -0.1571***	&	0.0019	&	         0.0000\\
Year=2008                                    	&	        -0.0780***	&	0.0017	&	0.0000	&	        -0.1558***	&	0.0019	&	         0.0000\\
Year=2009                                    	&	        -0.0798***	&	0.0017	&	0.0000	&	        -0.1556***	&	0.002	&	         0.0000\\
Year=2010                                    	&	        -0.0818***	&	0.0017	&	0.0000	&	        -0.1547***	&	0.002	&	         0.0000\\
Zero papers/patents cited (1=Yes)            	&	         0.9791***	&	         0.0001	&	         0.0000	&	         0.9024***	&	         0.0003	&	         0.0000\\
Number of papers/patents cited               	&	        -0.0006***	&	         0.0000	&	         0.0000	&	        -0.0006***	&	         0.0000	&	         0.0000\\
Number of new papers/patents        	&	         0.0000***	&	         0.0000	&	         0.0000	&	         0.0000***	&	         0.0000	&	         0.0000\\
Mean number of papers/patents cited          	&	         0.0007***	&	         0.0000	&	         0.0000	&	         0.0008***	&	         0.0000	&	         0.0000\\
Mean number of authors/inventors per paper/patent	&	         0.0083***	&	         0.0001	&	         0.0000	&	         0.0177***	&	         0.0014	&	         0.0000\\
Number of unlinked references                	&	         0.0004***	&	         0.0000	&	         0.0000	&		&		&	\\
Constant                                     	&	         0.0261***	&	0.0017	&	         0.0000&	         0.1260***	&	0.003	&	         0.0000\\	\midrule
Subfield fixed effects                       	&	            Yes   	&	               	&	               	&	            Yes   	&	               	&	               \\\midrule
N                                            	&	22456096	&	               	&	               	&	2926923	&	               	&	               \\
R2                                           	&	0.7776	&	               	&	               	&	0.518	&	               	&	\\\bottomrule

\end{tabular}

\begin{tablenotes}
\item \emph{Notes:} Estimates are from ordinary least squares (OLS) regressions with robust standard errors. The dependent variable is the CD$_5$ index. Following HATWG's recommendations, the models include a binary control variable for \emph{Zero papers/patents cited (1=Yes)}, addressing concerns about the potential influence of 0-bcite documents on the observed decline in disruptiveness. Additionally, both models incorporate adjustments for many potential changes in publication, citation, and authorship practices, using the exact variables from the original PLF analysis. These include adjustments at the field $\times$ year level---\emph{Number of new papers/patents}, \emph{Mean number of papers/patents cited}, \emph{Mean number of authors/inventors per paper/patent}---and at the paper/patent level---\emph{Number of papers/patents cited}, \emph{Number of unlinked references}. Subfield fixed effects are included to account for time-invariant differences across fields. The reference categories for the year indicators are 1945 (papers) and 1980 (patents). Each coefficient is tested against the null hypothesis of being equal to 0 using a two-sided t-test. These results suggest that the observed decline in disruptiveness is robust, even under HATWG's proposed regression adjustments. 
\item *$p<0.05$, **$p<0.01$, ***$p<0.001$. 
\end{tablenotes}

\end{threeparttable}
\end{table}

}%

\clearpage
\newgeometry{margin=1in}

\section{Mathematical Properties and Behavior of the CD Index}\label{sec:MathematicalProperties}

\subsection{Overview of the CD Index}\label{sec:MathematicalPropertiesOverview}
The primary indicator of disruptiveness employed in PLF, and also referenced in HATWG and this commentary, is the CD index \citep{funk2017dynamic}. While this measure has been described and validated extensively elsewhere \citep{li2024displacing, leibel2024we, bornmann2021convergent, bornmann2020disruptive, lin2023remote, wang2021science, wu2019large, li2024productive}, we provide a brief overview here to establish a common reference point and fix notation.

Let $f$ denote a focal work for which we would like to calculate the CD index. Denote the set of papers that cite $f$ by $C_f$. Denote the set of papers that $f$ cites (the backwards citations) as $B_f$. Let $C_B$ denote the set of papers that cite the backwards citations of $f$. 

Given these definitions, we can define $J$-type citations to the focal paper $J_f$ as the papers that cite both $f$ and at least one of $f$'s backwards citations $J_f = C_f \cap C_B$. Simplifying notation slightly, let $n_J(f) = |J_f|$ denote the number of papers that make these $J$-type citations to the focal paper.

We can then define $I$-type citations to $f$--the papers that cite only the focal paper but not any of its backward citations--as the residual set of papers $I_f = C_f - J_f$. Similarly, we denote the size of this set by $n_I(f) = |I_f|$.

Finally, the collection of $K$-type citations $K_f$ are those that cite the focal paper's backwards citations but do not cite $f$ directly. This is again defined as a residual from $J_f$: $K_f = C_B - J_f$. Again we simplify notation and denote the size of this set by $n_K(f) = |K_f|$. 

We may then define the CD index for a focal paper $f$ as:
\begin{equation}
    \mathrm{CD}(f) = \frac{n_I(f) - n_J(f)}{n_I(f) + n_J(f) + n_K(f)}.
\end{equation}

In the mathematical analyses that follow, we focus primarily on a single (arbitrary) focal paper, so we will simplify notation slightly to make $f$ implicit, denoting the CD index as: 

\begin{equation}
    \mathrm{CD}(f) = \frac{n_I - n_J}{n_I + n_J + n_K}.
\end{equation} 
\begin{equation}
    \mathrm{CD} = \frac{n_I - n_J}{n_I + n_J + n_K}.
\end{equation}

Empirical applications of the CD index typically define a ``forward window'' of relevance, denoted by $t$, to limit the time period for observing future citations to the focal paper and its references \citep{leibel2024we}. This forward window ensures comparability across papers published at different times, as it equalizes the opportunity for accumulating citations. Following PLF, a 5-year forward window is adopted, and the resulting metric is denoted $\mathrm{CD}_5$. For simplicity, the subscript $t$ is omitted in the mathematical analyses below, as the specific length of the forward window is not relevant to the theoretical properties being examined.

\subsection{Randomly Rewired Citation Networks}

The Monte Carlo random rewiring approach taken in PLF and HATWG preserves the in- and out-degree of all nodes within the citation network in addition to the citing and cited years of each edge. For simplicity, we analyze the rewiring behavior between two years $t_b < t_c$. Let $n_c$ denote the number of papers in year $t_c$. The argument outlined below can be repeated for each pair of citing/cited years to arrive at the expected behavior of the entire Monte Carlo rewiring process. 

The random rewiring results in a citation structure approximated by a configuration model, though there is some nuance derived from the fact that the rewiring is done in a bipartite manner between citing and cited papers. Consider two papers $i$ and $j$ where paper $j$ was published in year $t_c$ and paper $i$ in $t_b$. Let $c_i$ denote the number of citations received by paper $i$ and $b_j$ the number of backwards citations made by paper $j$. In other words, $c_i$ is the in-degree of paper/node $i$ and $b_j$ is the out-degree of paper/node $j$. An in- and out-degree preserving reshuffling of the network will result in $i$ and $j$ being connected with probability $$p_{ij} = \frac{c_i b_j}{2m}$$ for large number of citations in the network $m$. This is a well-known property of the configuration model \citep{newman2018networks}.

As noted in HATWG's commentary, $J$-type citations are papers that form a triadic relationship with the focal paper and at least one of its backwards citations. We can show that the probability of $J$-type citations forming will go to zero as the number of papers in the citation network becomes large. This can be shown by analyzing the probability of triangle formation within a randomly rewired citation networks with degree preservation.

To calculate the probability $p_{hij}$ that $j$ is a common neighbor of a pair of papers $h$ and $i$, we take the product of each of their independent probabilities of being connected, accounting for decrease in probability of $j$ making one connection before the other:

\begin{align}
    p_{hij} &= \sum\limits_{j}\left(\frac{c_ib_j}{2m}\right)\left( \frac{c_h (b_j - 1)}{2m} \right) \\ 
    &= \left(\frac{c_ic_h}{2m}\right) \sum\limits_{j}\frac{b_j(b_j - 1)}{\langle b \rangle n} \\ 
    &= \left(\frac{c_ic_h}{2m}\right) \frac{\langle b^2 \rangle - \langle b \rangle}{\langle b \rangle} \\
    &= \left( \frac{c_i c_h}{\langle b \rangle n_c} \right) \frac{\langle b^2 \rangle - \langle b \rangle}{\langle b \rangle}
\end{align} where $\langle b \rangle$ = $2m / n_c$ is the average number of citations made by papers in year $t_c$. 

In other words, the probability that $j$ co-cites papers $h$ and $i$ is given by the probability that $i$ and $h$ would both be cited weighted by the degree distribution of all citing papers. 

As the number of citing papers $n_c$ (or the number of citations between the years $m$) increases, the probability of a $J$-type citation forming $p_{hij}$ decreases across any pair of co-cited papers $i$ and $h$.

In real citation networks, the number of papers in a year far exceeds the average number of citations made by each paper \citep{milojevic2015quantifying, bornmann2015growth, wang2021science}, and the second moment is bounded. The average number of citations received by papers is also small, on the order of 9 in the PLF WoS data, so any product $c_i c_h$ will be much smaller than $n_c$ in expectation as well. Taken together, this implies $p_{hij}$--and therefore the presence of $J$-type papers--will be vanishingly small for randomly rewired networks.


\subsection{Mathematical Properties of the CD Index in Large, Randomly Rewired Citation Networks}\label{sec:MathematicalPropertiesLargeRandom}

Because $J$-type papers will be rare in degree-preserved random citation networks, the CD index will be driven almost purely by $I$- and $K$-type citations. In fact, due to the vanishing presence of $J$-type citations, any aggregate or temporal patterns in the CD index must be due to the effect of $K$-type citations picking up on changes in citation behavior. 

Recall that the term $n_I + n_J = |C_f|$ in the denominator of the CD index must be the same in the real, observed citation network as in the randomly rewired networks. This is due to the fact that the randomly rewired networks are in- and out-degree preserving and therefore preserve citation counts to each paper. Because $J$-type papers are extremely unlikely to appear in large-scale citation networks that have been edge rewired, we have that $n_J \to 0$ as $n_c \to \infty$, and for an arbitrary focal paper $f$ in a randomly rewired network:

\begin{align}
    \lim\limits_{n_c \to \infty} \mathrm{CD}(f) &= \frac{n_I}{n_I + n_K} \\
    &= \frac{|C_f|}{|C_f| + n_K}.
\end{align}

We can rewrite $K$-type citations as the sum of the citations received by the backwards cites of $f$, resulting in:

\begin{align}
    \lim\limits_{n_c \to \infty} \mathrm{CD}(f) &= \frac{n_I}{n_I + \sum\limits_{i \in B_f} c_i}.
\end{align}

\begin{figure}[htbp]\centering
\centering
\includegraphics[width=1\textwidth]{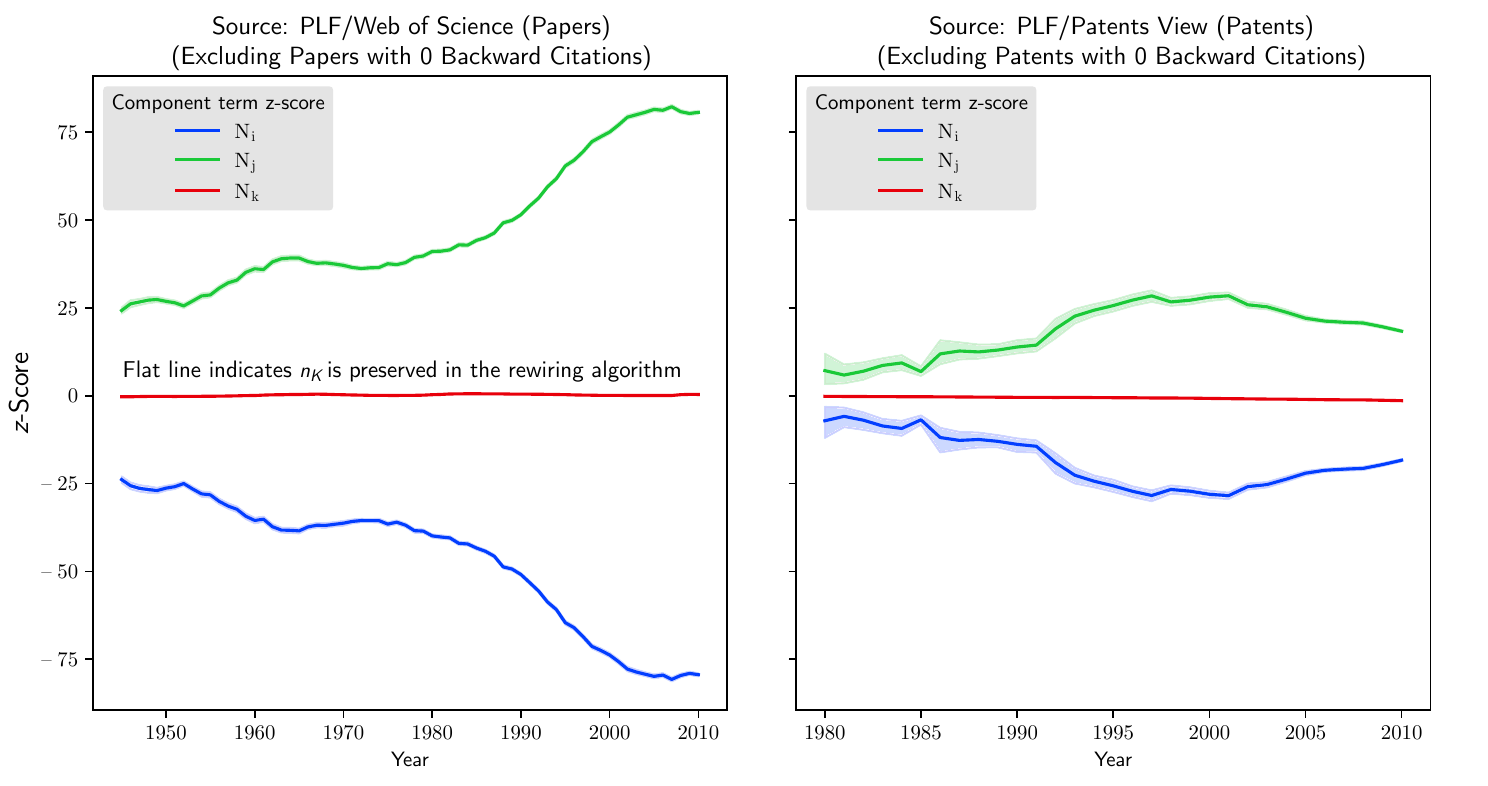}
\caption{\textbf{Comparison of Disruptiveness Component Terms in Observed/Randomly Rewired Citation Networks.} This figure demonstrates why HATWG find a decline in mean CD$_5$ across both observed and randomly rewired networks by showing that $n_K$ is effectively preserved in the rewiring process. The figure displays $z$-scores comparing three CD$_5$ components---$n_I$ (number of works citing the focal work), $n_J$ (number of works citing both the focal work and its references), and $n_K$ (number of works citing only the references) between observed and rewired citation networks. All 0-bcite works are excluded. Left and right panels correspond to Web of Science (papers) and Patents View (patents) datasets, respectively. The $z$-scores measure the deviations of the observed component values (for individual papers/patents) from the expected values in the rewired networks. Consistent with Sec.~\ref{sec:MathematicalProperties}, $n_K$ exhibits a flat trend, indicating that $n_K$ is effectively preserved by the rewiring algorithm. This preservation of $n_K$ makes HATWG's mean comparison approach misleading, because a key component of the CD$_5$ index remains identical (by design) across the observed and rewired networks. Instead, one must `net-out' the temporal citation effects in order to properly compare the CD index on observed data to randomly rewired null models (c.f., \cite{newman2001structure, uzzi2005collaboration, opsahl2008prominence, uzzi2013atypical, mukherjee2016new, yang2022gender, kedrick2024conceptual}), as was done in PLF's original $z$-score analysis. Shaded bands correspond to 95\% confidence intervals.}
\label{fig:RandomNetworkComponents}
\end{figure}

\begin{figure}[h!]
\centering
\includegraphics[width=\textwidth]{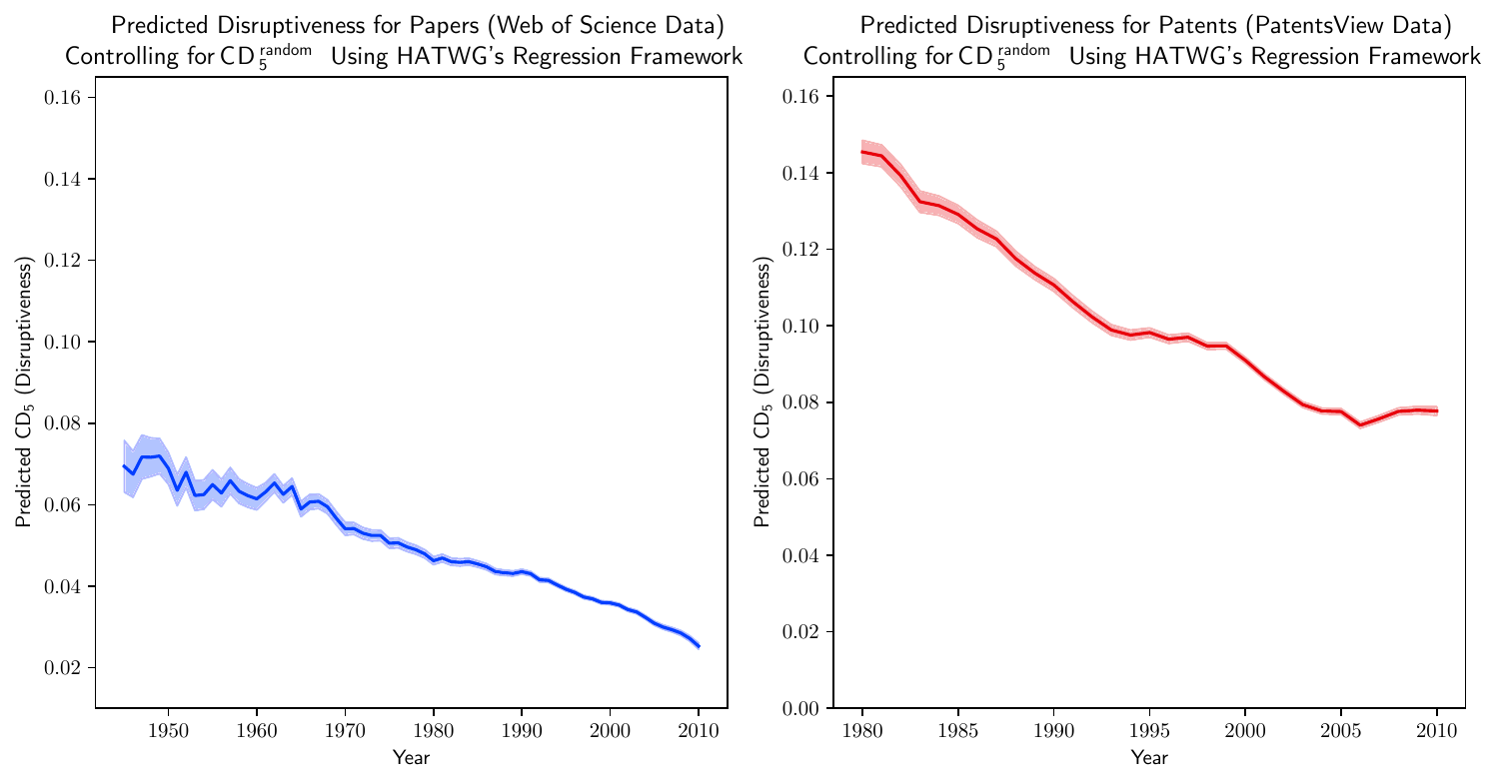}
\caption{\textbf{Persistent Decline in Disruptiveness of Papers and Patents After Adjusting for Disruptiveness in Randomly Rewired Citation Networks.} This figure visualizes the predicted values of the CD$_5$ index (disruptiveness) for papers (left panel) and patents (right panel), based on the regression results in Table~\ref{table:RandomRegressionAdjustment}. The analysis is presented as an alternative approach to PLF's $z$-score analysis (see their Extended Data Figure 8), which was designed to `net-out' the disruptiveness attributable to changes in citation practices over time at the level of individual papers/patents. For each paper/patent, the regression model includes control (CD$_5^{random}$) for the disruptiveness calculated for the same paper/patent in 10 randomly rewired citation networks. In addition, the models include a 0-bcite dummy variable, following HATWG, along with the full suite of control variables included in PLF's original regression model (see their Extended Data Figure 8) and field fixed effects. The shaded regions represent 95\% confidence intervals for the predicted values. Even after these extensive adjustments, the analysis demonstrates a persistent, statistically significant decline in CD$_5$ for both papers and patents, thereby corroborating PLF's $z$-score analysis (Extended Data Figure 8) and providing a further test of the robustness of the observed decline in disruptiveness. Shaded bands correspond to 95\% confidence intervals.}
\label{fig:RandomNetworkRegressionPlot}
\end{figure}

\begin{figure}[htbp]\centering
\centering
\includegraphics[width=1.0\textwidth]{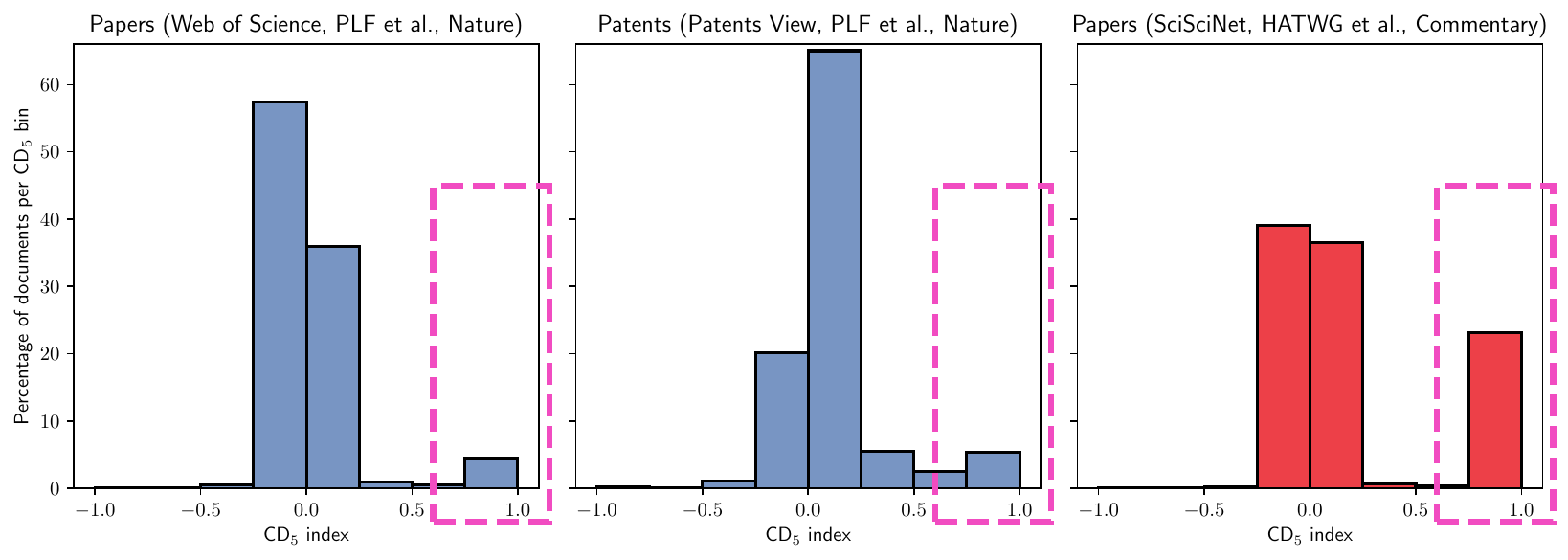}
\caption{\textbf{Severe Overrepresentation of CD=1 Works in HATWG's SciSciNet Data.} This figure compares the distribution of CD index values between PLF's datasets (papers in Web of Science and patents in Patents View) and HATWG's SciSciNet data. The plots reveal a dramatic difference in the proportion of documents with CD=1, which are central to HATWG's critique. Specifically, in PLF, the proportion of CD=1 documents is 4.3\% for Web of Science papers and 4.9\% for patents. In contrast, HATWG's SciSciNet data contains 23.1\% of documents with CD=1. This corresponds to 4.4 to 5.4 times more CD=1 documents in HATWG's data compared to PLF's. Such substantial inflation raises serious concerns about the quality of HATWG's data, especially as it relates to the content of their critique. Because CD=1 works are central to their argument that 0-bcite works should be excluded, understanding the source of this overrepresentation of CD=1 documents in HATWG's dataset is essential.}
\label{fig:Distributions}
\end{figure}

Therefore, on large rewired citation networks, the CD index will be almost surely driven by aggregate citation patterns on the network. By extension, any temporal trends in the CD index on randomly rewired networks will reflect changes in these citation patterns, namely, the average number of citations given and received within each year. These predictions are verified empirically in Fig.~\ref{fig:RandomNetworkComponents}.

Because temporal trends in aggregate citation patterns are known to exist \citep{chu2021slowed, varga2019shorter}, trends in randomly-rewired CD index measurements will also exist. Therefore, one must `net-out' the temporal citation effects in order to properly compare the CD index on observed data to randomly rewired null models (c.f., \cite{newman2001structure, uzzi2005collaboration, opsahl2008prominence, uzzi2013atypical, mukherjee2016new, yang2022gender, kedrick2024conceptual}). This is precisely the approach PLF used in their original paper \citep{park2023papers}, where $z$-scores were used to compare each paper's observed CD$_5$ value against the mean CD$_5^{random}$ value for the same paper in 10 randomly rewired citation networks.

Alternatively, as noted in the main text, if one wishes to follow HATWG's approach and visually compare means between observed and random networks, an alternative disruption metric must be used that is unaffected by the preservation of $n_K$ in the rewiring. The CD$_5^{noK}$ index, a previously developed and validated variation on the CD$_5$ index that excludes $n_K$,  fulfills this requirement \citep{bornmann2020disruption, bornmann2020disruptive, leibel2024we, leydesdorff2020proposal, deng2023enhancing}. As shown in Fig.~\ref{fig:RandomNetworkComparisonCDnoK}, analysis using CD$_5^{noK}$ reveals a persistent decline in disruptiveness within observed networks, while rewired networks maintain a stable trend.

\clearpage
\newgeometry{margin=0.5in}

\thispagestyle{empty}

\clearpage
\section{Persistent Decline in Disruptiveness Relative to Randomly Rewired Citation Networks}\label{sec:ObservedVsRandom}

\thispagestyle{empty}

\clearpage

{

\thispagestyle{empty}

\setlength{\tabcolsep}{12pt}

\renewcommand{\arraystretch}{0.7}

\begin{table}[ht!]
\scriptsize
\centering
\caption{Trends in Disruptiveness Accounting for Disruptiveness in Comparable Random Networks}
\label{table:RandomRegressionAdjustment}
\begin{threeparttable}
\begin{tabular}{lccccccccc}
\toprule
& \multicolumn{3}{c}{Papers (Web of Science)} & \multicolumn{3}{c}{Patents (Patents View)} \\\cmidrule(lr){2-4} \cmidrule(lr){5-7}
&	              b   	&	             Robust SE	&	              p-value	&	              b   	&	             Robust SE	&	              p-value\\\midrule
Year=1946                                    &        -0.0019   &         0.0043&         0.6502&                  &               &               \\
Year=1947                                    &         0.0023   &         0.0042&         0.5846&                  &               &               \\
Year=1948                                    &         0.0022   &         0.0040&         0.5761&                  &               &               \\
Year=1949                                    &         0.0025   &         0.0038&         0.5112&                  &               &               \\
Year=1950                                    &        -0.0005   &         0.0037&         0.8893&                  &               &               \\
Year=1951                                    &        -0.0059   &         0.0037&         0.1127&                  &               &               \\
Year=1952                                    &        -0.0015   &         0.0037&         0.6906&                  &               &               \\
Year=1953                                    &        -0.0072   &         0.0037&         0.0513&                  &               &               \\
Year=1954                                    &        -0.0070   &         0.0037&         0.0567&                  &               &               \\
Year=1955                                    &        -0.0045   &         0.0037&         0.2197&                  &               &               \\
Year=1956                                    &        -0.0066   &         0.0036&         0.0680&                  &               &               \\
Year=1957                                    &        -0.0035   &         0.0036&         0.3232&                  &               &               \\
Year=1958                                    &        -0.0061   &         0.0035&         0.0811&                  &               &               \\
Year=1959                                    &        -0.0072*  &         0.0035&         0.0386&                  &               &               \\
Year=1960                                    &        -0.0080*  &         0.0035&         0.0205&                  &               &               \\
Year=1961                                    &        -0.0063   &         0.0034&         0.0618&                  &               &               \\
Year=1962                                    &        -0.0041   &         0.0034&         0.2233&                  &               &               \\
Year=1963                                    &        -0.0069*  &         0.0033&         0.0392&                  &               &               \\
Year=1964                                    &        -0.0050   &         0.0033&         0.1380&                  &               &               \\
Year=1965                                    &        -0.0105** &         0.0033&         0.0016&                  &               &               \\
Year=1966                                    &        -0.0088** &         0.0033&         0.0081&                  &               &               \\
Year=1967                                    &        -0.0086** &         0.0033&         0.0093&                  &               &               \\
Year=1968                                    &        -0.0099** &         0.0033&         0.0026&                  &               &               \\
Year=1969                                    &        -0.0128***&         0.0033&         0.0001&                  &               &               \\
Year=1970                                    &        -0.0153***&         0.0033&         0.0000&                  &               &               \\
Year=1971                                    &        -0.0153***&         0.0033&         0.0000&                  &               &               \\
Year=1972                                    &        -0.0164***&         0.0033&         0.0000&                  &               &               \\
Year=1973                                    &        -0.0170***&         0.0033&         0.0000&                  &               &               \\
Year=1974                                    &        -0.0170***&         0.0033&         0.0000&                  &               &               \\
Year=1975                                    &        -0.0189***&         0.0033&         0.0000&                  &               &               \\
Year=1976                                    &        -0.0188***&         0.0033&         0.0000&                  &               &               \\
Year=1977                                    &        -0.0198***&         0.0032&         0.0000&                  &               &               \\
Year=1978                                    &        -0.0205***&         0.0032&         0.0000&                  &               &               \\
Year=1979                                    &        -0.0215***&         0.0032&         0.0000&                  &               &               \\
Year=1980                                    &        -0.0232***&         0.0032&         0.0000&                  &               &               \\
Year=1981                                    &        -0.0225***&         0.0032&         0.0000&        -0.0010   &         0.0019&         0.5851\\
Year=1982                                    &        -0.0234***&         0.0032&         0.0000&        -0.0062** &         0.0019&         0.0011\\
Year=1983                                    &        -0.0236***&         0.0032&         0.0000&        -0.0130***&         0.0019&         0.0000\\
Year=1984                                    &        -0.0234***&         0.0032&         0.0000&        -0.0140***&         0.0018&         0.0000\\
Year=1985                                    &        -0.0240***&         0.0032&         0.0000&        -0.0163***&         0.0017&         0.0000\\
Year=1986                                    &        -0.0247***&         0.0032&         0.0000&        -0.0201***&         0.0017&         0.0000\\
Year=1987                                    &        -0.0258***&         0.0032&         0.0000&        -0.0228***&         0.0017&         0.0000\\
Year=1988                                    &        -0.0261***&         0.0032&         0.0000&        -0.0279***&         0.0017&         0.0000\\
Year=1989                                    &        -0.0263***&         0.0032&         0.0000&        -0.0317***&         0.0016&         0.0000\\
Year=1990                                    &        -0.0259***&         0.0032&         0.0000&        -0.0348***&         0.0016&         0.0000\\
Year=1991                                    &        -0.0264***&         0.0032&         0.0000&        -0.0392***&         0.0016&         0.0000\\
Year=1992                                    &        -0.0279***&         0.0032&         0.0000&        -0.0431***&         0.0016&         0.0000\\
Year=1993                                    &        -0.0280***&         0.0032&         0.0000&        -0.0465***&         0.0016&         0.0000\\
Year=1994                                    &        -0.0291***&         0.0032&         0.0000&        -0.0479***&         0.0016&         0.0000\\
Year=1995                                    &        -0.0302***&         0.0033&         0.0000&        -0.0472***&         0.0016&         0.0000\\
Year=1996                                    &        -0.0310***&         0.0033&         0.0000&        -0.0489***&         0.0016&         0.0000\\
Year=1997                                    &        -0.0321***&         0.0033&         0.0000&        -0.0484***&         0.0016&         0.0000\\
Year=1998                                    &        -0.0325***&         0.0033&         0.0000&        -0.0507***&         0.0016&         0.0000\\
Year=1999                                    &        -0.0334***&         0.0033&         0.0000&        -0.0507***&         0.0016&         0.0000\\
Year=2000                                    &        -0.0335***&         0.0033&         0.0000&        -0.0545***&         0.0016&         0.0000\\
Year=2001                                    &        -0.0341***&         0.0033&         0.0000&        -0.0588***&         0.0016&         0.0000\\
Year=2002                                    &        -0.0352***&         0.0033&         0.0000&        -0.0625***&         0.0017&         0.0000\\
Year=2003                                    &        -0.0358***&         0.0033&         0.0000&        -0.0660***&         0.0017&         0.0000\\
Year=2004                                    &        -0.0371***&         0.0033&         0.0000&        -0.0677***&         0.0017&         0.0000\\
Year=2005                                    &        -0.0385***&         0.0033&         0.0000&        -0.0679***&         0.0017&         0.0000\\
Year=2006                                    &        -0.0395***&         0.0033&         0.0000&        -0.0714***&         0.0017&         0.0000\\
Year=2007                                    &        -0.0401***&         0.0033&         0.0000&        -0.0697***&         0.0017&         0.0000\\
Year=2008                                    &        -0.0409***&         0.0033&         0.0000&        -0.0678***&         0.0018&         0.0000\\
Year=2009                                    &        -0.0423***&         0.0033&         0.0000&        -0.0675***&         0.0018&         0.0000\\
Year=2010                                    &        -0.0441***&         0.0033&         0.0000&        -0.0677***&         0.0019&         0.0000\\
CD$_5^{random}$                                    &         0.3073***&         0.0015&         0.0000&         0.4302***&         0.0010&         0.0000\\
Zero papers/patents cited (1=Yes)            &         0.6882***&         0.0015&         0.0000&         0.5359***&         0.0010&         0.0000\\
Constant                                     &        -0.0141***&         0.0032&         0.0000&         0.0419***&         0.0028&         0.0000\\\midrule
Controls                       &            Yes   &               &               &            Yes   &               &               \\
Subfield fixed effects                       &            Yes   &               &               &            Yes   &               &               \\
\midrule
N                                            &       56537044   &               &               &       29183917   &               &               \\
R2                                           &         0.7874   &               &               &         0.5801   &               &               \\
\bottomrule
\end{tabular}

\begin{tablenotes}
\item \emph{Notes:} This table reports regression results analyzing trends in the CD$_5$ index (disruptiveness) for papers in Web of Science and patents in Patents View, using a framework adapted from HATWG's proposed regression model. In addition to the controls (e.g., number of papers/patents cited, field $\times$ year level adjustments for number of new papers/patents, mean number of papers/patents cited, mean number of authors/inventors per paper/patent) used in PLF and a dummy variable for 0-bcite documents (as proposed by HATWG), this model includes a predictor for the CD index in randomly ``rewired'' copies of the underlying citation networks (\emph{CD$_5^{random}$}). Due to the large number of papers in WoS and the associated computational burden, this analysis is based on a 25\% random sample. Following PLF and HATWG, \emph{CD$_5^{random}$} values are derived for each paper/patent from ten independently rewired citation networks, designed to preserve critical structural properties of the observed networks, including in-degree, out-degree, citation age distribution, and the number of publications per year. Each paper is represented by ten rows in the dataset, corresponding to its values across the ten random networks. Standard errors are clustered by paper (or patent) to account for within-paper/patent correlation across the repeated observations. In unreported analyses, we find similar results when aggregating and controlling for the mean CD$_5$ value across the randomized networks for each individual paper/patent. By controlling for the \emph{CD$_5^{random}$} index at the level of the individual paper, this model provides a highly conservative test of whether the observed trends in CD$_5$ exceed those predicted by changes in network structure alone. The persistence of a statistically significant decline in CD$_5$ after these extensive adjustments provides strong evidence that the observed trends reflect substantive changes in disruptiveness rather than data artifacts.
\item *$p<0.05$, **$p<0.01$, ***$p<0.001$.
\end{tablenotes}

\end{threeparttable}
\end{table}

}%

\clearpage
\section{Severe Overrepresentation of CD=1 Works in HATWG's SciSciNet Data}\label{sec:DisparitiesCD1}

\thispagestyle{empty}

\clearpage
\newgeometry{margin=1in}
\section{Problematic Exclusion of ALL CD=1 Works in HATWG}\label{sec:CD1AllGone}

In this section, we address a critical methodological choice made by HATWG in their critique---the exclusion of \textbf{ALL} papers and patents with a CD index value of 1. This approach is evident in their analyses across multiple figures (see their Fig.~1, Fig.~A1, Fig.~A3, Fig.~S4, Fig.~S6, Fig.~S7, Fig.~S8). HATWG's arguments pertain specifically to CD=1 works with zero backward citations (0-bcites) due to alleged metadata issues, yet their exclusion consists of all CD=1 works, including those with non-zero backward citations that are unaffected by the problems they highlight. This methodological decision is concerning for several reasons.

First, HATWG possess the data necessary to distinguish between CD=1 works with zero backward citations (the focus of their critique) and those with non-zero backward citations. The justification for excluding all CD=1 works, rather than isolating those implicated by their argument, remains unclear.  

Second, this broad exclusion appears designed to artificially flatten the temporal trend. By removing all CD=1 entries rather than just those with zero backward citations, their analysis suppresses the decline in disruptiveness beyond what would be justified by their concerns about 0-bcite works.

Finally, in a study of scientific and technological innovation, the decision to exclude all CD=1 works is particularly troubling. Excluding these works, especially without a clear and compelling theoretical justification, will result in misleading or inaccurate conclusions about trends in disruptiveness over time.

This issue aligns with broader concerns discussed in Sec.~\ref{sec:0BciteTheory}, where we emphasized the need for theoretical justification for excluding observations in scientometric analysis. HATWG's blanket removal of all CD=1 works extends far beyond their stated concern with 0-bcite works, undermining the methodological rigor of their analysis.

\clearpage
\newgeometry{margin=0.5in}

\clearpage
\section{Severe Overrepresentation of 0-Bcite Works in HATWG's SciSciNet Data}\label{sec:Compare0Bcite}

{

\footnotesize

\thispagestyle{empty}

\setlength{\tabcolsep}{2pt}

\renewcommand{\arraystretch}{0.9}

\begin{table}[ht!]
\centering
\caption{Matching Zero-Backward-Citation SciSciNet Papers in HATWG's PDF Search with WoS Data}
\label{table:HolstPDFWoSLink}
\footnotesize
\begin{threeparttable}
\begin{tabular}{ccccccc}\toprule
\multicolumn{2}{c}{SciSciNet} & \multicolumn{3}{c}{Web of Science}  \\
\cmidrule(lr){1-2} \cmidrule(lr){3-5}
PaperID & \shortstack{References \\ Found} & \shortstack{Accession\\Number} & \shortstack{Document\\Type} & \shortstack{References\\Found} & \shortstack{References Found \\ in HATWG \\ PDF search} & Notes \\
\midrule
1818446669 & 0 & WOS:A1990CY87600025 & Note & 25 & 25 &  Match \\
288441233 & 0 & WOS:A1994PN93400010 & Note & 3 & 3 &  Match \\
2418732624 & 0 & WOS:A1991GE10100092 & Meeting Abstract & 6 & 6 &  Match \\
2123154878 & 0 & WOS:000254648000303 & Meeting Abstract & 0 & 0 &  Match \\
2123154878 & 0 & WOS:000254648000305 & Meeting Abstract & 0 & 0 &  Match \\
2123154878 & 0 & WOS:000254648000304 & Meeting Abstract & 0 & 0 &  Match \\
2123154878 & 0 & WOS:000254648000306 & Meeting Abstract & 0 & 0 &  Match \\
1985373775 & 0 & WOS:A1996VU38900014 & Letter & 5 & 5 &  Match \\
75971907 & 0 & WOS:A1986C741900840 & Letter & 2 & 2 &  Match \\
2117751013 & 0 & WOS:000226788400001 & Editorial Material & 35 & 35 &  Match \\
2024560885 & 0 & WOS:A1975W167700018 & Editorial Material & 3 & 3 &  Match \\
1657310252 & 0 & WOS:A1987K014000012 & Editorial Material & 0 & 0 &  Match \\
1657310252 & 0 & WOS:A1987K014000012 & Editorial Material & 0 & 0 &  Match \\
2418744057 & 0 & WOS:A1987J624500004 & Article & 61 & 61 &  Match \\
1968249834 & 0 & WOS:A1995QM77100006 & Article & 53 & 53 &  Match \\
2313058271 & 0 & WOS:A1973R527500001 & Article & 41 & 41 &  Match \\
1991296186 & 0 & WOS:A1993LR60800001 & Article & 39 & 39 &  Match \\
2039319856 & 0 & WOS:A1997WT21800006 & Article & 37 & 37 &  Match \\
2412811714 & 0 & WOS:000087989700011 & Article & 32 & 32 &  Match \\
1977359555 & 0 & WOS:000242534500013 & Article & 31 & 31 &  Match \\
1970177916 & 0 & WOS:A1993KN63700014 & Article & 30 & 30 &  Match \\
1816860758 & 0 & WOS:000279050000004 & Article & 28 & 28 &  Match \\
27985486 & 0 & WOS:000264035700004 & Article & 27 & 27 &  Match \\
2405115874 & 0 & WOS:A1986D012100033 & Article & 24 & 24 &  Match \\\
2069640366 & 0 & WOS:A1992JQ54400002 & Article & 22 & 22 &  Match \\\
2090072051 & 0 & WOS:A1991FA69200003 & Article & 19 & 19 &  Match \\\
257670102 & 0 & WOS:000265309100037 & Article & 17 & 17 &  Match \\\
2016946988 & 0 & WOS:A19679679500004 & Article & 17 & 17 &  Match \\
2023511737 & 0 & WOS:000083201000058 & Article & 15 & 15 &  Match \\
2412560019 & 0 & WOS:000080737700010 & Article & 15 & 15 &  Match \\
2328439434 & 0 & WOS:A1969Y416500005 & Article & 14 & 14 &  Match \\
2027584436 & 0 & WOS:000236465600012 & Article & 13 & 13 &  Match \\
1522126398 & 0 & WOS:A1977DM76100017 & Article & 12 & 12 &  Match \\
1989925682 & 0 & WOS:000175612400016 & Article & 11 & 11 &  Match \\
1411236576 & 0 & WOS:A1972M376800019 & Article & 9 & 9 &  Match \\\
288606890 & 0 & WOS:A19633338A00007 & Article & 6 & 6 &  Match \\
1989942175 & 0 & WOS:A1990ED53400014 & Article & 6 & 6 &  Match \\
2419118622 & 0 & WOS:000182574600017 & Article & 5 & 5 &  Match \\
1004243738 & 0 & WOS:A1981MS11100010 & Article & 3 & 3 &  Match \\
1982053324 & 0 & WOS:A1995RV73700028 & Article & 65 & 1+ &  Match \\
2081547537 & 0 & WOS:A1969E216500002 & Article & 45 & 1+ &  Match \\\midrule
2461652553 & 0 & WOS:000071602800007 & Article & 77 & 51 &  non-English (excl. in PLF) \\
1978957683 & 0 & WOS:A1990DX92800011 & Article & 45 & 46 &  HATWG miscount\\
2010293280 & 0 & WOS:A1990DJ05000012 & Article & 44 & 42 &  (43) HATWG + WoS miscount\\
1976019262 & 0 & WOS:A1997XH87000033 & Article & 15 & 14 &  WoS miscount \\\midrule
1968371615 & 0 & WOS:A1990CX89600009 & Article & 0 & 14 &  non-English (excl. in PLF)\\
128280131 & 0 & WOS:A1990CM79900012 & Article & 0 & 1+ &  non-English (excl. in PLF)\\
2410752987 & 0 & WOS:A1989CB72200006 & Review & 0 & 182 & non-Article (excl. in PLF) \\
\bottomrule
\end{tabular}

\begin{tablenotes}
\item \emph{Notes:} This table compares reference data from HATWG's manual PDF search (see HATWG, Table S4) of a random sample of SciSciNet papers with zero recorded backward citations to the corresponding reference data in Web of Science (WoS). We were able to identify 48 of HATWG's 100 SciSciNet papers on our WoS data. The ``Match'' column indicates whether the references found in WoS align exactly with HATWG's manual counts. Cases of ``non-English'' or ``non-Article'' documents, excluded in HATWG's analysis, are flagged in the Notes column. To ensure higher metadata quality, non-English language research articles were excluded from PLF, following common scientometric practices. This analysis demonstrates that WoS consistently provides higher-quality and more complete reference data compared to SciSciNet.
\end{tablenotes}

\end{threeparttable}
\end{table}

}%

\clearpage
{

\setlength{\tabcolsep}{2pt}

\renewcommand{\arraystretch}{0.9}

\begin{table}[ht!]
\centering
\caption{Comparison of Backward Citation Coverage in SciSciNet and Web of Science}
\label{table:SciSciNetWoSContingencyTable}
\footnotesize
\begin{threeparttable}

\begin{tabular}{lccc}
\toprule
 & \multicolumn{3}{c}{Backward Citations in Web of Science} \\
 & Found (Yes) & Not Found (No) & Total \\
\cmidrule(lr){2-2}\cmidrule(lr){3-3}\cmidrule(lr){4-4}
Backward Citations in SciSciNet & & & \\
Found (Yes)      & 17,959,938 (76.3\%) & 280,311 (1.2\%) & 18,240,249 (77.4\%) \\
Not Found (No)   & 4,501,366 (19.1\%)  & 811,701 (3.4\%) & 5,313,067 (22.6\%) \\
\cmidrule(r){1-1}\cmidrule(lr){2-2}\cmidrule(lr){3-3}\cmidrule(lr){4-4}
Total            & 22,461,304 (95.4\%) & 1,092,012 (4.6\%) & 23,553,316 (100.0\%) \\
\bottomrule
\end{tabular}

\begin{tablenotes}
\item \emph{Notes:} This table examines the recording of references for papers in SciSciNet and Web of Science. The sample is limited to papers that were included in HATWG's analytical sample and that could be identified in both databases (1945-2010 period). Matching across databases was done based on (1) DOI, (2) PubMed ID, and (3) exact match on ISSN, publication year, volume, issue, and first page. Rows indicate whether backward citations are recorded in SciSciNet, and columns indicate the same for Web of Science. The focus is on the proportion of papers with 0 backward citations recorded, which HATWG argue could be the result of metadata errors. The analysis shown in the table above demonstrates that WoS provides significantly more complete and reliable reference data than SciSciNet. Specifically, 22.6\% of SciSciNet papers lack backward citations, compared to only 4.6\% in WoS. Furthermore, backward citations are recorded in SciSciNet but not in WoS in just 1.2\% of cases, while references are found in WoS but not in SciSciNet in 19.1\% of cases. We note that for this analysis, because our focus is on the proper recording of metadata across databases, we include ``unlinked'' references in the analysis of WoS, are references made by papers indexed by WoS to papers that are not themselves indexed in WoS. These references were controlled for by PLF in their original analysis (see their ``Methods'').
\end{tablenotes}

\end{threeparttable}
\end{table}

}%

\clearpage
\newgeometry{margin=1in}

\subsection{Departures from Standard Practices in HATWG's 0-Bcite Patent Analysis}\label{sec:0BcitePatentErrors}
In our commentary on HATWG, we focus primarily on their assessments of the sources of 0-bcite papers. However, we also identify significant concerns regarding their manual check of PDFs for 0-bcite patents (see their Table S4). Generally, the quality of metadata for patent citations is significantly higher than that for papers. Since 1976, the US Patent and Trademark Office (USPTO) has recorded patent documents in electronic, machine-readable form. Additionally, US patents are managed by a single administrative authority (the USPTO), their document structure and citation format are standardized, and citations are reviewed by examiners with substantial legal implications. 

Given these factors, HATWG's conclusion that ``98\% of the patent sample\ldots do make references in their original PDF'' (p. 3) (and therefore are recorded as making 0-bcites due to metadata errors) is striking and merits closer scrutiny. To that end, we carefully reviewed HATWG's analysis of a random sample of 100 patents using their original PDFs. Among the patents reviewed, only a single citation (in patent \#6,552,498) met the inclusion criteria for our study but was excluded.

The discrepancy stems in large part from HATWG's departure from established practices in patent citation analysis. As detailed in our paper, we followed standard practices from prior literature by focusing specifically on utility patents granted by the USPTO (see PLF, ``Methods''). Our analysis excluded design patents, plant patents, foreign patent documents, ungranted applications, and citations to these documents. These exclusions reflect well-documented understanding that different patent types exhibit distinct citation patterns, and that citations to foreign patents and applications are inconsistently recorded during our study period.

U.S. patents often cite various types of documents, including prior patents granted by the USPTO, patent applications submitted to the USPTO, foreign patents, and ``other'' documents such as scientific literature. Standard practice across multiple fields---scientometrics, the economics of innovation, and technology strategy and management---emphasizes citations to prior patented inventions, specifically utility patents. The methodological foundation for this focus was established in the seminal work \emph{Patents, Citations, and Innovations: A Window on the Knowledge Economy} by Adam B. Jaffe and Manuel Trajtenberg \citep{jaffe2002patents}, particularly in Chapter 13 . This book has defined the methodological approach for a generation of patent research. In addition to this foundational work, numerous high-quality studies published in leading journals follow the same approach of focusing on utility patents (e.g., \citep{ke2020technological, singh2021technological, corredoira2018federal, arora2021matching, atallah2006indirect, arts2018text, kovacs2021categories}). These studies underscore the validity of focusing exclusively on utility patents in the analysis of patent citations.

There are several reasons why analyses in the literature focus specifically on utility patents and their citations. Citations to different types of documents vary substantially by field and over time, introducing significant heterogeneity. For example, U.S. Patent and Trademark Office policies only allowed citations to patent applications starting in the early 2000s. This trend is evident in HATWG's Table S4, where no missing ``A1'' citations are documented prior to the early 2000s (e.g., \citep{johnson2003forced}). Furthermore, metadata quality for non-granted patent prior art is often incomplete or entirely absent. The U.S. Patent Office does not index foreign patents, and obtaining consistent and reliable data (e.g., grant dates, patent types) across numerous foreign patent offices would be highly challenging, if not impossible.

Citations under the ``Other'' category, which include references to scientific literature, technical manuals, and a broad range of other documents, present additional complications. These references are often cited in inconsistent formats, making accurate parsing and analysis difficult. Including citations to such a wide range of documents, as done in HATWG's Table S4 analysis, deviates significantly from established practices in the field. This approach introduces problematic and unobserved heterogeneity into analyses and risks undermining the credibility of any resulting findings. Such inclusions are highly unconventional in the context of the established scientometric literature and could invite substantial criticism from the research community.

\clearpage
\newgeometry{margin=0.5in}

\section{Mappings of Fields and Document Types Across Data Sources}\label{sec:Mappings}

{

\footnotesize

\thispagestyle{empty}

\begin{table}[htbp]\centering
\begin{threeparttable}
\caption{Mapping Between Fields in Web of Science and SciSciNet}
\label{table:FieldsMapping}
\begin{tabular}{p{0.35\textwidth}p{0.35\textwidth}}
\toprule

\multicolumn{1}{c}{\shortstack{Web of Science field}}	&	\multicolumn{1}{c}{\shortstack{SciSciNet field}} \\\midrule
 Humanities	&	art 	\\
 Humanities	&	history 	\\
 Humanities	&	philosophy 	\\
 Life sciences	&	biology 	\\
 Life sciences	&	environmental science 	\\
 Life sciences	&	medicine 	\\
 Physical sciences	&	chemistry 	\\
 Physical sciences	&	geology 	\\
 Physical sciences	&	mathematics 	\\
 Physical sciences	&	physics 	\\
 Social sciences	&	business 	\\
 Social sciences	&	economics 	\\
 Social sciences	&	geography 	\\
 Social sciences	&	political science 	\\
 Social sciences	&	psychology 	\\
 Social sciences	&	sociology 	\\
 Technology	&	computer science 	\\
 Technology	&	engineering 	\\
 Technology	&	materials science 	\\
 \bottomrule
\end{tabular}

\begin{tablenotes}
\item \emph{Notes:} This table presents the mapping of fields between SciSciNet and Web of Science data (referred to as ``Research Areas'' in Web of Science). The mapping facilitates comparisons between the results reported in PLF \citep{park2023papers} (based on WoS) and those derived from HATWG's SciSciNet data.
\end{tablenotes}

\end{threeparttable}
\end{table}

}%

\clearpage

{

\footnotesize

\thispagestyle{empty}
\setlength{\tabcolsep}{3pt} 
\renewcommand{\arraystretch}{0.9} 

\begin{table}[htbp]\centering
\centering
\caption{Mapping of WoS and Dimensions Document Types to Meta Categories}
\label{table:MetaCategories}
\footnotesize
\begin{threeparttable}
\begin{tabular}{>{\raggedright\arraybackslash}p{0.2\textwidth}%
                >{\raggedright\arraybackslash}p{0.3\textwidth}%
                >{\raggedright\arraybackslash}p{0.3\textwidth}}
\toprule
Meta Category & WoS Categories & Dimensions Categories \\
\midrule

Research articles & 
Article & 
RESEARCH\_ARTICLE \\\addlinespace[0.25cm]

Reviews & 
Book Review, Review, Art Exhibit Review, Database Review, Film Review, Music Score Review, Music Performance Review, Record Review, Software Review, Hardware Review, TV Review Radio Review, Theater Review, Dance Performance Review & 
REVIEW\_ARTICLE, BOOK\_REVIEW \\\addlinespace[0.25cm]

Books and proceedings & 
Bibliography, Book, Book Chapter, Proceedings Paper, Meeting Abstract & 
CONFERENCE\_PAPER, REFERENCE\_WORK, CONFERENCE\_ABSTRACT, RESEARCH\_CHAPTER, OTHER\_BOOK\_CONTENT \\\addlinespace[0.25cm]

Editorial and commentary & 
Abstract of Published Item, Chronology, Correction, Editorial Material, Excerpt, Letter, News Item, Note, Reprint, Discussion, Biographical-Item, Item About an Individual & 
OTHER\_JOURNAL\_CONTENT, LETTER\_TO\_EDITOR, EDITORIAL, CORRECTION\_ERRATUM \\\addlinespace[0.25cm]

All others & 
Fiction Creative Prose, Poetry, Script, Music Score & 
N/A \\

\bottomrule
\end{tabular}

\begin{tablenotes}
\item \emph{Notes:} This table presents the mapping of document types from Web of Science and Dimensions databases to a common set of meta categories used in this study. The mapping addresses challenges in comparing data across databases with differing and often limited categorization schemes, such as SciSciNet, which lacks the granularity of Dimensions---including the ability to subset to research articles. Research articles were the focus of PLF's analysis, following established scientometric conventions. These meta categories enable consistent cross-database analyses by aligning similar document types under broader, standardized classifications. For ``N/A'', there is no corresponding category in Dimensions.
\end{tablenotes}

\end{threeparttable}
\end{table}


}

\clearpage

\section{Inventory of Non-Research Articles in HATWG's Analytical Sample}\label{sec:nonresearch}

{
\footnotesize
\thispagestyle{empty}
\setlength{\tabcolsep}{3pt} 
\renewcommand{\arraystretch}{0.9} 
\begin{table}[htbp]\centering
\centering
\caption{Inventory of non-research articles in HATWG's analytical sample}
\label{table:HATWGNonResearch}
\footnotesize
\begin{threeparttable}

\begin{tabular}{>{\raggedright\arraybackslash}p{0.30\textwidth}%
                >{\raggedright\arraybackslash}p{0.45\textwidth}%
                >{\raggedright\arraybackslash}r%
                >{\raggedright\arraybackslash}r}
\toprule
Category & Includes & N & \% \\
\midrule
Editorial and commentary & 
Abstract of Published Item; Biographical-Item; Chronology; Correction; Correction, Addition; Discussion; Editorial Material; Excerpt; Item About an Individual; Letter; News Item; Note; Reprint & 
2,827,654 & 12.0\% \\\addlinespace[0.25cm]
Reviews & & & \\\addlinespace[0.1cm]
\hspace{0.3cm}of books, media, and artistic performances & 
Art Exhibit Review; Book Review; Dance Performance Review; Film Review; Music Performance Review; Music Score Review; Record Review; Theater Review; TV Review, Radio Review & 
254,125 & 1.1\% \\\addlinespace[0.15cm]
\hspace{0.3cm}of technical products & 
Database Review; Hardware Review; Software Review & 
2,480 & $<$0.1\% \\\addlinespace[0.25cm]
Books and proceedings & 
Bibliography; Book; Book Chapter; Meeting Abstract; Proceedings Paper & 
1,547,125 & 6.6\% \\\addlinespace[0.25cm]
Artistic works & 
Fiction, Creative Prose; Music Score; Poetry; Script & 
752 & $<$0.1\% \\\addlinespace[0.25cm]
\midrule
\textbf{Total} & & \textbf{4,632,136} & \textbf{19.7\%} \\
\bottomrule
\end{tabular}
\begin{tablenotes}
\item \emph{Notes:} Counts based on HATWG's SciSciNet analytical sample (1945--2010) matched to Web of Science (N=23,553,316 of 44,374,787 total, 53.1\%). Percentages calculated relative to matched sample. Excludes WoS document types `Article' and `Review'. Review articles, while not primary research, cite prior literature by design and are excluded from this inventory.

\end{tablenotes}

\end{threeparttable}
\end{table}
}

\clearpage
{
\scriptsize
\thispagestyle{empty}
\begin{table}[htbp]
\centering
\caption{Selected zero-backward-citation works in HATWG's analytical sample}
\label{table:HATWGNonScientific}
\begin{threeparttable}

\begin{tabular}{>{\raggedright\arraybackslash}p{0.95\textwidth}}
\toprule
\\[-0.8em]

\textbf{\textit{For Dummies} (N=456)} \\*
\textit{MySpace For Dummies} (117571868); 
\textit{AOLTV For Dummies} (2414210322); 
\textit{Sex For Dummies} (2799015865); 
\textit{Massage For Dummies} (400896340); 
\textit{Sushi For Dummies} (639253819); 
\textit{Bartending For Dummies} (3192141389); 
\textit{Adoption For Dummies} (617758150); 
\textit{Breastfeeding For Dummies} (562425042); 
\textit{Divorce For Dummies} (1539101869); 
\textit{Bipolar Disorder For Dummies} (2798797046); 
\textit{Schizophrenia For Dummies} (2899036383); 
\textit{Menopause For Dummies} (1767328172); 
\textit{Christianity For Dummies} (630558203); 
\textit{Islam For Dummies} (613833845); 
\textit{Buddhism For Dummies} (603249656); 
\textit{Existentialism For Dummies} (2419050770); 
\textit{World Poverty For Dummies} (561815778); 
\textit{NASCAR For Dummies} (599739661); 
\textit{Curling For Dummies} (630845378); 
\textit{Ballet For Dummies} (2385249983); 
\textit{Raising Goats For Dummies} (3002816966); 
\textit{Writing A Romance Novel For Dummies} (612862026); 
\textit{Year 2000 Solutions For Dummies} (2473980037); 
and 433 others. \\\addlinespace[0.35em]

\textbf{\textit{Complete Idiot's Guide} (N=88)} \\*
\textit{The Complete Idiot's Guide to Being Psychic} (562017520); 
\textit{The Complete Idiot's Guide to Body Language} (565740101); 
\textit{The Complete Idiot's Guide to Geocaching} (90545040); 
\textit{The Complete Idiot's Guide to Etiquette} (2154972865); 
\textit{The Complete Idiot's Guide to American History} (1551674056); 
\textit{The Complete Idiot's Guide to Economics} (619047683); 
\textit{The Complete Idiot's Guide to Calculus} (577851795); 
and 81 others. \\\addlinespace[0.35em]

\textbf{\textit{Dr. Seuss} (N=26)} \\*
\textit{The Cat in the Hat Comes Back} (1518726047); 
\textit{Yertle the Turtle and Other Stories} (381946459); 
\textit{Oh Say Can You Say} (578741411); 
\textit{If I Ran the Circus} (601666773); 
\textit{My Many Colored Days} (611153699); 
\textit{McElligot's Pool} (1484400156); 
\textit{The Cat's Quizzer} (632450922); 
and 19 others. \\\addlinespace[0.35em]

\textbf{\textit{Curious George} (N=24)} \\*
\textit{Curious George Goes Camping} (371092128); 
\textit{Curious George's Dinosaur Discovery} (428842577); 
\textit{Curious George Plays Baseball} (565966362); 
\textit{Curious George Goes to an Ice Cream Shop} (575876783); 
\textit{Curious George Flies a Kite} (590794719); 
\textit{Curious George Goes to a Chocolate Factory} (647398705); 
and 18 others. \\\addlinespace[0.35em]

\textbf{\textit{Roald Dahl} (N=23)} \\*
\textit{Charlie and the Chocolate Factory: A Play} (607175040); 
\textit{James and the Giant Peach} (1547452223); 
\textit{George's Marvelous Medicine} (1489337093); 
\textit{The Magic Finger} (1510088679); 
\textit{Boy: Tales of Childhood} (1536241376); 
\textit{The Giraffe and the Pelly and Me} (1529390462); 
and 17 others. \\\addlinespace[0.35em]

\textbf{\textit{Berenstain Bears} (N=19)} \\*
\textit{The Berenstain Bears Get in a Fight} (368963323); 
\textit{The Berenstain Bears and the Truth} (598447605); 
\textit{The Berenstain Bears and Too Much TV} (653296235); 
\textit{The Berenstain Bears and the Bully} (655902378); 
\textit{Bears in the Night} (1538149355); 
\textit{Old Hat, New Hat} (1540868065); 
and 13 others. \\\addlinespace[0.35em]

\textbf{\textit{Nancy Drew} (N=17)} \\*
\textit{The Moonstone Castle Mystery} (578500696); 
\textit{The Clue of the Black Keys} (596507366); 
\textit{The Ghost of Blackwood Hall} (613970181); 
\textit{The Mystery of the Fire Dragon} (653775448); 
\textit{The E-mail Mystery} (638759924); 
and 12 others. \\\addlinespace[0.35em]

\textbf{\textit{Magic Tree House} (N=13)} \\*
\textit{Polar Bears Past Bedtime} (612503150); 
\textit{Viking Ships at Sunrise} (623144355); 
\textit{Civil War on Sunday} (1481662738); 
\textit{Tonight on the Titanic} (1523345185); 
\textit{Mummies in the Morning} (1547911488); 
\textit{Pirates Past Noon} (1594101699); 
and 7 others. \\\addlinespace[0.35em]

\textbf{\textit{Magic School Bus} (N=10)} \\*
\textit{The Magic School Bus Inside the Human Body} (382735495); 
\textit{The Magic School Bus Inside the Earth} (637269955); 
\textit{The Magic School Bus In the Time of the Dinosaurs} (646952559); 
\textit{The Magic School Bus Lost in the Solar System} (1585207758); 
\textit{The Magic School Bus on the Ocean Floor} (649880876); 
and 5 others. \\\addlinespace[0.35em]

\textbf{\textit{Goosebumps} (N=9)} \\*
\textit{Welcome to Dead House} (379604605); 
\textit{The Haunted Mask} (571243051); 
\textit{The Werewolf of Fever Swamp} (626530652); 
\textit{The Girl Who Cried Monster} (1555521325); 
\textit{It Came from Beneath the Sink} (1571863817); 
and 4 others. \\\addlinespace[0.35em]

\textbf{\textit{Captain Underpants} (N=5)} \\*
\textit{The Adventures of Captain Underpants} (631792803); 
\textit{The Adventures of Super Diaper Baby} (633508564); 
\textit{Captain Underpants and the Preposterous Plight of the Purple Potty People} (597090066); 
and 2 others. \\\addlinespace[0.35em]

\textbf{\textit{Diary of a Wimpy Kid} (N=3)} \\*
\textit{Diary of a Wimpy Kid} (2798629682); 
\textit{Diary of a Wimpy Kid: Dog Days} (2481473575); 
\textit{Diary of a Wimpy Kid: The Ugly Truth} (3022792873). \\[0.3em]

\bottomrule
\end{tabular}
\begin{tablenotes}
\item \emph{Notes:} While investigating the prevalence of zero-backward-citation works in HATWG's sample, we discovered substantial non-scientific content. A few simple title searches revealed the examples above; SciSciNet Paper IDs are provided for verification. This list is illustrative, not exhaustive. Additional works identified include \textit{Clifford the Big Red Dog}, \textit{American Girl}, \textit{Boxcar Children}, \textit{Fear Street}, \textit{Hardy Boys}, and 64 O'Reilly technical guides.
\end{tablenotes}

\end{threeparttable}
\end{table}
}

\clearpage
\section{Inventory of Studies Documenting Trends in Disruptiveness}\label{sec:LitReview}
\addtocontents{toc}{\protect\setcounter{tocdepth}{2}}

Table~\ref{table:LitReview} summarizes over 100 studies examining trends in disruptiveness across scientific papers, patents, products, and other creative domains. Studies were identified through forward citation searches from seed articles, supplemented by searches of Dimensions, Google Scholar, arXiv, and similar databases. We included studies reporting original empirical findings on temporal trends, regardless of whether they documented decline, stability, or increase. Importantly, we focused narrowly on studies using established disruption indicators (e.g., CD, DI, D) or explicitly examining trends in ``disruption'' or ``disruptiveness.'' We did not include the vast literatures on related constructs---novelty, originality, breakthrough innovation, business dynamism, or economic growth---even though many of these document similar declines. Our inventory of 105 studies therefore reflects a conservative, focused scope. Full methodological details and complete citations are available in Wu et al.\cite{wu2026inventory} (see also \url{https://arxiv.org/abs/2602.05140}).

Several patterns emerge from this literature. First, evidence of decline appears across multiple data sources. Declining disruptiveness has been documented across large open-access databases including OpenAlex, Microsoft Academic Graph, and SciSciNet, as well as commercial databases with higher metadata quality such as Web of Science, Scopus, and Dimensions. 

Second, evidence of decline does not depend on the specific metric. The literature has produced dozens of indicators intended to capture disruptiveness---CD, DI, D, and numerous variants---each developed with particular improvements in mind, each with specific strengths and weaknesses. Yet similar patterns appear across operationalizations, consistent with a substantive phenomenon rather than an artifact of any particular measurement approach.

Third, and critically, evidence of decline does not depend on bibliometric or citation data. Researchers have documented declining disruptiveness using text-based measures of patent novelty, firm displacement rates derived from Census and Compustat data, product similarity networks for automobiles and mobile phones, audio features, and deep learning analysis of images. These approaches avoid citation data entirely, yet report patterns consistent with declining disruptiveness.

Fourth, while the aggregate decline is robust, the literature has identified important exceptions. Several studies document rebounds in particular domains or stable and increasing disruption in specific areas, including fetal surgery, AI research after 2014, highly disruptive patents after 2008 (especially in IT), and machine-learning-related economics papers after 2010. These exceptions are theoretically informative---if declining disruption were a mechanical artifact of citation inflation, database expansion, or network growth, we would expect uniform decline across all domains, as mechanical artifacts do not generate structured exceptions. The existence of rebounds and increases---often in young or rapidly evolving fields---is consistent with theoretical expectations and with our original finding on the conservation of highly disruptive work, which has been independently replicated.

\newcounter{inventry}
\setcounter{inventry}{106}

\begin{longtable}{@{}r p{15cm} c@{}}
\caption{Inventory of studies documenting trends in disruptiveness. Adapted from Wu et al.\cite{wu2026inventory}\\\textbf{(For full citations, see: \url{https://arxiv.org/abs/2602.05140}.)}} \label{table:LitReview} \\
\toprule
\addlinespace[0.5em]
\endfirsthead
\multicolumn{3}{c}{\tablename\ \thetable\ -- \textit{Continued from previous page \textbf{(For full citations, see: \url{https://arxiv.org/abs/2602.05140}.)}}} \\
\toprule
\addlinespace[0.5em]
\endhead
\bottomrule
\endlastfoot

\multicolumn{3}{c}{\large\textsc{2026} (2 studies)} \\[0.8em]
\setcounter{inventry}{104}\refstepcounter{inventry}\label{inv:cao2025mm}\textcolor{metgray}{105} & \nociteinv{cao2025mm}\textbf{\href{https://doi.org/10.1007/s11023-025-09755-8}{\color{black}Cao \& Cai (2026)}}, \emph{\emph{Minds Mach}} \newline
{\small\color{metgray} AI patents, 1976--2020. $\cdot$ Data: USPTO} \newline
``A noticeable decline over the past two decades\ldots across nearly all AI subcategories'' (Fig. 2). AI ``entered the track of consolidation.'' & {\color{DeclineRed}\Large$\bm{\downarrow}$} \\
\addlinespace[1.2em]

\setcounter{inventry}{103}\refstepcounter{inventry}\label{inv:liu2025ipm}\textcolor{metgray}{104} & \nociteinv{liu2025ipm}\textbf{\href{https://doi.org/10.1016/j.ipm.2025.104382}{\color{black}Liu et al. (2026)}}, \emph{\emph{Inf Process Manag}} \newline
{\small\color{metgray} 647K papers, 1893--2020. $\cdot$ Data: APS} \newline
``Overall disruptiveness\ldots exhibited a consistent decline starting from the 1940s'' (Fig. 5A). Newcomers more disruptive (Fig. 5B) but science increasingly reliant on established elites (Fig. 5C).  & {\color{DeclineRed}\Large$\bm{\downarrow}$} \\
\addlinespace[1.2em]

\\[1.5em]
\multicolumn{3}{c}{\large\textsc{2025} (32 studies)} \\[0.8em]
\setcounter{inventry}{102}\refstepcounter{inventry}\label{inv:becerra2024ajhp}\textcolor{metgray}{103} & \nociteinv{becerra2024ajhp}\textbf{\href{https://doi.org/10.1093/ajhp/zxae309}{\color{black}Becerra et al. (2025)}}, \emph{\emph{Am J Health-Syst Pharm}} \newline
{\small\color{metgray} Pharmacy, 1954--2014. $\cdot$ Data: PubMed} \newline
``Despite the many disruptive papers, there was a lack of more recent papers, indicating the need for modern innovation in pharmacy'' (Fig. 4).  & {\color{DeclineRed}\Large$\bm{\downarrow}$} \\
\addlinespace[1.2em]

\setcounter{inventry}{101}\refstepcounter{inventry}\label{inv:boot2025jar}\textcolor{metgray}{102} & \nociteinv{boot2025jar}\textbf{\href{https://doi.org/10.1111/1475-679x.12580}{\color{black}Boot \& Vladimirov (2025)}}, \emph{\emph{J Account Res}} \newline
{\small\color{metgray} U.S. patents, AIPA (2000) policy shock. $\cdot$ Data: USPTO} \newline
Exploiting AIPA for causal inference, reports ``the nature of patented innovation has changed, with the proportion of nondisruptive patents increasing substantially.'' Multiple disruption proxies---citation and text-based---show decline (Fig. 5). & {\color{DeclineRed}\Large$\bm{\downarrow}$} \\
\addlinespace[1.2em]

\setcounter{inventry}{100}\refstepcounter{inventry}\label{inv:chen2025chi}\textcolor{metgray}{101} & \nociteinv{chen2025chi}\textbf{\href{https://doi.org/10.1145/3706598.3713917}{\color{black}Chen \& Li (2025)}}, \emph{CHI} \newline
{\small\color{metgray} HCI, 1982--2023. $\cdot$ Data: ACM DL, OpenAlex} \newline
``The growth rate of disruptive works does not match up with that of the explosion of published papers'' (Figs. 2, 3). Robustness checks (Appendix C, D) further validate decline. & {\color{DeclineRed}\Large$\bm{\downarrow}$} \\
\addlinespace[1.2em]

\setcounter{inventry}{99}\refstepcounter{inventry}\label{inv:cheng2025wp}\textcolor{metgray}{100} & \nociteinv{cheng2025wp}\textbf{\href{https://doi.org/10.51408/issi2025_009}{\color{black}Cheng \& Ma (2025)}}, \emph{ISSI} \newline
{\small\color{metgray} Scientometrics, 1955--2024. $\cdot$ Data: OpenAlex, WoS} \newline
Using Main Path Analysis, concludes ``the field of scientometrics is experiencing a decline in disruption'' (Figs. 7--8).  & {\color{DeclineRed}\Large$\bm{\downarrow}$} \\
\addlinespace[1.2em]

\setcounter{inventry}{98}\refstepcounter{inventry}\label{inv:cheng2025s}\textcolor{metgray}{99} & \nociteinv{cheng2025s}\textbf{\href{https://doi.org/10.1007/s11192-025-05285-3}{\color{black}Cheng et al. (2025)}}, \emph{\emph{Scientometrics}} \newline
{\small\color{metgray} Earth Sciences, 2000--2019. $\cdot$ Data: WoS} \newline
Statistical analysis reveals declining disruption over time; Disruption$\sim$Year: r=-0.227, p$<$0.001 (Table 2). & {\color{DeclineRed}\Large$\bm{\downarrow}$} \\
\addlinespace[1.2em]

\setcounter{inventry}{97}\refstepcounter{inventry}\label{inv:deng2025wp}\textcolor{metgray}{98} & \nociteinv{deng2025wp}\textbf{\href{https://doi.org/10.1038/s43588-025-00808-7}{\color{black}Deng et al. (2025)}}, \emph{\emph{Nat Comput Sci}} \newline
{\small\color{metgray} 1900--2020, 90M papers. $\cdot$ Data: APS, SciSciNet} \newline
Confirms that ``both $d^r$ [disruption of prior work] and $d^c$ [disruption by future work] decline with publication year'' (Fig. S2c), attributing this to ``the growing prevalence of consolidatory work in modern science.'' While the fraction of ``persistently disruptive'' papers is rising (Fig. 5c), this is driven by the collapse of $d^c$; papers today are ``more likely to be consolidated than to be disrupted,'' even as their own ability to disrupt the past ($d^r$) steadily fades (Figs. 5d, S10b). & {\color{DeclineRed}\Large$\bm{\downarrow}$} \\
\addlinespace[1.2em]

\setcounter{inventry}{96}\refstepcounter{inventry}\label{inv:fang2025wp}\textcolor{metgray}{97} & \nociteinv{fang2025wp}\textbf{\href{https://arxiv.org/abs/2510.03240v2}{\color{black}Fang \& Evans (2025)}}, \emph{arXiv} \newline
{\small\color{metgray} 1950--2023, 23M/19M papers. $\cdot$ Data: OpenAlex, WoS} \newline
Provides evidence for declining disruptiveness by showing ``foundational\ldots work within fields has declined in recent years'' (Fig. 4). The ``Foundation Index'' captures work that ``eclipses prior ideas'' and is thus ``closely aligned with the original Disruption Index.'' They further note declining ``generalization'' (cross-field combinations), suggesting a shift toward synthesizing rather than disrupting existing ideas. & {\color{DeclineRed}\Large$\bm{\downarrow}$} \\
\addlinespace[1.2em]

\setcounter{inventry}{95}\refstepcounter{inventry}\label{inv:he2024bwp}\textcolor{metgray}{96} & \nociteinv{he2024bwp}\textbf{\href{https://doi.org/10.1016/j.joi.2025.101703}{\color{black}He \& Lee (2025)}}, \emph{\emph{J Informetr}} \newline
{\small\color{metgray} Automobiles, 2014--2023, 4,114 car models. $\cdot$ Data: Edmunds} \newline
Introduces a Product Disruption Index based on product similarity networks, avoiding bibliometric data entirely. Reports ``a clear downward trend\ldots mirroring the overarching trend of diminishing innovation disruptiveness over time'' (Fig. 4b). ``Earlier [car] models\ldots are more likely to be disruptive.'' & {\color{DeclineRed}\Large$\bm{\downarrow}$} \\
\addlinespace[1.2em]

\setcounter{inventry}{94}\refstepcounter{inventry}\label{inv:jeong2025s}\textcolor{metgray}{95} & \nociteinv{jeong2025s}\textbf{\href{https://doi.org/10.1007/s11192-025-05395-y}{\color{black}Jeong \& Lee (2025)}}, \emph{\emph{Scientometrics}} \newline
{\small\color{metgray} 1994--2023, 12,554 mobile products. $\cdot$ Data: GSMArena} \newline
Applies CD index on product similarity network (shared OS, camera, chipset). ``Products with the highest disruption index for all four companies are their initial product lines'' (Fig. 10). ``Most firms introduce more disruptive products in early stages\ldots followed by gradual decline.'' & {\color{DeclineRed}\Large$\bm{\downarrow}$} \\
\addlinespace[1.2em]

\setcounter{inventry}{93}\refstepcounter{inventry}\label{inv:jiang2025jsp}\textcolor{metgray}{94} & \nociteinv{jiang2025jsp}\textbf{\href{https://doi.org/10.3138/jsp-2023-0056}{\color{black}Jiang et al. (2025)}}, \emph{\emph{J Sch Publ}} \newline
{\small\color{metgray} Asian journals, 2010--2019. $\cdot$ Data: OpenCitations} \newline
Uses a Journal Disruption Index (JDI) to evaluate journals from China, India, S. Korea, finding an upward trend 2010--2019. But the authors also note ``disruptive innovation level\ldots still far from Nature and Science,'' implying catching up from lower baseline.  & {\color{IncreaseGreen}\Large$\bm{\uparrow}$} \\
\addlinespace[1.2em]

\setcounter{inventry}{92}\refstepcounter{inventry}\label{inv:jurowetzki2025as}\textcolor{metgray}{93} & \nociteinv{jurowetzki2025as}\textbf{\href{https://doi.org/10.1007/s00146-024-02171-z}{\color{black}Jurowetzki et al. (2025)}}, \emph{\emph{AI Soc}} \newline
{\small\color{metgray} AI research, 2000--2021, 1.7M papers, 143K researchers. $\cdot$ Data: OpenAlex} \newline
Finds academia-to-industry transitions are rising (Fig. 2c), and that post-transition, researchers show ``reduced disruptive influence'' (Table 2). ``As researchers spend more time in industry, they produce less cited, disruptive, and novel science.'' & {\color{DeclineRed}\Large$\bm{\downarrow}$} \\
\addlinespace[1.2em]

\setcounter{inventry}{91}\refstepcounter{inventry}\label{inv:li2025wp}\textcolor{metgray}{92} & \nociteinv{li2025wp}\textbf{\href{https://arxiv.org/abs/2506.15959v4}{\color{black}Li et al. (2025)}}, \emph{Working Paper} \newline
{\small\color{metgray} 1965--2020, 41M papers. $\cdot$ Data: OpenAlex} \newline
Documents ``the decline of disruption in science over time'' (Sec. 4.1). Proportion of highly disruptive papers (top 5\%) fell from $\sim$50\% to below 20\% (Fig. 2). & {\color{DeclineRed}\Large$\bm{\downarrow}$} \\
\addlinespace[1.2em]

\setcounter{inventry}{90}\refstepcounter{inventry}\label{inv:li2025awp}\textcolor{metgray}{91} & \nociteinv{li2025awp}\textbf{\href{https://arxiv.org/abs/2512.03723v1}{\color{black}Li et al. (2025)}}, \emph{arXiv} \newline
{\small\color{metgray} 1965--2020, 41M papers. $\cdot$ Data: OpenAlex} \newline
Finds the ``share of disruptive papers (D $>$ 0) declines from nearly 50\% in 1965 to under 20\% by 2020'' (Figs. 14--15). However, ``absolute number of highly disruptive papers has remained stable over the past six decades\ldots consistent with the 'conservation of highly disruptive work' (Park et al., 2023).'' (Fig. 7) & {\color{DeclineRed}\Large$\bm{\downarrow}$} \\
\addlinespace[1.2em]

\setcounter{inventry}{89}\refstepcounter{inventry}\label{inv:li2025re}\textcolor{metgray}{90} & \nociteinv{li2025re}\textbf{\href{https://doi.org/10.1093/reseval/rvaf026}{\color{black}Li et al. (2025)}}, \emph{\emph{Res Eval}} \newline
{\small\color{metgray} Social sciences, 1997--2017, 1.09M papers. $\cdot$ Data: OpenAlex, WoS} \newline
``Figure 2a presents the declining trend of the CD5 over the last 20 years. This downtrend is similar across disciplines.'' The authors further report after 2007 ``the growth rate of disruptive innovative papers was flat'' while consolidating papers grew rapidly (Fig. 2c). Validated with two text-based measures of disruption, showing results not driven by citation-data artifacts.  & {\color{DeclineRed}\Large$\bm{\downarrow}$} \\
\addlinespace[1.2em]

\setcounter{inventry}{88}\refstepcounter{inventry}\label{inv:lyu2025wp}\textcolor{metgray}{89} & \nociteinv{lyu2025wp}\textbf{Lyu (2025)}, \emph{Working Paper} \newline
{\small\color{metgray} U.S. patents, 1977--2006, 308K patents. $\cdot$ Data: USPTO} \newline
Shows steady decline in disruptiveness (Fig. 2). ``Overall impact of knowledge accumulation on innovation is found to be negative.'' Results supported by regression analysis and formal model (Table 4).  & {\color{DeclineRed}\Large$\bm{\downarrow}$} \\
\addlinespace[1.2em]

\setcounter{inventry}{87}\refstepcounter{inventry}\label{inv:naude2025pone}\textcolor{metgray}{88} & \nociteinv{naude2025pone}\textbf{\href{https://doi.org/10.1371/journal.pone.0323297}{\color{black}Naud{\'e} (2025)}}, \emph{\emph{PLoS ONE}} \newline
{\small\color{metgray} Business and Entrepreneurship, 1961--2020, 68K papers. $\cdot$ Data: MAG} \newline
Reports ``the average disruption score of papers have declined by a factor of 36 between the 1960s and the 2010s'' (Fig. 2), and ``the share of papers with a D$>$0 has consistently declined over time'' (Fig. 3).  & {\color{DeclineRed}\Large$\bm{\downarrow}$} \\
\addlinespace[1.2em]

\setcounter{inventry}{86}\refstepcounter{inventry}\label{inv:park2025wp}\textcolor{metgray}{87} & \nociteinv{park2025wp}\textbf{\href{https://arxiv.org/abs/2503.00184v1}{\color{black}Park et al. (2025)}}, \emph{arXiv} \newline
{\small\color{metgray} 1945--2010 (papers, 22M), 1980--2010 (patents, 2.9M). $\cdot$ Data: SciSciNet, USPTO, WoS} \newline
Responds to Holst et al.'s critique that declining disruptiveness is an artifact of zero-backward-citation papers. Using critics' own dataset, metric, and regression model, finds ``statistically and practically significant declines\ldots that equal major benchmark transformations in science'' (Figs. 1--2, p$<$0.001). Decline magnitude equals the gap between an average paper and a Nobel winner. Critics' own regression (their Table S1) shows significant decline---``a finding they neither acknowledge nor interpret.'' & {\color{DeclineRed}\Large$\bm{\downarrow}$} \\
\addlinespace[1.2em]

\setcounter{inventry}{85}\refstepcounter{inventry}\label{inv:petersen2025joi}\textcolor{metgray}{86} & \nociteinv{petersen2025joi}\textbf{\href{https://doi.org/10.1016/j.joi.2024.101605}{\color{black}Petersen et al. (2025)}}, \emph{\emph{J Informetr}} \newline
{\small\color{metgray} 1995--2015, 7.8M papers. $\cdot$ Data: SciSciNet} \newline
Argues that after controls, trends are ``at the level of noise.'' However, the study covers 1995--2015, a period that falls after the steepest decline documented in larger-scale studies spanning 1945--1990. Additionally, their preprocessing excludes papers with fewer than 10 references and more than 1,000 citations---restrictions that may bias against finding disruptive work. & {\color{StableGray}\Large$\bm{\rightarrow}$} \\
\addlinespace[1.2em]

\setcounter{inventry}{84}\refstepcounter{inventry}\label{inv:qu2024rdm}\textcolor{metgray}{85} & \nociteinv{qu2024rdm}\textbf{\href{https://doi.org/10.1111/radm.12685}{\color{black}Qu et al. (2025)}}, \emph{\emph{R\&D Manage}} \newline
{\small\color{metgray} Electric communication industry, 1995--2019, 29.6K patents. $\cdot$ Data: USPTO} \newline
Documents a significant negative effect of year on disruption ($\beta$ = -0.0592, p$<$0.01, Table 2). & {\color{DeclineRed}\Large$\bm{\downarrow}$} \\
\addlinespace[1.2em]

\setcounter{inventry}{83}\refstepcounter{inventry}\label{inv:shinichi2025wp}\textcolor{metgray}{84} & \nociteinv{shinichi2025wp}\textbf{\href{https://arxiv.org/abs/2505.08284v1}{\color{black}Shinichi \& Matsui (2025)}}, \emph{arXiv} \newline
{\small\color{metgray} Visual Art (Ukiyo-e), 11,000 images. $\cdot$ Data: Ukiyo-e images} \newline
Using a novel deep learning adaptation of the Disruption Index (DI) to quantify the disruptiveness of images, the authors find ``the disruptiveness of works diminishes over the data period, consistent with the dynamics\ldots of scientific works'' (Fig. 8a). However, within specific schools ``disruptiveness\ldots remained stable'' (Fig. 8b). This divergence, combined with the use of non-citation data, suggests that downward trends in DI reported for scientific domains are not driven by mechanical artifacts. & {\color{DeclineRed}\Large$\bm{\downarrow}$} \\
\addlinespace[1.2em]

\setcounter{inventry}{82}\refstepcounter{inventry}\label{inv:stack2025wp}\textcolor{metgray}{83} & \nociteinv{stack2025wp}\textbf{Stack (2025)}, \emph{Working Paper} \newline
{\small\color{metgray} 1900--2000, all fields. $\cdot$ Data: Dimensions} \newline
Reports ``disruptiveness as measured by the CD index across all research fields has been falling since the 19th century'' (Fig. 2.2). & {\color{DeclineRed}\Large$\bm{\downarrow}$} \\
\addlinespace[1.2em]

\setcounter{inventry}{81}\refstepcounter{inventry}\label{inv:tonde2025wp}\textcolor{metgray}{82} & \nociteinv{tonde2025wp}\textbf{Tonde (2025)}, \emph{Thesis} \newline
{\small\color{metgray} 1960--2020, 52.9M papers. $\cdot$ Data: MAG} \newline
``Our results indicate a steady decline in CD index scores, suggesting that more recent publications tend to reinforce existing knowledge frameworks rather than introduce transformative innovations.'' (Figs. 3.14--3.19). & {\color{DeclineRed}\Large$\bm{\downarrow}$} \\
\addlinespace[1.2em]

\setcounter{inventry}{80}\refstepcounter{inventry}\label{inv:vaughn2025jps}\textcolor{metgray}{81} & \nociteinv{vaughn2025jps}\textbf{\href{https://doi.org/10.1016/j.jpedsurg.2025.162338}{\color{black}Vaughn et al. (2025)}}, \emph{\emph{J Pediatr Surg}} \newline
{\small\color{metgray} Fetal surgery, 1975--2021. $\cdot$ Data: MAG, PubMed} \newline
``While disruption across scientific fields has been on the decline since the turn of the millennium\ldots fetal surgery tells a different story. Here, disruption is not only thriving but also increasing over time'' (Fig. 2). This exception---in a young, evolving field---suggests the decline observed elsewhere is not a mechanical artifact, since such artifacts would appear uniformly across domains. & {\color{IncreaseGreen}\Large$\bm{\uparrow}$} \\
\addlinespace[1.2em]

\setcounter{inventry}{79}\refstepcounter{inventry}\label{inv:wang2025hssc}\textcolor{metgray}{80} & \nociteinv{wang2025hssc}\textbf{\href{https://doi.org/10.1057/s41599-025-06038-6}{\color{black}Wang \& Maynard (2025)}}, \emph{\emph{Humanit Soc Sci Commun}} \newline
{\small\color{metgray} U.S. patents, 1976--2021, 3.7M. $\cdot$ Data: USPTO} \newline
Mean CD5 disruption falls steadily toward zero by 2010 across all gender collaboration types, reinforcing evidence of a broad decline in technological disruption (Fig. 3N). & {\color{DeclineRed}\Large$\bm{\downarrow}$} \\
\addlinespace[1.2em]

\setcounter{inventry}{78}\refstepcounter{inventry}\label{inv:yang2025ipm}\textcolor{metgray}{79} & \nociteinv{yang2025ipm}\textbf{\href{https://doi.org/10.1016/j.ipm.2025.104171}{\color{black}Yang (2025)}}, \emph{\emph{Inf Process Manag}} \newline
{\small\color{metgray} 1950--2020, 1.4M scientists, 27.6M papers. $\cdot$ Data: MAG} \newline
Topic switching is a key driver of disruption; a 1SD increase in switching makes papers 39\% more likely to be disruptive (Fig. 6). Yet ``real switch scores decline over time'' (Figs. 3--5). As switching declines, so does disruption. & {\color{DeclineRed}\Large$\bm{\downarrow}$} \\
\addlinespace[1.2em]

\setcounter{inventry}{77}\refstepcounter{inventry}\label{inv:yang2025jis}\textcolor{metgray}{78} & \nociteinv{yang2025jis}\textbf{\href{https://doi.org/10.1177/01655515251330614}{\color{black}Yang (2025)}}, \emph{\emph{J Inf Sci}} \newline
{\small\color{metgray} 1960--2020, 38.9M papers. $\cdot$ Data: MAG} \newline
Disruptive potential ``drops by 62.5\%, from 0.40 to 0.15'' (Fig. 1f), as ``disruption---measured by how much research challenges existing scientific paradigms---has been on the decline.'' & {\color{DeclineRed}\Large$\bm{\downarrow}$} \\
\addlinespace[1.2em]

\setcounter{inventry}{76}\refstepcounter{inventry}\label{inv:yang2025rp}\textcolor{metgray}{77} & \nociteinv{yang2025rp}\textbf{\href{https://doi.org/10.1016/j.respol.2025.105295}{\color{black}Yang (2025)}}, \emph{\emph{Res Policy}} \newline
{\small\color{metgray} 1980--2017, 6M+ patents. $\cdot$ Data: USPTO} \newline
``The CD index has experienced a steady decline over time'' and ``average CD5 has shown a clear decline since 1980'' (Fig. 4d). The author compares the citation-based CD index to a text-based ``breakthrough'' measure (KI index), finding that the citation-based decline is steady while the text-based measure fluctuates with economic cycles. & {\color{DeclineRed}\Large$\bm{\downarrow}$} \\
\addlinespace[1.2em]

\setcounter{inventry}{75}\refstepcounter{inventry}\label{inv:yang2025jdis}\textcolor{metgray}{76} & \nociteinv{yang2025jdis}\textbf{\href{https://doi.org/10.2478/jdis-2025-0029}{\color{black}Yang et al. (2025)}}, \emph{\emph{J Data Inf Sci}} \newline
{\small\color{metgray} 1950--2016, 29M articles. $\cdot$ Data: MAG, OpenAlex} \newline
Shows disruptiveness (CD index) declined from 0.09 in 1950 to 0.03 in 2016 (Figure 5c). Addresses the ``disruptiveness-citation puzzle'', i.e., highly disruptive papers are increasingly failing to garner high citation impact. & {\color{DeclineRed}\Large$\bm{\downarrow}$} \\
\addlinespace[1.2em]

\setcounter{inventry}{74}\refstepcounter{inventry}\label{inv:yu2025wp}\textcolor{metgray}{75} & \nociteinv{yu2025wp}\textbf{\href{https://arxiv.org/abs/2504.09589v2}{\color{black}Yu et al. (2025)}}, \emph{arXiv} \newline
{\small\color{metgray} 1950--2024, 114M papers. $\cdot$ Data: OECD Patents, OpenAlex, SciSciNet, WoS} \newline
The authors ``find a systematic decline in disruption over time, consistent with prior findings'' (Figs. 2C, 4B). This conclusion holds across multiple large-scale databases, including Web of Science (Figs. S19C, S21B), OpenAlex (Figs. 2C, 4B), and SciSciNet (Figs. S10C, S12B). The study conducts extensive robustness checks to rule out confounds from  citation inflation (Fig. S1, S7) and zero-reference works. Further finds that knowledge independence has declined substantially over time (Fig. 1C), yet ``knowledge independence strongly predicts disruption'' (Fig. 2, Table 1), providing an additional indirect, mechanism-driven explanation for the decline. & {\color{DeclineRed}\Large$\bm{\downarrow}$} \\
\addlinespace[1.2em]

\setcounter{inventry}{73}\refstepcounter{inventry}\label{inv:zhang2025s}\textcolor{metgray}{74} & \nociteinv{zhang2025s}\textbf{\href{https://doi.org/10.1007/s11192-025-05319-w }{\color{black}Zhang et al. (2025)}}, \emph{\emph{Scientometrics}} \newline
{\small\color{metgray} Synchrotron research, 95,720 articles. $\cdot$ Data: OpenAlex} \newline
In models controlling for team characteristics, publication year is a significant negative predictor of disruption (all p$<$0.001). Consistent across DI3 (Table 3) and DI5 robustness checks (Tables 5--6). & {\color{DeclineRed}\Large$\bm{\downarrow}$} \\
\addlinespace[1.2em]

\setcounter{inventry}{72}\refstepcounter{inventry}\label{inv:zheng2025wp}\textcolor{metgray}{73} & \nociteinv{zheng2025wp}\textbf{\href{https://arxiv.org/abs/2509.06212v1}{\color{black}Zheng \& Hou (2025)}}, \emph{arXiv} \newline
{\small\color{metgray} 1960--2020. $\cdot$ Data: SciSciNet} \newline
Identifies four knowledge-production modes; the elite-driven mode, which ``performed most prominently'' on disruption, declines from 59.0\% of papers in the 1960s to 10.4\% in the 2020s (Fig. 4b; Fig. S4). & {\color{DeclineRed}\Large$\bm{\downarrow}$} \\
\addlinespace[1.2em]

\setcounter{inventry}{71}\refstepcounter{inventry}\label{inv:zhou2024wp}\textcolor{metgray}{72} & \nociteinv{zhou2024wp}\textbf{\href{https://doi.org/10.1016/j.eswa.2025.127355 }{\color{black}Zhou et al. (2025)}}, \emph{\emph{Expert Syst Appl}} \newline
{\small\color{metgray} 1976--2022, 8.3M patents. $\cdot$ Data: USPTO} \newline
The authors conclude that ``in terms of scientific value\ldots patents are becoming less disruptive'', attributed to a ``dilution effect.'' & {\color{DeclineRed}\Large$\bm{\downarrow}$} \\
\addlinespace[1.2em]

\\[1.5em]
\multicolumn{3}{c}{\large\textsc{2024} (34 studies)} \\[0.8em]
\setcounter{inventry}{70}\refstepcounter{inventry}\label{inv:brauer2024wp}\textcolor{metgray}{71} & \nociteinv{brauer2024wp}\textbf{Br{\"a}uer (2024)}, \emph{Working Paper} \newline
{\small\color{metgray} Global patents, 1990--2010. $\cdot$ Data: PATSTAT} \newline
Documents declining disruptiveness (Figs. 2, 3), then develops a calibrated model attributing the decline to firms ``consciously preventing disruptive innovation,'' explaining ``52\% of the decline of disruptive innovation until 2010'' (Fig. 6). & {\color{DeclineRed}\Large$\bm{\downarrow}$} \\
\addlinespace[1.2em]

\setcounter{inventry}{69}\refstepcounter{inventry}\label{inv:cao2024ic}\textcolor{metgray}{70} & \nociteinv{cao2024ic}\textbf{\href{https://doi.org/10.1007/978-3-031-57860-1_23}{\color{black}Cao et al. (2024)}}, \emph{iConference} \newline
{\small\color{metgray} Scientometrics research, 6,562 papers. $\cdot$ Data: OpenAlex, WoS} \newline
Disruptiveness declines across all four major topics, with ``a decreasing trend in the disruptive indicator\ldots suggesting a diminishing generation of novel knowledge'' (Figs. 4b, 7).  & {\color{DeclineRed}\Large$\bm{\downarrow}$} \\
\addlinespace[1.2em]

\setcounter{inventry}{68}\refstepcounter{inventry}\label{inv:coles2024wp}\textcolor{metgray}{69} & \nociteinv{coles2024wp}\textbf{Coles (2024)}, \emph{PsyArXiv} \newline
{\small\color{metgray} 1811--2015, 29M papers. $\cdot$ Data: MAG} \newline
``Big team science efforts tend to build upon---vs. disrupt---pre-existing lines of thinking'' (Fig. 3). Rising prevalence of big teams (Fig. 1) implies compositional shift away from disruption. & {\color{DeclineRed}\Large$\bm{\downarrow}$} \\
\addlinespace[1.2em]

\setcounter{inventry}{67}\refstepcounter{inventry}\label{inv:copara2024wp}\textcolor{metgray}{68} & \nociteinv{copara2024wp}\textbf{\href{https://arxiv.org/abs/2406.18792v4}{\color{black}Copara et al. (2024)}}, \emph{arXiv} \newline
{\small\color{metgray} Biomedicine, 2014--2021, 33M papers. $\cdot$ Data: PubMed} \newline
Develops a novel disruptiveness measure for concepts. Concludes ``disruptiveness, influence, and usefulness decrease over time'' (Fig. 2); holds statistically across all top-level MeSH categories (p$<$0.001). Authors note, ``Park and colleagues also corroborate the decreasing trend of disruptiveness, although their analysis was based on scientific articles, in contrast to ours, which is based on MeSH concepts.'' & {\color{DeclineRed}\Large$\bm{\downarrow}$} \\
\addlinespace[1.2em]

\setcounter{inventry}{66}\refstepcounter{inventry}\label{inv:fernandez2024villaverde}\textcolor{metgray}{67} & \nociteinv{fernandez2024villaverde}\textbf{Fern{\'a}ndez-Villaverde et al. (2024)}, \emph{Working Paper} \newline
{\small\color{metgray} 1970--2019. $\cdot$ Data: Census, Compustat, USPTO} \newline
Provides evidence of declining disruption (``creative destruction'') using economic indicators (e.g., firm entry) rather than bibliometrics (Fig. 1b). The authors explicitly state that ``creative destruction\ldots [has] fallen substantially in the US over the past twenty years'', confirming the trend is visible in market behavior independent of citation data. & {\color{DeclineRed}\Large$\bm{\downarrow}$} \\
\addlinespace[1.2em]

\setcounter{inventry}{65}\refstepcounter{inventry}\label{inv:he2024awp}\textcolor{metgray}{66} & \nociteinv{he2024awp}\textbf{\href{https://arxiv.org/abs/2412.19662v1}{\color{black}He et al. (2024)}}, \emph{arXiv} \newline
{\small\color{metgray} 1980--2010, 3.4M patents. $\cdot$ Data: USPTO} \newline
``The average CD index displays a downward trend over time'' with field-level declines 79.2\%--99.6\% (Fig. 3a). & {\color{DeclineRed}\Large$\bm{\downarrow}$} \\
\addlinespace[1.2em]

\setcounter{inventry}{64}\refstepcounter{inventry}\label{inv:holst2024wp}\textcolor{metgray}{65} & \nociteinv{holst2024wp}\textbf{\href{https://arxiv.org/abs/2402.14583v1}{\color{black}Holst et al. (2024)}}, \emph{arXiv} \newline
{\small\color{metgray} Papers and patents (artifact critique). $\cdot$ Data: DBLP, SciSciNet, USPTO} \newline
Using models explicitly designed to correct alleged biases, the authors still find negative and statistically significant year effects on disruptiveness for papers (p$<$0.01) and patents (p$<$0.01) across all specifications (Table S1). & {\color{DeclineRed}\Large$\bm{\downarrow}$} \\
\addlinespace[1.2em]

\setcounter{inventry}{63}\refstepcounter{inventry}\label{inv:huang2024wp}\textcolor{metgray}{64} & \nociteinv{huang2024wp}\textbf{\href{https://arxiv.org/abs/2404.14228v1}{\color{black}Huang \& Li (2024)}}, \emph{arXiv} \newline
{\small\color{metgray} MOEA/D research, 2009--2022. $\cdot$ Data: WoS} \newline
Using text- and citation-based indicators of disruption, the authors find that ``MOEA/D research is becoming less disruptive over time'' (Fig. 15). & {\color{DeclineRed}\Large$\bm{\downarrow}$} \\
\addlinespace[1.2em]

\setcounter{inventry}{62}\refstepcounter{inventry}\label{inv:kedrick2024nhb}\textcolor{metgray}{63} & \nociteinv{kedrick2024nhb}\textbf{\href{https://doi.org/10.1038/s41562-024-01957-x}{\color{black}Kedrick et al. (2024)}}, \emph{\emph{Nat Hum Behav}} \newline
{\small\color{metgray} Physics and social sciences, 1965--2016, 2.6M papers. $\cdot$ Data: APS, WoS} \newline
Conceptual churn predicts disruption ($\beta$ $\approx$ 0.17--0.33, p$<$0.001). Core concept churn declined 15.96--16.82\% (Fig. 2). ``Increasing rigidity'' suggests ``innovative potential of a field declines over time.'' & {\color{DeclineRed}\Large$\bm{\downarrow}$} \\
\addlinespace[1.2em]

\setcounter{inventry}{61}\refstepcounter{inventry}\label{inv:lee2024wp}\textcolor{metgray}{62} & \nociteinv{lee2024wp}\textbf{\href{https://arxiv.org/abs/2410.04493v1}{\color{black}Lee et al. (2024)}}, \emph{arXiv} \newline
{\small\color{metgray} U.S. case law, 1800--2000, 3.87M U.S. cases. $\cdot$ Data: Caselaw Access Project} \newline
Reports that D fell from 0.807 to 0.003 (Fig. 3). Authors conclude ``the legal system is becoming less disruptive.'' & {\color{DeclineRed}\Large$\bm{\downarrow}$} \\
\addlinespace[1.2em]

\setcounter{inventry}{60}\refstepcounter{inventry}\label{inv:li2024pnas}\textcolor{metgray}{61} & \nociteinv{li2024pnas}\textbf{\href{https://doi.org/10.1073/pnas.2322462121}{\color{black}Li et al. (2024)}}, \emph{\emph{PNAS}} \newline
{\small\color{metgray} Scientific careers, cohorts 1960--1990. $\cdot$ Data: APS, SciSciNet} \newline
Holding productivity constant, later cohorts are systematically less disruptive---1990 entrants underperform 1960 cohorts at the same output levels (Figs. 2C, S1C, Table S1); ``the proportion of new papers that continue existing scientific ideas increases and the proportion of disruptive ideas decreases.'' & {\color{DeclineRed}\Large$\bm{\downarrow}$} \\
\addlinespace[1.2em]

\setcounter{inventry}{59}\refstepcounter{inventry}\label{inv:li2024wp}\textcolor{metgray}{60} & \nociteinv{li2024wp}\textbf{\href{https://doi.org/10.48550/arXiv.2402.16839}{\color{black}Li et al. (2024)}}, \emph{arXiv} \newline
{\small\color{metgray} 1965--2020, 41M papers. $\cdot$ Data: MAG} \newline
Disruptive papers ``declined from 50\% to 25\%'' (Fig. 1a, Fig. S3a). Authors address artifact concerns and conclude the trend ``is unlikely to be a pure artifact'' (S2.6). & {\color{DeclineRed}\Large$\bm{\downarrow}$} \\
\addlinespace[1.2em]

\setcounter{inventry}{58}\refstepcounter{inventry}\label{inv:macher2024rp}\textcolor{metgray}{59} & \nociteinv{macher2024rp}\textbf{\href{https://doi.org/10.1016/j.respol.2024.104992}{\color{black}Macher et al. (2024)}}, \emph{\emph{Res Policy}} \newline
{\small\color{metgray} U.S. patents, 1976--2016, 5.3M patents. $\cdot$ Data: USPTO} \newline
The authors evaluate trends in disruptiveness in patents after adding in pre-1976 references and patent applications. Finds that the decline persists, though more modestly, and a rebound in the number of highly disruptive patents after 2008, especially in IT (Figs. 5--6).  & {\color{DeclineRed}\Large$\bm{\downarrow}$\hspace{-3pt}}{\color{IncreaseGreen}\Large$\bm{\uparrow}$} \\
\addlinespace[1.2em]

\setcounter{inventry}{57}\refstepcounter{inventry}\label{inv:martini2024wp}\textcolor{metgray}{58} & \nociteinv{martini2024wp}\textbf{\href{https://doi.org/10.17863/CAM.111428}{\color{black}Martini et al. (2024)}}, \emph{Working Paper} \newline
{\small\color{metgray} Database software, 1983--1993, 51 patents. $\cdot$ Data: Patbase} \newline
Finds ``the average technological disruptiveness\ldots decreased over the years. The same trend can be found for radicalness and originality.'' Concludes ``the inventive step of the patents appears to decrease continuously'' (Fig. 2). & {\color{DeclineRed}\Large$\bm{\downarrow}$} \\
\addlinespace[1.2em]

\setcounter{inventry}{56}\refstepcounter{inventry}\label{inv:petersen2024qss}\textcolor{metgray}{57} & \nociteinv{petersen2024qss}\textbf{\href{https://doi.org/10.1162/qss_a_00333}{\color{black}Petersen et al. (2024)}}, \emph{\emph{Quant Sci Stud}} \newline
{\small\color{metgray} 1945--2012, 30M papers. $\cdot$ Data: MAG} \newline
Replicates decline in disruptiveness reported by other studies, but attributes it to ``citation inflation.'' Computational model reproduces decline. However, the simulation is not compared to empirical trends, and whether decline persists after the proposed citation inflation adjustments are made remains unclear. & {\color{DeclineRed}\Large$\bm{\downarrow}$} \\
\addlinespace[1.2em]

\setcounter{inventry}{55}\refstepcounter{inventry}\label{inv:risha2024wp}\textcolor{metgray}{56} & \nociteinv{risha2024wp}\textbf{\href{https://arxiv.org/abs/2304.14518v2}{\color{black}Risha et al. (2024)}}, \emph{arXiv} \newline
{\small\color{metgray} 1960--2020, 2.3M scientists, 29.7M articles. $\cdot$ Data: MAG} \newline
Documents ``the death of Renaissance scientist''---proportion of generalists dropped from 55\% to 40\%. Since ``generalist teams are 31\% more likely to produce highly disruptive papers'' while ``specialist teams' likelihood of producing highly disruptive work declines by 22\%,``rising specialization contributes to declining disruption. & {\color{DeclineRed}\Large$\bm{\downarrow}$} \\
\addlinespace[1.2em]

\setcounter{inventry}{54}\refstepcounter{inventry}\label{inv:shinichi2024wp}\textcolor{metgray}{55} & \nociteinv{shinichi2024wp}\textbf{\href{https://doi.org/10.2139/ssrn.4706101}{\color{black}Shinichi \& Matsui (2024)}}, \emph{Working Paper} \newline
{\small\color{metgray} Visual Art (Ukiyo-e), 11,000 images. $\cdot$ Data: Ukiyo-e images} \newline
Analyzing visual content (VGGNet features) rather than citations, this study finds that ``the overall creativity of Ukiyo-e has been in decline since the late 1840s.'' The Disruption Index drops over time (Figs. 11--12), with early masters scoring systematically higher than later generations (Tables 4--5), indicating that measurement of declining disruption in creative domains is not an artifact of citation data.  & {\color{DeclineRed}\Large$\bm{\downarrow}$} \\
\addlinespace[1.2em]

\setcounter{inventry}{53}\refstepcounter{inventry}\label{inv:singh2024epjds}\textcolor{metgray}{54} & \nociteinv{singh2024epjds}\textbf{\href{https://doi.org/10.1140/epjds/s13688-024-00451-8}{\color{black}Singh et al. (2024)}}, \emph{\emph{EPJ Data Sci}} \newline
{\small\color{metgray} 1992--2018, 1.5M arXiv articles. $\cdot$ Data: arXiv} \newline
``Explorers'' more disruptive than ``exploiters'' (Fig. 5, p$<$0.001). Exploration peaks mid-career; ``senior scientists tend to exploit more'' (Fig. 4c). Combined with workforce aging, consistent with declining disruption. & {\color{DeclineRed}\Large$\bm{\downarrow}$} \\
\addlinespace[1.2em]

\setcounter{inventry}{52}\refstepcounter{inventry}\label{inv:sixt2024qss}\textcolor{metgray}{53} & \nociteinv{sixt2024qss}\textbf{\href{https://doi.org/10.1162/qss_a_00328}{\color{black}Sixt \& Pasin (2024)}}, \emph{\emph{Quant Sci Stud}} \newline
{\small\color{metgray} 1945--2010, 79M papers. $\cdot$ Data: Dimensions} \newline
Replicates decline using data from Dimensions database, across all fields of science, using CD index. Authors ``reproduce the decline of disruptive papers over time, which is one of the central results of Park et al. (2023)'' (Fig. 5). & {\color{DeclineRed}\Large$\bm{\downarrow}$} \\
\addlinespace[1.2em]

\setcounter{inventry}{51}\refstepcounter{inventry}\label{inv:tang2024ic}\textcolor{metgray}{52} & \nociteinv{tang2024ic}\textbf{\href{https://doi.org/10.1007/978-3-031-57860-1_27}{\color{black}Tang et al. (2024)}}, \emph{iConference} \newline
{\small\color{metgray} AI research, 1950--2019 (646,100 papers). $\cdot$ Data: MAG} \newline
The ``average disruptiveness percentile\ldots significantly and monotonically declines'' from 1950--2014, but ``since 2014\ldots [it] exhibit[s] startling upward trends'' as ``AI steps into the deep learning stage'' (Fig. 2b). & {\color{DeclineRed}\Large$\bm{\downarrow}$\hspace{-3pt}}{\color{IncreaseGreen}\Large$\bm{\uparrow}$} \\
\addlinespace[1.2em]

\setcounter{inventry}{50}\refstepcounter{inventry}\label{inv:tang2024joi}\textcolor{metgray}{51} & \nociteinv{tang2024joi}\textbf{\href{https://doi.org/10.1016/j.joi.2024.101570}{\color{black}Tang et al. (2024)}}, \emph{\emph{J Informetr}} \newline
{\small\color{metgray} AI researchers, 1961--2023, 665K papers. $\cdot$ Data: Scopus} \newline
Uses a novel, semantic-disruptiveness measure---``Disruptive Knowledge Content'' (Dkc) index---thereby avoiding concerns with citation data). Finds ``papers were becoming less disruptive over time,'' and that ``scholars who started their scientific research earlier tend to have higher disruptiveness'' (Fig. 6). & {\color{DeclineRed}\Large$\bm{\downarrow}$} \\
\addlinespace[1.2em]

\setcounter{inventry}{49}\refstepcounter{inventry}\label{inv:wang2024joi}\textcolor{metgray}{50} & \nociteinv{wang2024joi}\textbf{\href{https://doi.org/10.1016/j.joi.2024.101493}{\color{black}Wang et al. (2024)}}, \emph{\emph{J Informetr}} \newline
{\small\color{metgray} 3D printing patents, 1997--2019. $\cdot$ Data: PatSnap} \newline
Peak disruption in early years; ``disruption degrees of nodes in the main path show an overall decreasing trend'' (Fig. 10).  & {\color{DeclineRed}\Large$\bm{\downarrow}$} \\
\addlinespace[1.2em]

\setcounter{inventry}{48}\refstepcounter{inventry}\label{inv:wang2024eist}\textcolor{metgray}{49} & \nociteinv{wang2024eist}\textbf{\href{https://doi.org/10.1016/j.eist.2024.100836}{\color{black}Wang et al. (2024)}}, \emph{\emph{Environ Innov Soc Transit}} \newline
{\small\color{metgray} Climate mitigation, 1988--2017, 248K patents. $\cdot$ Data: USPTO} \newline
``CCMT patents parallel the pattern of utility patents, showing a marked decline [in disruption]'' (Fig. 5f).  & {\color{DeclineRed}\Large$\bm{\downarrow}$} \\
\addlinespace[1.2em]

\setcounter{inventry}{47}\refstepcounter{inventry}\label{inv:wu2024s}\textcolor{metgray}{48} & \nociteinv{wu2024s}\textbf{\href{https://doi.org/10.1007/s11192-024-05126-9}{\color{black}Wu et al. (2024)}}, \emph{\emph{Scientometrics}} \newline
{\small\color{metgray} Pharmaceuticals, 1940s--2018, 1.9M patents. $\cdot$ Data: MAG, PATSTAT} \newline
Shows technological disruptiveness declining (Fig. 3); authors note ``technological disruptiveness is diminishing over time.'' & {\color{DeclineRed}\Large$\bm{\downarrow}$} \\
\addlinespace[1.2em]

\setcounter{inventry}{46}\refstepcounter{inventry}\label{inv:yang2024bqss}\textcolor{metgray}{47} & \nociteinv{yang2024bqss}\textbf{\href{https://doi.org/10.1162/qss_a_00323}{\color{black}Yang \& Deng (2024)}}, \emph{\emph{Quant Sci Stud}} \newline
{\small\color{metgray} 1887--2010, 712 Nobel papers. $\cdot$ Data: MAG} \newline
Shows disruptiveness (CD index) declining (Fig. 2c); ``even for scientific breakthroughs, their disruptive potential diminishes over time.'' & {\color{DeclineRed}\Large$\bm{\downarrow}$} \\
\addlinespace[1.2em]

\setcounter{inventry}{45}\refstepcounter{inventry}\label{inv:yang2024joi}\textcolor{metgray}{46} & \nociteinv{yang2024joi}\textbf{\href{https://doi.org/10.1016/j.joi.2024.101588}{\color{black}Yang \& Wang (2024)}}, \emph{\emph{J Informetr}} \newline
{\small\color{metgray} Medicine, 1980--2021, 5M+ papers. $\cdot$ Data: MAG} \newline
Team freshness declining---from 0.41 to 0.13 (Fig. 3g); freshness of newcomers down 77\%. Freshness positively associated with disruption (Table 1, p$<$0.001). Declining freshness consistent with declining disruption. & {\color{DeclineRed}\Large$\bm{\downarrow}$} \\
\addlinespace[1.2em]

\setcounter{inventry}{44}\refstepcounter{inventry}\label{inv:yang2024wp}\textcolor{metgray}{45} & \nociteinv{yang2024wp}\textbf{\href{https://arxiv.org/abs/2311.16360v2}{\color{black}Yang \& Youn (2024)}}, \emph{arXiv} \newline
{\small\color{metgray} U.S. patents, 1981--2010. $\cdot$ Data: Google Patents, USPTO} \newline
Patent ``curvature'' rises over time (Fig. 2c); the authors find that ``inventions arising at high curvatures tend to be less disruptive'' (Table 1: $\beta$ = -0.002, p$<$0.001), implying declining disruption. & {\color{DeclineRed}\Large$\bm{\downarrow}$} \\
\addlinespace[1.2em]

\setcounter{inventry}{43}\refstepcounter{inventry}\label{inv:yang2024aqss}\textcolor{metgray}{44} & \nociteinv{yang2024aqss}\textbf{\href{https://doi.org/10.1162/qss_a_00335}{\color{black}Yang et al. (2024)}}, \emph{\emph{Quant Sci Stud}} \newline
{\small\color{metgray} STEM fields, 1980--2016, 3.2M papers. $\cdot$ Data: MAG} \newline
The probability of producing a disruptive paper (CD$>$0) declines steadily over time, falling from $\sim$25--30\% in 1980 to $\sim$15\% by 2016 for both male- and female-led teams (Fig. 3i). & {\color{DeclineRed}\Large$\bm{\downarrow}$} \\
\addlinespace[1.2em]

\setcounter{inventry}{42}\refstepcounter{inventry}\label{inv:yang2024s}\textcolor{metgray}{43} & \nociteinv{yang2024s}\textbf{\href{https://doi.org/10.1007/s11192-024-04987-4}{\color{black}Yang et al. (2024)}}, \emph{\emph{Scientometrics}} \newline
{\small\color{metgray} 1950--2017, 30M papers. $\cdot$ Data: SciSciNet} \newline
Teams grow older and less age-diverse over time (Fig. 1), and both shifts are associated with lower disruption (Table 4). & {\color{DeclineRed}\Large$\bm{\downarrow}$} \\
\addlinespace[1.2em]

\setcounter{inventry}{41}\refstepcounter{inventry}\label{inv:yang2023s}\textcolor{metgray}{42} & \nociteinv{yang2023s}\textbf{\href{https://doi.org/10.1007/s11192-023-04889-x}{\color{black}Yang et al. (2024)}}, \emph{\emph{Scientometrics}} \newline
{\small\color{metgray} Physics, 1980--2010, 300K papers. $\cdot$ Data: APS} \newline
Original DI shows declining trend (Fig. 7c). Rescaled (RDI) flat by design---removes temporal trends.  & {\color{DeclineRed}\Large$\bm{\downarrow}$} \\
\addlinespace[1.2em]

\setcounter{inventry}{40}\refstepcounter{inventry}\label{inv:yu2024s}\textcolor{metgray}{41} & \nociteinv{yu2024s}\textbf{\href{https://doi.org/10.1007/s11192-024-05184-z }{\color{black}Yu et al. (2024)}}, \emph{\emph{Scientometrics}} \newline
{\small\color{metgray} Quantum information, Nanomaterials, and Synthetic biology, 2000--2022, 1M papers. $\cdot$ Data: WoS} \newline
Rising interdisciplinary diversity is negatively associated with disruptiveness (Fig. 2), consistent with declining disruption overall, though field-level trends vary. Quantum information and Synthetic biology decline modestly (with a brief rebound around 2013); Nanomaterials spikes in 2002--2003 before stabilizing (Supp. Fig. 3). & {\color{MixedBlue}\Large$\bm{\updownarrow}$} \\
\addlinespace[1.2em]

\setcounter{inventry}{39}\refstepcounter{inventry}\label{inv:zhang2024joi}\textcolor{metgray}{40} & \nociteinv{zhang2024joi}\textbf{\href{https://doi.org/10.1016/j.joi.2024.101520 }{\color{black}Zhang et al. (2024)}}, \emph{\emph{J Informetr}} \newline
{\small\color{metgray} Physics papers, 1925--2016, 137K papers; robustness in Sociology and Psychology. $\cdot$ Data: APS} \newline
Year-disruption correlation: r=-0.102, p$<$0.001 (Table 3). Publication year remains negative predictor in regressions controlling for team characteristics (Tables 4--5). Sociology shows stronger decline ($\beta$=-0.191); Psychology mixed. & {\color{DeclineRed}\Large$\bm{\downarrow}$} \\
\addlinespace[1.2em]

\setcounter{inventry}{38}\refstepcounter{inventry}\label{inv:zhang2024wp}\textcolor{metgray}{39} & \nociteinv{zhang2024wp}\textbf{\href{https://arxiv.org/abs/2405.14543v2}{\color{black}Zhang et al. (2024)}}, \emph{arXiv} \newline
{\small\color{metgray} 1950s--2010s, 300K+ careers. $\cdot$ Data: SciSciNet} \newline
``Scientists nowadays make fewer disruptive contributions than those in the early years'' (Figs. 7c--d).  & {\color{DeclineRed}\Large$\bm{\downarrow}$} \\
\addlinespace[1.2em]

\setcounter{inventry}{37}\refstepcounter{inventry}\label{inv:zhang2024s}\textcolor{metgray}{38} & \nociteinv{zhang2024s}\textbf{\href{https://doi.org/10.1007/s11192-024-05134-9 }{\color{black}Zhang et al. (2024)}}, \emph{\emph{Scientometrics}} \newline
{\small\color{metgray} Chinese listed firm patents. $\cdot$ Data: Chinese patents} \newline
Compares D5 to Park et al. (2023): ``results indicate high consistency and correlation\ldots consistent [with an] overall downward trend'' (Fig. 3). Extends declining disruption evidence beyond U.S. to Chinese patent data. & {\color{DeclineRed}\Large$\bm{\downarrow}$} \\
\addlinespace[1.2em]

\\[1.5em]
\multicolumn{3}{c}{\large\textsc{2023} (17 studies)} \\[0.8em]
\setcounter{inventry}{36}\refstepcounter{inventry}\label{inv:bentley2023acs}\textcolor{metgray}{37} & \nociteinv{bentley2023acs}\textbf{\href{https://doi.org/10.1142/s0219525923500066}{\color{black}Bentley et al. (2023)}}, \emph{\emph{Adv Complex Syst}} \newline
{\small\color{metgray} Life sciences, 1930--2015. $\cdot$ Data: PubMed} \newline
Observes steep decline 1930--1950. Disruption ``close to zero for the 50 years after about 1970'' with ``modest but consistent growth\ldots  since 2000'' (Fig. 2). & {\color{DeclineRed}\Large$\bm{\downarrow}$\hspace{-3pt}}{\color{IncreaseGreen}\Large$\bm{\uparrow}$} \\
\addlinespace[1.2em]

\setcounter{inventry}{35}\refstepcounter{inventry}\label{inv:brauer2023wp}\textcolor{metgray}{36} & \nociteinv{brauer2023wp}\textbf{Br{\"a}uer (2023)}, \emph{Working Paper} \newline
{\small\color{metgray} Global patents, 1980--2010. $\cdot$ Data: PATSTAT} \newline
Using novel disruption metric (Citation Year Gap), tracks ``aggregate evolution of disruptive innovation'', showing CYG ``declines from -1 to almost -5'' across IPC classes, with the decline ``widespread'' (Figs. 2, 3). Concludes ``technology fields trend even further towards incrementalism.'' & {\color{DeclineRed}\Large$\bm{\downarrow}$} \\
\addlinespace[1.2em]

\setcounter{inventry}{34}\refstepcounter{inventry}\label{inv:gebhart2023wp}\textcolor{metgray}{35} & \nociteinv{gebhart2023wp}\textbf{\href{https://arxiv.org/abs/2308.16363v1}{\color{black}Gebhart \& Funk (2023)}}, \emph{arXiv} \newline
{\small\color{metgray} Physics, 1893--2019. $\cdot$ Data: APS} \newline
Measurement-robust decline. Using 7 mathematically distinct disruption metrics, the authors find a consistent downward trend on 6 of 7 measures (Fig. 8), ruling out metric-specific artifacts. & {\color{DeclineRed}\Large$\bm{\downarrow}$} \\
\addlinespace[1.2em]

\setcounter{inventry}{33}\refstepcounter{inventry}\label{inv:kaltenberg2023rp}\textcolor{metgray}{34} & \nociteinv{kaltenberg2023rp}\textbf{\href{https://doi.org/10.1016/j.respol.2022.104629}{\color{black}Kaltenberg et al. (2023)}}, \emph{\emph{Res Policy}} \newline
{\small\color{metgray} U.S. patents, 1976--2017, 3M patents, 1.2M inventors. $\cdot$ Data: USPTO} \newline
Inventor age negatively associated with disruption; ``disruptiveness of the average inventor's patents from their twenties is roughly three standard deviations greater than the average from their 60s and 70s'' (Fig. 9). Inventor workforce is also aging, consistent with declining disruptiveness. & {\color{DeclineRed}\Large$\bm{\downarrow}$} \\
\addlinespace[1.2em]

\setcounter{inventry}{32}\refstepcounter{inventry}\label{inv:leahey2023asr}\textcolor{metgray}{33} & \nociteinv{leahey2023asr}\textbf{\href{https://doi.org/10.1177/00031224231168074}{\color{black}Leahey et al. (2023)}}, \emph{\emph{Am Sociol Rev}} \newline
{\small\color{metgray} Citation Classics, 1931--1987, 2.5K papers. $\cdot$ Data: WoS} \newline
``Publication year is inversely related to the CD index, so recent papers tend to be less disruptive than older papers'' (Table 3). Robust to field, journal, author controls.  & {\color{DeclineRed}\Large$\bm{\downarrow}$} \\
\addlinespace[1.2em]

\setcounter{inventry}{31}\refstepcounter{inventry}\label{inv:lee2023wp}\textcolor{metgray}{32} & \nociteinv{lee2023wp}\textbf{\href{https://arxiv.org/abs/2308.14724v1}{\color{black}Lee \& Kim (2023)}}, \emph{arXiv} \newline
{\small\color{metgray} Marketing, 1978--2022, 8.6K articles. $\cdot$ Data: OpenAlex} \newline
Conceptual research is more disruptive (Fig. 2, p<.01 for JM, JMR) ``but is declining in publication quantity.'' Table 3 provides direct evidence of declining disruption---year fixed effects increasingly negative relative to 1978--1982 (p<.01). & {\color{DeclineRed}\Large$\bm{\downarrow}$} \\
\addlinespace[1.2em]

\setcounter{inventry}{30}\refstepcounter{inventry}\label{inv:lin2023n}\textcolor{metgray}{31} & \nociteinv{lin2023n}\textbf{\href{https://doi.org/10.1038/s41586-023-06767-1}{\color{black}Lin et al. (2023)}}, \emph{\emph{Nature}} \newline
{\small\color{metgray} 20M papers, 4M patents. $\cdot$ Data: MAG, USPTO} \newline
Reports declining disruptiveness; decline accelerates post-2000 (Ext. Data Tables 1--2). Remote collaboration rising; remote teams less likely to engage in ``early-stage, conceptual tasks,'' contributing to decline. Effect persists within same individuals, pronounced for junior researchers.  & {\color{DeclineRed}\Large$\bm{\downarrow}$} \\
\addlinespace[1.2em]

\setcounter{inventry}{29}\refstepcounter{inventry}\label{inv:park2023n}\textcolor{metgray}{30} & \nociteinv{park2023n}\textbf{\href{https://doi.org/10.1038/s41586-022-05543-x}{\color{black}Park et al. (2023)}}, \emph{\emph{Nature}} \newline
{\small\color{metgray} 45M papers (1945--2010), 3.9M patents (1976--2010). $\cdot$ Data: USPTO, WoS} \newline
Disruptiveness declines over time across papers and patents (Fig. 2a--b). ``Science and technology are becoming less disruptive.'' Demonstrates robustness on 4 additional datasets (JSTOR, APS, MAG, PubMed), Monte Carlo simulations, alternative metrics (ED Figs. 6--8). Provides corroborating evidence using text-based (non-citation) metrics. Highly disruptive works stable in absolute terms (Fig. 4). & {\color{DeclineRed}\Large$\bm{\downarrow}$} \\
\addlinespace[1.2em]

\setcounter{inventry}{28}\refstepcounter{inventry}\label{inv:sheng2023s}\textcolor{metgray}{29} & \nociteinv{sheng2023s}\textbf{\href{https://doi.org/10.1007/s11192-023-04751-0}{\color{black}Sheng et al. (2023)}}, \emph{\emph{Scientometrics}} \newline
{\small\color{metgray} Biomedicine, 2001--2010, 1.3M articles. $\cdot$ Data: PubMed} \newline
Reports that disruption decreased from 2001 to 2010 (Fig. 3). ``The maximum value shows a descending trend. Based on different citation windows, the interquartile range\ldots also decreases gradually, indicating that the discrete degree of the D index gradually decreases each year.'' & {\color{DeclineRed}\Large$\bm{\downarrow}$} \\
\addlinespace[1.2em]

\setcounter{inventry}{27}\refstepcounter{inventry}\label{inv:shu2023wp}\textcolor{metgray}{28} & \nociteinv{shu2023wp}\textbf{\href{https://arxiv.org/abs/2312.05927v1}{\color{black}Shu \& Pan (2023)}}, \emph{arXiv} \newline
{\small\color{metgray} 1960--2010, 25M papers. $\cdot$ Data: MAG} \newline
Using a novel text-based measure, the authors find that ``stylized papers [i.e., work that deviates from established paradigms] exhibit a higher likelihood of demonstrating disruption'' (Fig. 3), yet there is a ``concerning decline in stylized research'' over time (Fig. 1). Science now actively suppresses styled work; stylized papers face longer review times and receive fewer citations (Fig. 3), consistent with declining disruptiveness. & {\color{DeclineRed}\Large$\bm{\downarrow}$} \\
\addlinespace[1.2em]

\setcounter{inventry}{26}\refstepcounter{inventry}\label{inv:spinellis2023pone}\textcolor{metgray}{27} & \nociteinv{spinellis2023pone}\textbf{\href{https://doi.org/10.1371/journal.pone.0294946}{\color{black}Spinellis (2023)}}, \emph{\emph{PLoS ONE}} \newline
{\small\color{metgray} 1945--2016, 50.9M papers. $\cdot$ Data: Crossref} \newline
Using novel data, shows that CD index drops from 0.134 to 0.001 (Fig. 2). ``The fall in the CD index is in line with recently published research reporting that papers are becoming less disruptive over time.'' Reported correlation with Park et al.: $\rho$ = 0.95, p$<$0.001. & {\color{DeclineRed}\Large$\bm{\downarrow}$} \\
\addlinespace[1.2em]

\setcounter{inventry}{25}\refstepcounter{inventry}\label{inv:zutphen2023wp}\textcolor{metgray}{26} & \nociteinv{zutphen2023wp}\textbf{Van Zutphen (2023)}, \emph{Thesis} \newline
{\small\color{metgray} Marketing and AI, 1980--2020. $\cdot$ Data: WoS} \newline
Both fields exhibit periods of declining disruptiveness, but ``while AI papers still appear to have declined in disruptiveness in the most recent years, Marketing papers appear to have reversed their downward trend'' (Fig. 13). & {\color{MixedBlue}\Large$\bm{\updownarrow}$} \\
\addlinespace[1.2em]

\setcounter{inventry}{24}\refstepcounter{inventry}\label{inv:wang2023wp}\textcolor{metgray}{25} & \nociteinv{wang2023wp}\textbf{\href{https://doi.org/10.2139/ssrn.4923505}{\color{black}Wang (2023)}}, \emph{Working Paper} \newline
{\small\color{metgray} Biomedicine and life sciences, 1950--2021, 8.7M papers. $\cdot$ Data: PubMed} \newline
Using a citation-based disruption measure, the author finds that ``the disruptive nature of discoveries has declined'' over time, for both average and ``star'' (top 5\%) papers (Fig. 3). & {\color{DeclineRed}\Large$\bm{\downarrow}$} \\
\addlinespace[1.2em]

\setcounter{inventry}{23}\refstepcounter{inventry}\label{inv:wu2023joi}\textcolor{metgray}{24} & \nociteinv{wu2023joi}\textbf{\href{https://doi.org/10.1016/j.joi.2023.101436}{\color{black}Wu et al. (2023)}}, \emph{\emph{J Informetr}} \newline
{\small\color{metgray} Economics (ML vs. non-ML), 2001--2020. $\cdot$ Data: MAG} \newline
Using a standard citation-based disruption index, ML-related economics papers are significantly more disruptive than non-ML work, with the gap widening after 2010 (Fig. 7). This pattern suggests that disruption can increase under methodological regime shifts, providing evidence against claims that observed declines are purely mechanical. & {\color{IncreaseGreen}\Large$\bm{\uparrow}$} \\
\addlinespace[1.2em]

\setcounter{inventry}{22}\refstepcounter{inventry}\label{inv:yang2023ipm}\textcolor{metgray}{23} & \nociteinv{yang2023ipm}\textbf{\href{https://doi.org/10.1016/j.ipm.2023.103420}{\color{black}Yang et al. (2023)}}, \emph{\emph{Inf Process Manag}} \newline
{\small\color{metgray} Physics, 1893--2010, 463K papers. $\cdot$ Data: APS} \newline
Uses Disruptive Citation measure (resistant to reference-count manipulation). Shows decline from $\sim$1960s onward (Fig. B1f). & {\color{DeclineRed}\Large$\bm{\downarrow}$} \\
\addlinespace[1.2em]

\setcounter{inventry}{21}\refstepcounter{inventry}\label{inv:yang2023issip}\textcolor{metgray}{22} & \nociteinv{yang2023issip}\textbf{Yang et al. (2023)}, \emph{ISSI} \newline
{\small\color{metgray} Information science, 1978--1989. $\cdot$ Data: WoS} \newline
Using Global Disruption Index (GDI), designed to address limitations of previous measures, reports ``average disruptiveness has declined\ldots papers are less likely to be disruptive'' (Fig. 5). Further, also observe that ``early-published papers tend to be more disruptive as they are more likely to challenge existing scientific paradigms.'' & {\color{DeclineRed}\Large$\bm{\downarrow}$} \\
\addlinespace[1.2em]

\setcounter{inventry}{20}\refstepcounter{inventry}\label{inv:zeng2023wp}\textcolor{metgray}{21} & \nociteinv{zeng2023wp}\textbf{\href{https://arxiv.org/abs/2305.03589v1}{\color{black}Zeng et al. (2023)}}, \emph{arXiv} \newline
{\small\color{metgray} Physics and five other disciplines. $\cdot$ Data: AMiner, APS, MAG} \newline
Documents sharp decline in disruptiveness over time, especially among highly cited papers---``disruptive papers in science are losing impact.'' Fraction with positive disruption (D5$>$0) also declines over time. & {\color{DeclineRed}\Large$\bm{\downarrow}$} \\
\addlinespace[1.2em]

\\[1.5em]
\multicolumn{3}{c}{\large\textsc{2022} (8 studies)} \\[0.8em]
\setcounter{inventry}{19}\refstepcounter{inventry}\label{inv:abu-omar2022cr}\textcolor{metgray}{20} & \nociteinv{abu-omar2022cr}\textbf{\href{https://doi.org/10.1016/j.crad.2022.07.003}{\color{black}Abu-Omar et al. (2022)}}, \emph{\emph{Clin Radiol}} \newline
{\small\color{metgray} Radiology, 1954--2014. $\cdot$ Data: PubMed} \newline
Downward trend (Fig. 3). ``Most papers published showed negligible or negative disruption scores.'' & {\color{DeclineRed}\Large$\bm{\downarrow}$} \\
\addlinespace[1.2em]

\setcounter{inventry}{18}\refstepcounter{inventry}\label{inv:becerra2022dcr}\textcolor{metgray}{19} & \nociteinv{becerra2022dcr}\textbf{\href{https://doi.org/10.1097/dcr.0000000000002118}{\color{black}Becerra et al. (2022)}}, \emph{\emph{Dis Colon Rectum}} \newline
{\small\color{metgray} Colorectal surgery, 1954--2014. $\cdot$ Data: PubMed} \newline
``The decade with the most articles was 1980 to 1989 and the decade with the least\ldots was 2000 to 2009'' (Fig. 4).  & {\color{DeclineRed}\Large$\bm{\downarrow}$} \\
\addlinespace[1.2em]

\setcounter{inventry}{17}\refstepcounter{inventry}\label{inv:bessen2022disruption}\textcolor{metgray}{18} & \nociteinv{bessen2022disruption}\textbf{\href{https://doi.org/10.2307/j.ctv2jn91q7.6}{\color{black}Bessen (2022)}}, \emph{Yale Univ Press} \newline
{\small\color{metgray} U.S. industries, 1970s--2010s. $\cdot$ Data: Compustat, NETS} \newline
Documents ``declining industrial disruption'' using non-citation-based measure. ``The likelihood that a top firm in any industry will be displaced by a rival is less than half of what it was in the late 1990s,'' as ``the rate of disruption sharply reversed trend around 2000'' (Fig. 1). & {\color{DeclineRed}\Large$\bm{\downarrow}$} \\
\addlinespace[1.2em]

\setcounter{inventry}{16}\refstepcounter{inventry}\label{inv:chen2022wp}\textcolor{metgray}{17} & \nociteinv{chen2022wp}\textbf{Chen et al. (2022)}, \emph{Working Paper} \newline
{\small\color{metgray} Scientific teams, 1961--2020. $\cdot$ Data: MAG} \newline
The authors show that ``the percentage of fundamental or disruptive scientific discoveries has declined'' in tandem with the rise of distributed teams (Fig. 1). While improved ICT mitigates this effect after 2010, the long-run decline in disruption persists. & {\color{DeclineRed}\Large$\bm{\downarrow}$} \\
\addlinespace[1.2em]

\setcounter{inventry}{15}\refstepcounter{inventry}\label{inv:cui2022wp}\textcolor{metgray}{16} & \nociteinv{cui2022wp}\textbf{\href{https://arxiv.org/abs/2202.04044v1}{\color{black}Cui et al. (2022)}}, \emph{arXiv} \newline
{\small\color{metgray} 1960--2015, 244M scholars. $\cdot$ Data: MAG, WoS} \newline
Shows the 1990s cohort had lower disruption than the 1960s at the same career age (Fig. S7B). ``As scientists age\ldots research is less likely to disrupt.'' Also shows the workforce is aging, further corroborating declining disruptiveness (Fig. S5). & {\color{DeclineRed}\Large$\bm{\downarrow}$} \\
\addlinespace[1.2em]

\setcounter{inventry}{14}\refstepcounter{inventry}\label{inv:peters2022wp}\textcolor{metgray}{15} & \nociteinv{peters2022wp}\textbf{Peters \& Walsh (2022)}, \emph{Working Paper} \newline
{\small\color{metgray} 1980--2012. $\cdot$ Data: Census LBD} \newline
Provides evidence of declining disruption using economic rather than bibliometric data. ``The decline in population growth has contributed to\ldots a decline in firm entry and in the rate of `creative destruction''' (Fig. 1, right panel). This suggests the trend is visible in market dynamics independent of citation metrics. & {\color{DeclineRed}\Large$\bm{\downarrow}$} \\
\addlinespace[1.2em]

\setcounter{inventry}{13}\refstepcounter{inventry}\label{inv:singh2022pone}\textcolor{metgray}{14} & \nociteinv{singh2022pone}\textbf{\href{https://doi.org/10.1371/journal.pone.0270131}{\color{black}Singh et al. (2022)}}, \emph{\emph{PLoS ONE}} \newline
{\small\color{metgray} 1986--2018, 175 fields, 1.5M arXiv articles. $\cdot$ Data: arXiv} \newline
Disruption highest in early ``creation'' stages, declines as fields mature (Fig. 4). ``Disruptiveness [is] fading.''  & {\color{DeclineRed}\Large$\bm{\downarrow}$} \\
\addlinespace[1.2em]

\setcounter{inventry}{12}\refstepcounter{inventry}\label{inv:you2022joi}\textcolor{metgray}{13} & \nociteinv{you2022joi}\textbf{\href{https://doi.org/10.1016/j.joi.2022.101294}{\color{black}You et al. (2022)}}, \emph{\emph{J Informetr}} \newline
{\small\color{metgray} 1996--2018, 48.6M papers. $\cdot$ Data: Scopus} \newline
Shows ``market share of questionable [predatory] publications has steadily increased'' (Fig. S1) and that ``questionable publications were less disruptive\ldots than their counterparts'' ($D \sim 0.05$ vs $\sim 0.14$, Fig. 4C), providing compositional evidence for declining disruption. & {\color{DeclineRed}\Large$\bm{\downarrow}$} \\
\addlinespace[1.2em]

\\[1.5em]
\multicolumn{3}{c}{\large\textsc{2021} (7 studies)} \\[0.8em]
\setcounter{inventry}{11}\refstepcounter{inventry}\label{inv:bornmann2021joi}\textcolor{metgray}{12} & \nociteinv{bornmann2021joi}\textbf{\href{https://doi.org/10.1016/j.joi.2021.101159}{\color{black}Bornmann \& Tekles (2021)}}, \emph{\emph{J Informetr}} \newline
{\small\color{metgray} Physics, 1980--2002. $\cdot$ Data: WoS} \newline
Validates multiple disruption indices against expert milestones. Notes ``a time dependency of disruption, with high disruptive values for older milestone papers'' (Fig. 3).  & {\color{DeclineRed}\Large$\bm{\downarrow}$} \\
\addlinespace[1.2em]

\setcounter{inventry}{10}\refstepcounter{inventry}\label{inv:chu2021pnas}\textcolor{metgray}{11} & \nociteinv{chu2021pnas}\textbf{\href{https://doi.org/10.1073/pnas.2021636118}{\color{black}Chu \& Evans (2021)}}, \emph{\emph{PNAS}} \newline
{\small\color{metgray} 1960--2014, 90M papers. $\cdot$ Data: WoS} \newline
Disruptive share drops from $\sim$49\% to 13\% as fields grow (Fig. 4). Concludes, ``newly published papers become unlikely to disrupt existing work.'' & {\color{DeclineRed}\Large$\bm{\downarrow}$} \\
\addlinespace[1.2em]

\setcounter{inventry}{9}\refstepcounter{inventry}\label{inv:grunvald2021apjcp}\textcolor{metgray}{10} & \nociteinv{grunvald2021apjcp}\textbf{\href{https://doi.org/10.31557/apjcp.2021.22.8.2385}{\color{black}Grunvald et al. (2021)}}, \emph{\emph{Asian Pac J Cancer Prev}} \newline
{\small\color{metgray} Breast cancer, 1954--2014. $\cdot$ Data: PubMed} \newline
Reports that the median publication years of the top 100 is 1977. ``Disruptive studies skewed towards being older'' (Fig. 3).  & {\color{DeclineRed}\Large$\bm{\downarrow}$} \\
\addlinespace[1.2em]

\setcounter{inventry}{8}\refstepcounter{inventry}\label{inv:hansdorfer2021prsgo}\textcolor{metgray}{9} & \nociteinv{hansdorfer2021prsgo}\textbf{\href{https://doi.org/10.1097/gox.0000000000003446}{\color{black}Hansdorfer et al. (2021)}}, \emph{\emph{Plast Reconstr Surg Glob Open}} \newline
{\small\color{metgray} Plastic surgery, 1954--2014. $\cdot$ Data: PubMed} \newline
The authors find the 1980s had the most disruptive papers, while the 2000s had the least (Fig. 4).  & {\color{DeclineRed}\Large$\bm{\downarrow}$} \\
\addlinespace[1.2em]

\setcounter{inventry}{7}\refstepcounter{inventry}\label{inv:horen2021jcs}\textcolor{metgray}{8} & \nociteinv{horen2021jcs}\textbf{\href{https://doi.org/10.1097/scs.0000000000007804}{\color{black}Horen et al. (2021)}}, \emph{\emph{J Craniofac Surg}} \newline
{\small\color{metgray} Craniofacial surgery, 1954--2014. $\cdot$ Data: PubMed} \newline
Finds among top disruptive publications, decade with the most was 1980s, 2000s was least (Fig. 4). 1960s highest average disruptiveness; 2000s the lowest.  & {\color{DeclineRed}\Large$\bm{\downarrow}$} \\
\addlinespace[1.2em]

\setcounter{inventry}{6}\refstepcounter{inventry}\label{inv:sullivan2021jps}\textcolor{metgray}{7} & \nociteinv{sullivan2021jps}\textbf{\href{https://doi.org/10.1016/j.jpedsurg.2021.02.002}{\color{black}Sullivan et al. (2021)}}, \emph{\emph{J Pediatr Surg}} \newline
{\small\color{metgray} Pediatric surgery, 1954--2014. $\cdot$ Data: PubMed} \newline
Among highly disruptive articles, ``1970s and 1980s were the most represented decades'' with decline thereafter (Fig. 3).  & {\color{DeclineRed}\Large$\bm{\downarrow}$} \\
\addlinespace[1.2em]

\setcounter{inventry}{5}\refstepcounter{inventry}\label{inv:zeng2021nhb}\textcolor{metgray}{6} & \nociteinv{zeng2021nhb}\textbf{\href{https://doi.org/10.1038/s41562-021-01084-x }{\color{black}Zeng et al. (2021)}}, \emph{\emph{Nat Hum Behav}} \newline
{\small\color{metgray} Physics, 1893--2010, 480K+ papers. $\cdot$ Data: APS} \newline
Studies disruption and multidisciplinary impact, finding that ``both indexes decrease with time.'' (Supp. Fig. 6). & {\color{DeclineRed}\Large$\bm{\downarrow}$} \\
\addlinespace[1.2em]

\\[1.5em]
\multicolumn{3}{c}{\large\textsc{2020} (2 studies)} \\[0.8em]
\setcounter{inventry}{4}\refstepcounter{inventry}\label{inv:bessen2020wp}\textcolor{metgray}{5} & \nociteinv{bessen2020wp}\textbf{\href{https://doi.org/10.2139/ssrn.3682745}{\color{black}Bessen et al. (2020)}}, \emph{Working Paper} \newline
{\small\color{metgray} U.S. industries, 1976--2017. $\cdot$ Data: Compustat, NETS} \newline
Non-citation measure shows ``declining industrial disruption.'' Firm displacement hazards ``rose from 1970 until the late 1990s'' then ``sharply reversed'' around 2000, with ``substantial declines in the displacement rate'' (Fig. 1). & {\color{DeclineRed}\Large$\bm{\downarrow}$} \\
\addlinespace[1.2em]

\setcounter{inventry}{3}\refstepcounter{inventry}\label{inv:falcao2020pismirc}\textcolor{metgray}{4} & \nociteinv{falcao2020pismirc}\textbf{\href{https://doi.org/10.5281/zenodo.4245355}{\color{black}Falc{\~a}o et al. (2020)}}, \emph{ISMIR} \newline
{\small\color{metgray} Forró music, 1945--2016, 27,352 songs. $\cdot$ Data: Forró music} \newline
Measuring disruption from raw audio features, the study shows that ``the higher disruptions of Forró are mostly concentrated on its first years\ldots between 1960 and 1970,'' with consolidation thereafter (Fig. 8). Because this approach avoids citation data entirely, the finding suggests that declining disruption in creative domains is not driven by citation-based artifacts. & {\color{DeclineRed}\Large$\bm{\downarrow}$} \\
\addlinespace[1.2em]

\\[1.5em]
\multicolumn{3}{c}{\large\textsc{2019} (2 studies)} \\[0.8em]
\setcounter{inventry}{2}\refstepcounter{inventry}\label{inv:bowen2019wp}\textcolor{metgray}{3} & \nociteinv{bowen2019wp}\textbf{\href{https://doi.org/10.2139/ssrn.3478503}{\color{black}Bowen et al. (2019)}}, \emph{Working Paper} \newline
{\small\color{metgray} U.S. patents, 1930--2010. $\cdot$ Data: USPTO} \newline
Text-based disruption. The ``average disruptive potential of U.S. patents has markedly declined since 1950,'' falling to ``roughly one quarter'' of its peak by 2010, even after adjusting for technology class and location (Fig. 2); the decline in technological disruptiveness indicates a widespread deceleration in vocabulary usage growth rates.'' & {\color{DeclineRed}\Large$\bm{\downarrow}$} \\
\addlinespace[1.2em]

\setcounter{inventry}{1}\refstepcounter{inventry}\label{inv:wu2019n}\textcolor{metgray}{2} & \nociteinv{wu2019n}\textbf{\href{https://doi.org/10.1038/s41586-019-0941-9}{\color{black}Wu et al. (2019)}}, \emph{\emph{Nature}} \newline
{\small\color{metgray} 1954--2014, 65M+ papers, patents, and software. $\cdot$ Data: GitHub, USPTO, WoS} \newline
Decline in disruption observed across all team sizes over time. ``Earlier cohorts are more disruptive than later cohorts'' (ED Fig. 3c). Small teams are more disruptive but have declined in prevalence (ED Fig. 8).  & {\color{DeclineRed}\Large$\bm{\downarrow}$} \\
\addlinespace[1.2em]

\\[1.5em]
\multicolumn{3}{c}{\large\textsc{2017} (1 studies)} \\[0.8em]
\setcounter{inventry}{0}\refstepcounter{inventry}\label{inv:funk2017ms}\textcolor{metgray}{1} & \nociteinv{funk2017ms}\textbf{\href{https://doi.org/10.1287/mnsc.2015.2366}{\color{black}Funk \& Owen-Smith (2017)}}, \emph{\emph{Manage Sci}} \newline
{\small\color{metgray} U.S. patents, 1977--2005, 2.9M patents. $\cdot$ Data: USPTO} \newline
Reports a decline in disruptiveness using the CD index, noting ``the average patent of today is less destabilizing than those of the late 1970s and early 1980s.'' Also finds ``significant (all p$<$0.001) negative correlations across categories between application and grant year and the CD5 index.'' & {\color{DeclineRed}\Large$\bm{\downarrow}$}

\end{longtable}

\end{document}